\documentclass{article}

\usepackage{arxiv}

\usepackage[utf8]{inputenc} 
\usepackage[T1]{fontenc}    
\usepackage{hyperref}       
\usepackage{url}            
\usepackage{booktabs}       
\usepackage{amsfonts}       
\usepackage{nicefrac}       
\usepackage{microtype}      
\usepackage{lipsum}
\usepackage{graphicx}
\graphicspath{ {./images/} }

\usepackage{amsmath}
\usepackage{bm}
\usepackage[subrefformat=parens,labelformat=parens]{subfig}
\usepackage{multirow}
\usepackage{comment}
\usepackage{overpic}
\usepackage{color}
\usepackage{xspace}

\newcommand{\openfoam}{Open\nolinebreak\hspace{-.2em}\nolinebreak\hspace{-.2em}FOAM\textsuperscript{\textregistered}\xspace}

\title{Data-driven reduced order modelling for patient-specific hemodynamics of coronary artery bypass grafts with physical and geometrical parameters}

\author{
 Pierfrancesco Siena \\
  MathLab, Mathematics Area, \\
  SISSA International School for Advanced Studies, \\
    Via Bonomea, 265, 34136, Trieste, Italy,\\
  \texttt{psiena@sissa.it} \\
   \And
 Michele Girfoglio \\
  MathLab, Mathematics Area, \\
  SISSA International School for Advanced Studies, \\
    Via Bonomea, 265, 34136, Trieste, Italy,\\
  \texttt{mgifogl@sissa.it} \\
  \And
 Francesco Ballarin \\
 Department of Mathematics and Physics,\\
 Catholic University of the Sacred Heart,\\
 Via Garzetta 48, 25133, Brescia, Italy, \\
  \texttt{francesco.ballarin@unicatt.it} \\
  \And
 Gianluigi Rozza \\
  MathLab, Mathematics Area, \\
  SISSA International School for Advanced Studies, \\
    Via Bonomea, 265, 34136, Trieste, Italy,\\
  \texttt{grozza@sissa.it} \\
}

\begin{document}
\maketitle
\begin{abstract}
In this work the development of a machine learning-based Reduced Order Model (ROM) for the investigation of hemodynamics in a patient-specific configuration of Coronary Artery Bypass Graft (CABG) is proposed.
 The computational domain is referred to left branches of coronary arteries  when a stenosis of the Left Main Coronary Artery  (LMCA) occurs.
The method extracts a reduced basis space from a collection of high-fidelity solutions via a Proper Orthogonal Decomposition (POD) algorithm and employs Artificial Neural Networks (ANNs) for the computation of the modal coefficients.
The Full Order Model (FOM) is represented by the incompressible Navier-Stokes equations discretized using a Finite Volume (FV) technique. 
Both physical and geometrical parametrization are taken into account, the former one related to the inlet flow rate and the latter one related to the stenosis severity. With respect to the previous works focused on the development of a ROM
framework for the evaluation of coronary artery disease, the novelties of our study include the use of the FV method in a patient-specific configuration, the use of a data-driven ROM technique and the mesh deformation strategy based on a Free Form Deformation (FFD) technique. 
The performance of our ROM approach is analyzed in terms of the error between full order and reduced order solutions as well as the speedup achieved at the online stage.

\end{abstract}

\section{Introduction and motivation}
\label{intro}
Coronary artery diseases are one of the main causes of death worldwide. When they occur, one or more coronary arteries are occluded causing a poor perfusion of oxygen-rich blood to the heart, leading to clinical complications such as heart attack and heart failure. Coronary artery bypass grafts (CABGs) surgery is still one of the most used procedures worldwide, although after some years blood supply often fails again, causing the need for reintervention. Many papers  \cite{Revault,Scott,Harling,Gaudino,Rosenblum,Rastan} study different CABG configurations in order to establish a good clinical treatment, especially in term of mid-long survival. 
However an isolated stenosis of the main trunk is rare, and there is not a sample of consolidated studies which allow to establish the most appropriate procedure to perform in this case. \\
Computational Fluid Dynamics (CFD) applied to the cardiovascular system \cite{Formaggia} represents a research area of significant importance and in recent times it had a strong impulse due to the increasing demand from the medical community for quantitative investigations of cardiovascular diseases.
For the problem at hand, a better understanding of the blood flow behaviour in grafts and graft junctions could aid in surgical planning of grafting and improve the lifetime of grafts. \\
High-fidelity numerical methods, among which finite element (FE) and finite volume (FV) methods, often referred to as full order models (FOMs), are commonly used for the solution of parameterized Navier-Stokes equations governing the blood flow dynamics, where the parameters are geometric features, boundary conditions and/or physical properties. However, for applications which require repeated model evaluations over a range of parameter values, FOMs are very expensive in terms of computational time and memory demand due to the large amount of degrees of freedom to be considered for a proper description of the flow system. In such a framework, reduced order models (ROMs) 
\cite{Rozza,rozza2008reduced,Huynh2012,degruyter,noor1981,quarternoniandrozza,manzoni2012,deparisandrozza} are applied to enable fast computations varying the parameters, as often required in the clinical context \cite{Ballarin,Lassila,Faggiano,Manzoni,Manzoni2,Pitton,BMQR,Zainib,Fevola}. \\
A further step forward in this scenario has been given by the development and diffusion of intelligent technologies. 
Many recent works use machine learning as an alternative to CFD simulations in order to reproduce hemodynamic parameters \cite{Mao,Kissas,Liang,Gharleghi,Su}. In all these works, CFD, that is used to compute the data set for training and testing, is seen as a black box whilst the machine learning algorithms are used to detect a nonlinear manifold that maps CFD inputs to their corresponding outputs of interest. 
Neural networks can in theory represent any functional relationship between inputs and outputs. However, many applications remain unexplored and this reinforces the need to carry out further studies in this area.

The main goal of this work is to propose a partnership between neural networks, ROM and CFD with the aim to lower the computational cost of the numerical simulations and at the same time to provide accurate predictions of the blood flow behaviour. In particular, the Proper Orthogonal Decomposition-Artificial Neural Network (POD-ANN) method \cite{Hesthaven,Hesthaven2}, where the POD is employed for the computation of the reduced basis and feedforward ANNs are adopted for the evaluation of the modal coefficients, is used to reconstruct in a fast and reliable way both primal variables (pressure and velocity) and derived quantities (the wall shear stress (WSS)). The method is applied to the investigation of hemodynamics in a CABG patient-specific configuration when an isolated stenosis of the Left Main Coronary Artery (LMCA) occurs at varying of the inlet boundary conditions and of the severy of the stenosis. So we extend what has been done in \cite{Siena} for the time reconstruction to a physical and geometrical settings. 
The POD-ANN method has been successfully used in several application fields ranging between automotive~\cite{Zancanaro2021}, casting \cite{shah2021} and combustion \cite{Hesthaven2}.

With respect to the previous investigations focused on the development of a ROM framework for the evaluation of coronary artery disease, the novelties of our investigation include: 
\begin{itemize}
    \item The use of the FV method in a patient-specific framework. Indeed, although the FV method is also adopted in \cite{Buoso}, the geometric database used consists of idealized geometries, i.e. obtained from the deformation of a three-dimensional straight pipe by means of a discrete empirical interpolation method (DEIM). Other recent works, e.g. \cite{Infantino,Infantino1}, employ the FV method for the development of ROMs for hemodynamic applications but its adoption in this environment is still rather unexplored, so this work contributes to fill this gap. In addition many commercial codes widely used from bioengineering community are based on FV schemes, therefore the combination of ROM and FV methods is particularly appealing in this field.
   \item The use of a data-driven ROM technique. In all the previous works \cite{Buoso,Zainib,Ballarin,Faggiano}, a standard POD-Galerkin method is adopted. Data-driven approaches are based only on data and do not require knowledge about the governing equations that describe the system. They are also non-intrusive, i.e. no modification of the simulation software is carried out. Typically data-driven methods are able to provide a computational speed-up larger than classic projection-based methods \cite{Infantino,Infantino1,pichi2021,papapicco,strazzullo,meneghetti}, so they are to be preferred by allowing real time simulations to be accessed in hospitals and
   operating rooms in a more efficient way. 
   \item The mesh deformation strategy. In \cite{Ballarin,Faggiano}, it is introduced a centerlines-based parametrization to efficiently handle geometrical variations for a wide range of patient-specific configurations of CABGs in a FE environment. This method allows an efficient variation of geometrical quantities of interest, such as stenosis severity. However it is hardly compatible with the FV formulation of the governing equations that is not written in a reference domain setting. Conversely, the Free Form Deformation (FFD) allows to act directly on the mesh. Another mesh deformation strategy consistent with the FV approximation is introduced in \cite{Stabile}: while our approach uses a Non-Uniform Rational Basis Spline (NURBS) parameterization, \cite{Stabile} is based on a Radial Basis Function (RBF) approach. 
   However, while RBF approach deforms the grid as a whole and therefore it does not preserve the original geometry, in the NURBS strategy all the vertices remains on the initial surface. 
   \end{itemize}
The rest of the paper is structured as follows. 
In Section \ref{sec:1}, the computational domain as well as Navier-Stokes equations, representing our FOM, are introduced. Section \ref{sec:2} presents our ROM approach. 
Then the numerical results are reported in Section \ref{sec:3}. Finally, in Section \ref{sec:4} we draw conclusions and discuss future perspectives of the current study.

\section{The full order model}
\label{sec:1}

\subsection{The computational domain}
\label{sec:2.3}
\begin{figure}
    \centering
	 \subfloat[][CABG virtual geometry.\label{CABG1:a}]{\includegraphics[width=.5\textwidth]{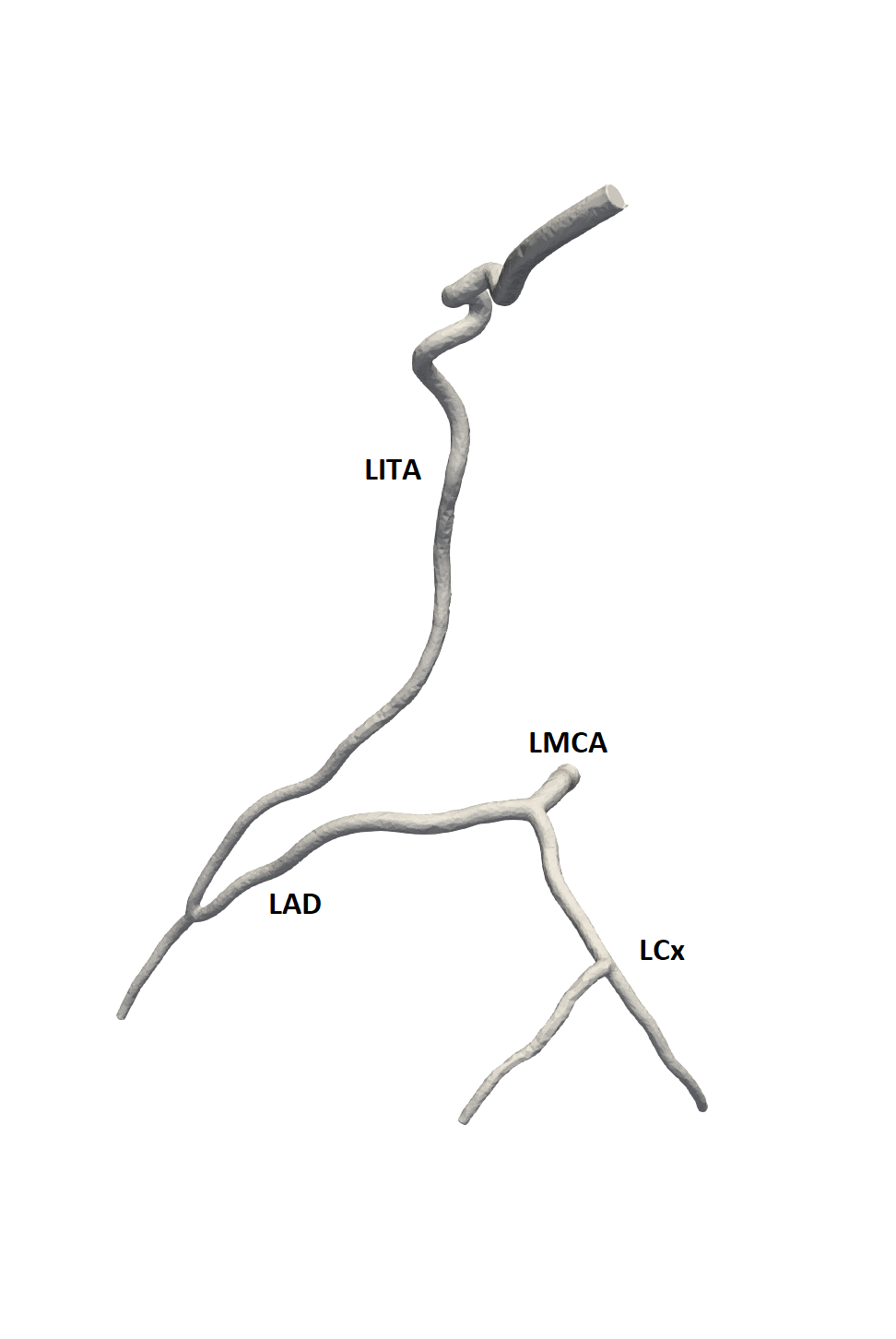}}
	\subfloat[][Internal mesh.\label{CABG1:b}]{\includegraphics[width=.45\textwidth]{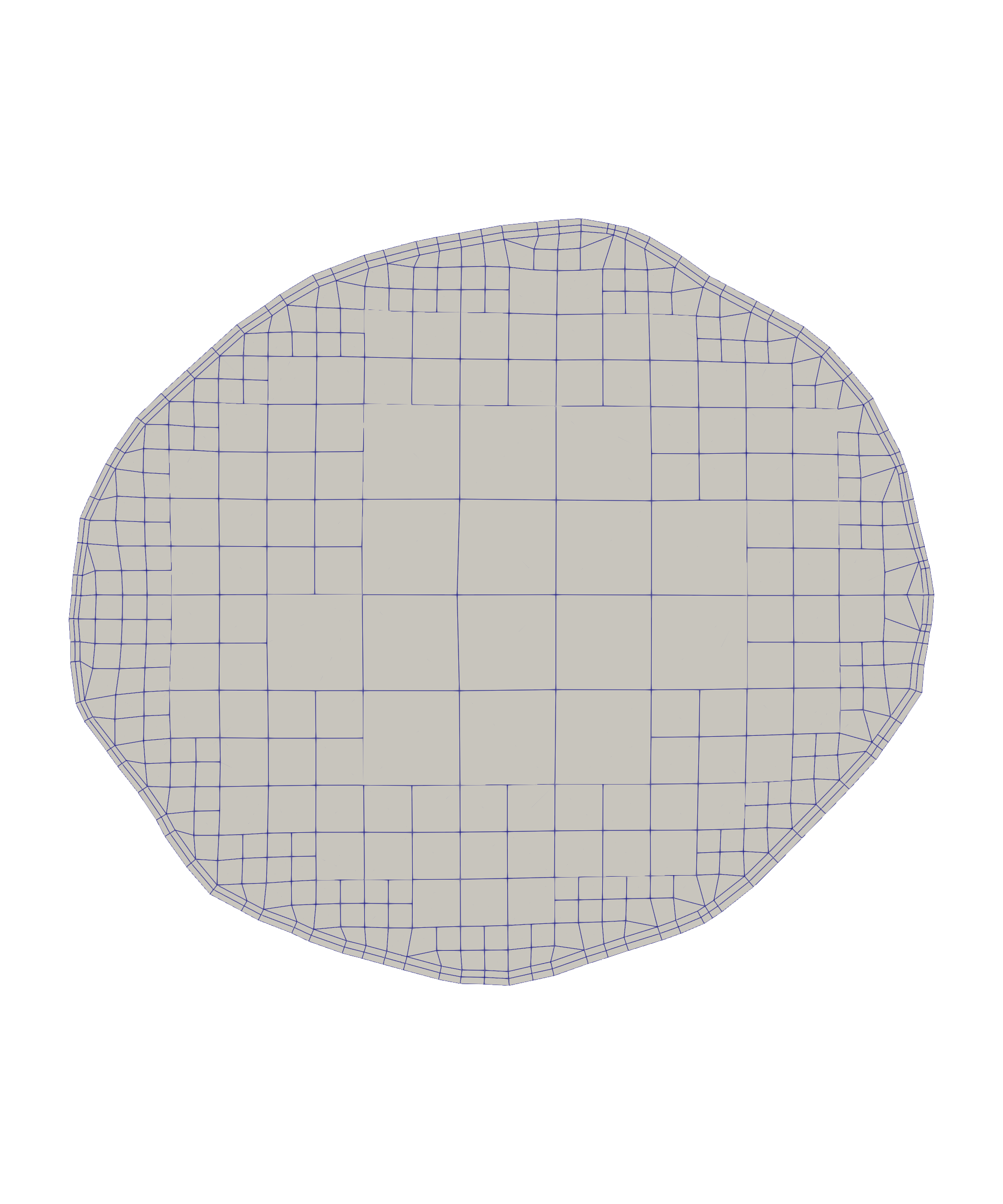}}
	\caption{Three-dimensional reconstruction of the CABG (a) and view of the mesh on an internal section (b).}
\label{CABG1}       
\end{figure}

The virtual geometry related to the CABG configuration at hand is shown in Figure \subref*{CABG1:a}. The model includes, beyond the LMCA, the left internal thoracic artery (LITA), the left anterior descending artery (LAD) and the left circumflex artery (LCx). 
The patient-specific computed tomography (CT) - scan data, from which the virtual geometry has been obtained by using the procedure explained in \cite{Ballarin}, have been provided by Ospedale Luigi Sacco in Milan. The mesh, displayed in Figure \subref*{CABG1:b}, has been built by using the mesh generation utility \emph{snappyHexMesh} available in \openfoam (\url{www.openfoam.org}). 
Its features are shown in Table \ref{grid}.
\begin{table}
\centering
\caption{Features of the mesh. }
\label{grid}       
\begin{tabular}{cccc}
\hline\noalign{\smallskip}
Number of cells & Min/max mesh size [$m$] & Average non-orthogonality [$^\circ$] & Max skewness  \\
\noalign{\smallskip}\hline\noalign{\smallskip}
986.278 & 
3.0e-5 - 4.3e-4
& 12.9 & 2.95 \\
\noalign{\smallskip}\hline
\end{tabular}
\end{table}
To introduce different severity of stenosis in the LMCA in a way that is compatible with the data-driven reduced order model which will be introduced next it is important to warp directly the mesh and not just the geometry, so that the same number of cells is present in all the deformed configurations. At this aim, FFD is performed by means of a NURBS volumetric parameterization \cite{Brujic}. 
The procedure consists of three main steps \cite{Amoiralis,Lamousin}: 
\begin{itemize}
    \item i) A parametric lattice of control points is constructed by means of a structured mesh placed around the region of the LMCA to be deformed. Then the control points are used to define a NURBS volume which contains the LMCA portion to be warped. 

    \item ii) 
    The octree algorithm \cite{Amoiralis} can be used to find a matching between the control points
of the lattice and the points of the computational domain: 
    \begin{enumerate}
        \item the parametric lattice is divided in eight subvolumes; 
        \item the coordinates of the vertices of each subvolume are compared against the coordinates of the portion of LMCA under consideration. This allows to identify the subvolume in which the computational domain is embedded; 
        \item the subvolumes are again divided and the procedure is repeated until a desirable accuracy is achieved.
    \end{enumerate}
    
    \item iii) The coordinates of the control points are modified, so that the parametric volume and consequently the LCMA portion are deformed.
\end{itemize}
\begin{figure}
	\centering
 	\subfloat[][Initial lattice.\label{mimmo:a}]{\includegraphics[width=.33\textwidth]{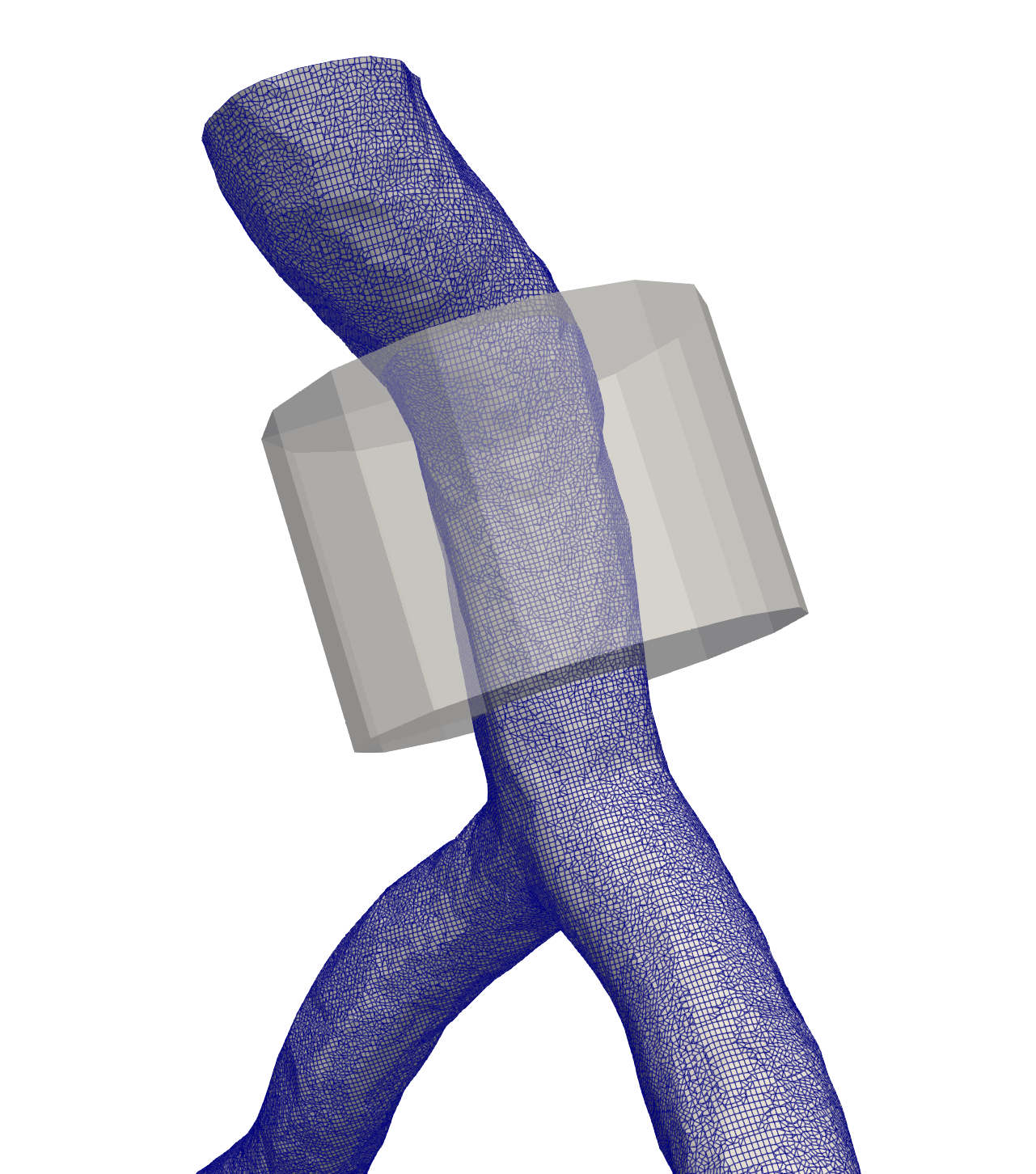}}
	\subfloat[][Deformed lattice.\label{mimmo:b}]{\includegraphics[width=.35\textwidth]{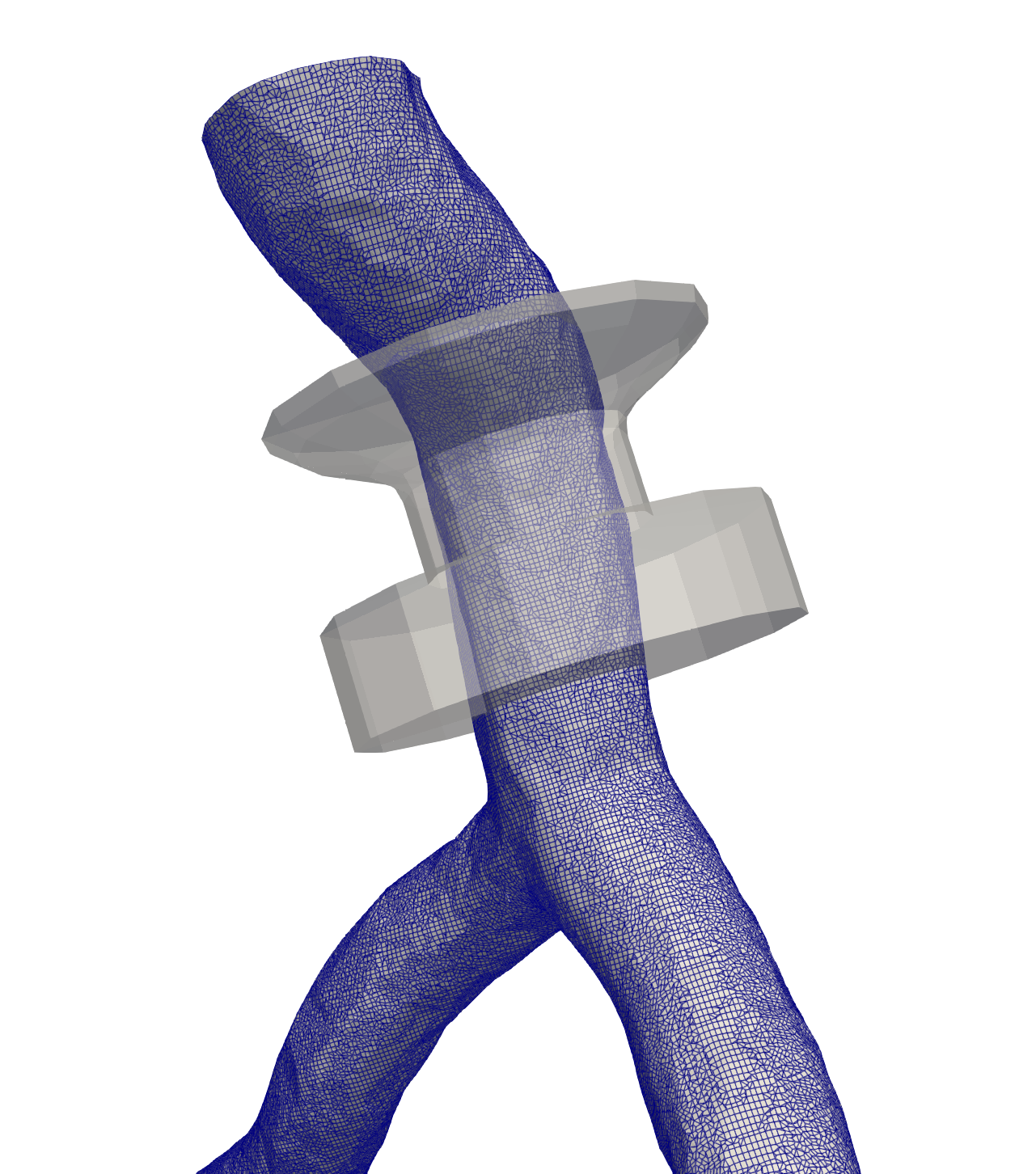}}
	\subfloat[][50\% stenosis.\label{mimmo:c}]{\includegraphics[width=.35\textwidth]{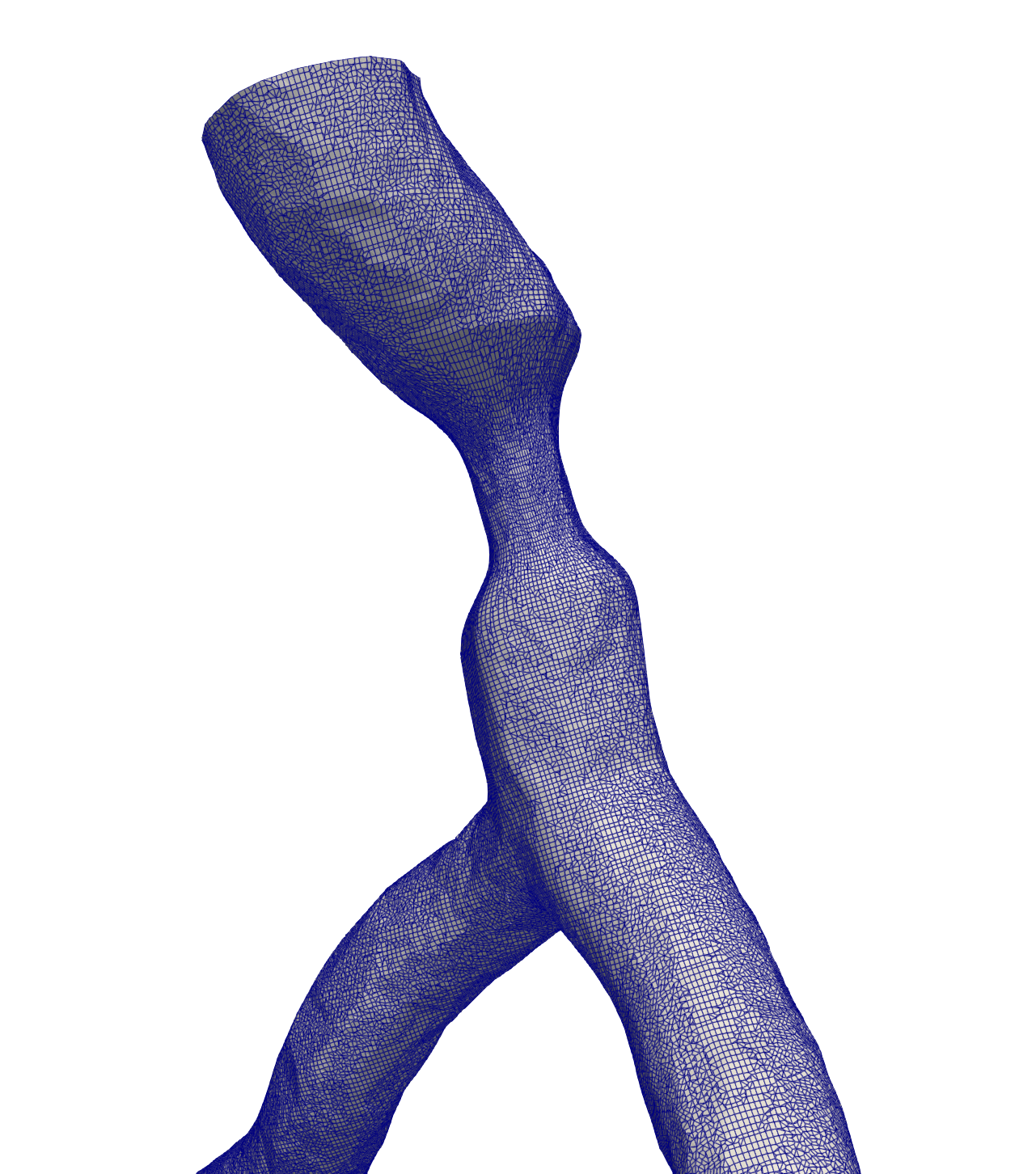}}
	\caption{Introduction of the stenosis in the LMCA using FFD technique.}
	\label{Mimmo}
\end{figure}
In Figure \subref*{mimmo:a} the lattice and the LMCA region where the stenosis is introduced are illustrated in their initial configurations, whilst in Figure \subref*{mimmo:b} and Figure \subref*{mimmo:c} the deformed lattice and a $50\%$ stenosis are displayed, respectively. \\
We highlight that the introduction of the stenosis in the LMCA does not significantly affect the mesh quality. 


\subsection{The Navier-Stokes equations}

Let us consider the dynamics of the blood flow in a patient-specific domain \\ $\Omega (\bm{\mu}) \subset \mathbb{R}^3$ over a cardiac cycle $(0,T]$, when the transient effects are passed: 
\begin{equation}
	\begin{cases} 
	\rho \partial_t \mathbf u (\bm{x}, t; \bm{\mu})+ \rho \nabla \cdot (\mathbf u  (\bm{x}, t; \bm{\mu}) \otimes \mathbf u  (\bm{x}, t; \bm{\mu})) - \nabla \cdot \mathbb T  (\bm{x}, t; \bm{\mu}) = 0, \\ 
	\nabla \cdot \mathbf u  (\bm{x}, t; \bm{\mu}) = 0, 
	\end{cases}  \label{N-S}
\end{equation}	
in $\Omega (\bm{\mu}) \times (0,T]$, where $\rho$ is the density, $\bm{u}$ is the velocity, $\partial_t$ denotes the time derivative and $\mathbb T$ is the Cauchy stress tensor. 
The vector $\bm{\mu} \in \mathcal{P} \subset \mathbb{R}^d$ represents a parameter vector in a $d$-dimensional parameter space $\mathcal{P}$ containing both physical and geometrical parameters of the problem.
For the sake of simplicity, from now on the dependance of the variables on $\bm{x}$, $t$, and $\bm{\mu}$ will be omitted.

In this work, we model the blood as a Newtonian fluid, so the constitutive relation for $\mathbb T$ is given by
	\begin{equation}
		\mathbb T =-p\mathbb I + 2\mu\mathbb{D(\mathbf{u})},
		\label{Newtonian}
	\end{equation}
where $p$ is the pressure, $\mu$ is the dynamic viscosity and $\mathbb{D(\mathbf u)}=\frac{\nabla \mathbf u + \nabla \mathbf u ^T}{2}$ is the strain rate tensor. By using equation \eqref{Newtonian}, the system \eqref{N-S} can be rewritten as 
\begin{equation}
	\begin{cases} 
	\partial_t \mathbf u + \nabla \cdot (\mathbf u \otimes \mathbf u) + \nabla P - \nu \Delta \mathbf{u} = 0, \\ 
	\nabla \cdot \mathbf u = 0, 
	\end{cases}  \label{N-S-2}
\end{equation}
in $\Omega \times (0,T]$, where $P=\frac{p}{\rho}$ is the kinematic pressure, i.e. the pressure divided by the density, and $\nu=\frac{\mu}{\rho}$ is the kinematic viscosity.

We also introduce the Wall Shear Stress (WSS) defined as follows:
\begin{equation}
\text{WSS} = \tau \cdot \mathbf{n},
\end{equation}
where $\tau = \nu (\nabla \mathbf{u} + \nabla \mathbf{u}^T)$ is the stress tensor and $\mathbf{n}$ is the unit normal outward vector.

Finally, in order to characterize the flow regime under consideration, we
define the Reynolds number as
\begin{equation}
Re = \dfrac{U L}{\nu},
\end{equation}
where $U$ and $L$ are characteristic macroscopic velocity and length, respectively. For a
blood flow in a cylindrical vessel, $U$ is the mean sectional velocity and $L$ is the
diameter.

\subsection{Boundary conditions}
\label{boundary_cond}
\begin{figure}
    \centering
	\includegraphics[width=0.8\linewidth]{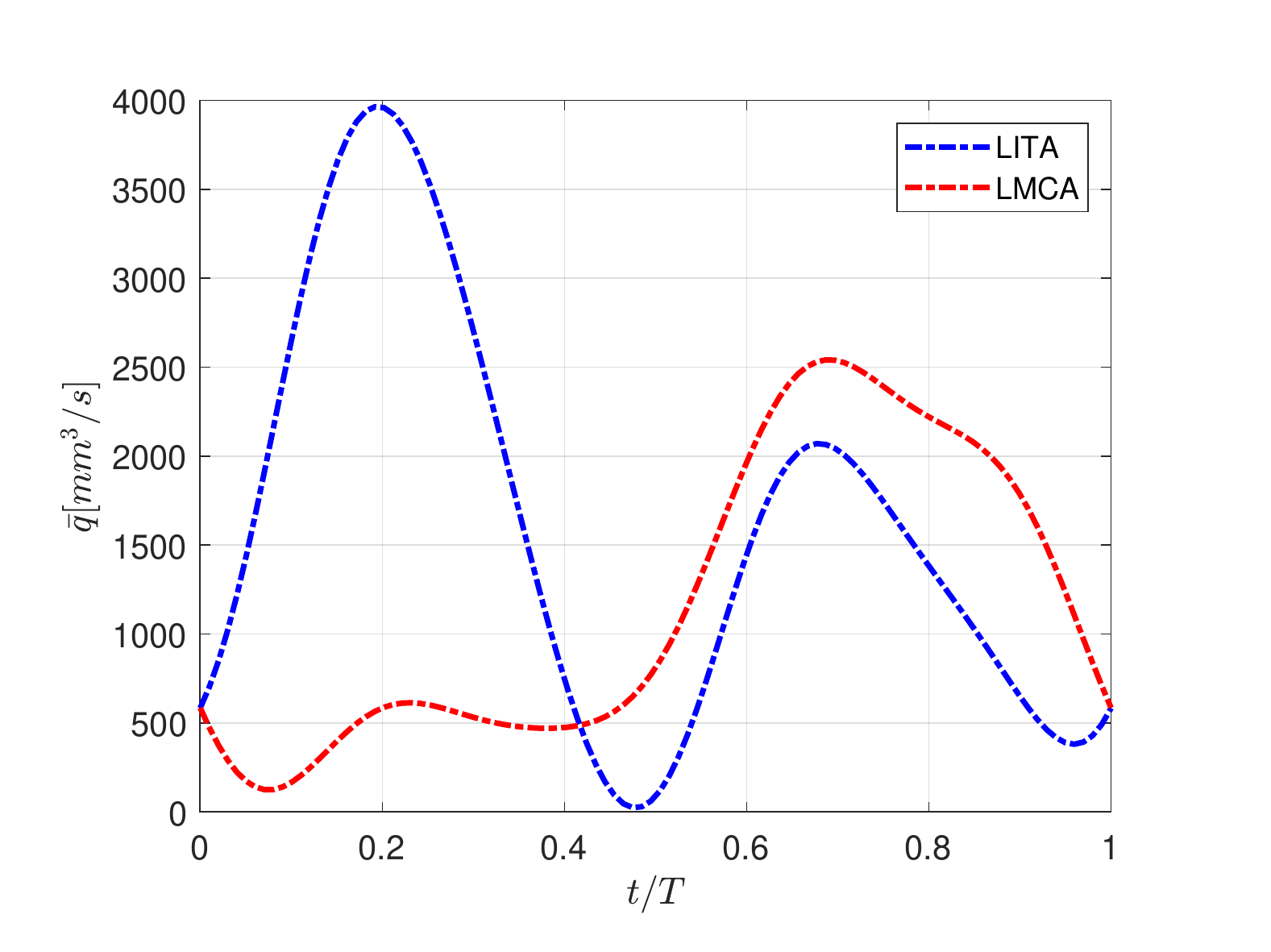}	
	\caption{Time evolution over the cardiac cycle of $\bar{q}_{\text{LMCA}}$ and $\bar{q}_{\text{LITA}}$ \cite{Keegan,Ishida}.}
\label{BC}       
\end{figure}

The boundary $\partial \Omega$ of our computational domain consists of:
\begin{itemize}
    \item two inflow, the LCMA and the LITA sections. We consider 
    a realistic flow rate waveform \cite{Keegan,Ishida}:
\begin{equation}
    q_i(t)=f^i\bar{q}_i(t), \quad \quad i=\text{LMCA},\text{LITA},
    \label{flowrate}
\end{equation}
where $f^i \in \left[ \frac{2}{3}, \frac{4}{3} \right] $ (see \cite{Keegan,Ishida,Verim}). The functions $\bar{q}_i(t)$ are represented in Figure \ref{BC}.

Moreover, since 
the stenosis severity influences the inlet flow, in order to enforce more realistic inflow conditions, it is relevant to scale flow rates as follows:
\begin{equation}
    \bar{Q}_{\text{LMCA}}^{\text{healthy}}=\bar{Q}_{\text{LMCA}}^{\text{stenosis}}+\bar{Q}_{\text{LITA}}^{\text{stenosis}}=G\bar{Q}_{\text{LMCA}}^{\text{healthy}}+C\bar{Q}_{\text{LITA}}^{\text{healthy}},
    \label{flux}
\end{equation}
where $\bar{Q}_i^{\text{healthy}}=\frac{1}{T}\int_0^T f^i\bar{q}_i(t)\, dt$, 
$G$ scales as the square of the stenosis severity diameter, and $C$ is consequently computed. Then, the inflow boundary conditions are given by:
\begin{equation}
    q_{\text{LMCA}}(t)=G f^i\bar{q}_i(t), \quad q_{\text{LITA}}(t)=C f^i\bar{q}_i(t);
    \label{flowrate2}
\end{equation}
    \item the vessel wall, on which we enforce a no slip
condition;
\vspace{0.3cm}
    \item two outflow, the LAD and LCx sections, on which we enforce homogeneous Neumann boundary conditions. We highlight that it would be necessary to set outlet boundary conditions able to capture as much as possible the physiology of vascular networks outside of the domain of the model (see, e.g.,\cite{Fevola,Infantino,Infantino1,Zainib}.
    However such treatment is out of scope of this work and will be taken into account in a future contribution.
\end{itemize}

\subsection{Time and space discretization}
To discretize in time the problem \eqref{N-S-2}, let us consider a time step $\Delta t \in \mathbb{R}^+$ such that $t^n=n \Delta t$, $n=0,1,\ldots, N_T$ and $T = N_T \Delta t$. Let $(\mathbf{u}^n,P^n)$ be the approximations of the velocity and the pressure at time $t^n$. We adopt a Backward Differentiation Formula of order 2 (BDF2):

\begin{equation}
\begin{split}
	\begin{cases} 
	\displaystyle{\frac{3}{2 \Delta t}} \mathbf{u}^{n+1}+ \nabla \cdot (\mathbf u^{n} \otimes \mathbf u^{n+1}) + \nabla P^{n+1}
	-\nu \Delta \mathbf u ^{n+1} = \mathbf{b}^{n+1},  &  \\
	\nabla \cdot \mathbf u^{n+1} =0, & 	
	\end{cases} 
	\label{N-S-time-discretized}
\end{split}
\end{equation}
where $\mathbf{b}^{n+1}=\frac{4\mathbf{u}^n-\mathbf{u}^{n-1}}{2\Delta t}$.\\
Concerning the space discretization of the problem \eqref{N-S-time-discretized}, we adopt a FV method within the C++ library \openfoam. 
The computational domain $\Omega$ is discretized into $N_h$ non-overlapping control volumes $\Omega_i$ with $i = 1,\dots, N_h$. 
Let $\bm A_j$ be the surface vector of each face $j$ of the control volume $\Omega_i$. 
Then the fully discretized form of problem \eqref{N-S-2} is given by

\begin{equation}
    \frac{3}{2 \Delta t} \mathbf{u}_i^{n+1}+\sum_j \phi_j \mathbf{u}_{i,j}^{n+1}-\nu\sum_j(\nabla \mathbf{u}_i^{n+1})_j\cdot \mathbf{A}_j+\sum_j P_{i,j}^{n+1}\mathbf{A}_j=\mathbf{b}_i^{n+1},
    \label{FV_form}
\end{equation}

\begin{equation}
    \sum_j (\nabla P^{n+1})_j\cdot \mathbf{A}_j = \sum_j  \Big( -\sum_j \phi_j \mathbf{u}_{j}^{n+1} + \nu \sum_j(\nabla \mathbf{u}^{n+1})_j\cdot \mathbf{A}_j + \mathbf{b}^{n+1} \Big)_j \cdot \mathbf{A}_j,
    \label{FV_form1}
\end{equation}
where $\phi_j = \bm{u}_j^{n} \cdot \bm{A}_j$ is the convective flux associated to $\bm{u}^{n}$ through face $j$ of the control volume $\Omega_i$, $\bm{u}_{i,j}^{n+1}$ and $P_{i,j}^{n+1}$ indicate the velocity and pressure associated to the centroid of face $j$ normalized with respect to $\Omega_i$,  and $\bm{u}_i^{n+1}$ and $\bm{b}_i^{n+1}$ are the average velocity and the source term in the control volume $\Omega_i$. 

For more details, we refer the reader to \cite{Siena,Girfogl}. 

\section{The reduced order model}
\label{sec:2}
In this work we use the POD-ANN method consisting of the following stages: 
\begin{itemize}
	\item \emph{Offline}: a reduced basis space is built by applying POD to a database of high-fidelity solutions obtained by solving the FOM for different values of physical and/or geometrical parameters. Once the reduced basis space is computed, we project the original snapshots onto such a space by obtaining the corresponding parameter dependent modal coefficients. Then the training of the neural networks to approximate the map between parameters and modal coefficients is carried out. This stage is computationally expensive, however it only needs to be performed once. 
	\item \emph{Online}: for any new parameter value, we approximate the new coefficients by using the trained neural network and the reduced solution is obtained as a linear combination of the POD basis functions multiplied by modal coefficients. During this stage, it is possible to explore the parameter space at a significantly reduced cost.
\end{itemize}

\subsection{The proper orthogonal decomposition}
\label{sec:2.1}
The POD is one of the most widely-used techniques to compress data and extract an optimal set of orthonormal basis in the least-squares sense. There are two main strategies for the construction of the reduced basis: greedy algorithms \cite{Hesthaven,Bang} and the POD method. The former allows to minimize the number of snapshots to be computed. However, a major drawback of greedy algorithms is that they are based on an a posteriori estimate of the projection error, which is often difficult to compute in practical applications. For this reason, in this paper, we opt for the
POD method, which although often requires a larger number of snapshots, is in turn more general.\\

Let $\mathcal{K}= \{\bm{\mu}_1, \dots, \bm{\mu}_{N_k}\}$ be a finite dimensional training set of samples chosen inside the parameter space $\mathcal{P}$ and for each time instance $t_k \in \{t_1, \dots, t_{N_t}\} \subset (0, T]$. We solve the FOM for each $\bm{\mu}_k \in \mathcal{K}  \subset \mathcal{P}$. The total number of snapshots $N_s$ is given by $N_s = N_k \cdot N_t$.
After computing the full order solutions, 
they are stored into a matrix $\mathcal{S}_{\Phi}\in\mathbb{R}^{N_{h} \times N_s}$ 
in a column-wise sense, i.e.:
\begin{equation*}
    \mathcal{S}_{\Phi} = \begin{bmatrix}
    \Phi_1(t_1, \bm{\mu}_1) & \cdots & \Phi_1(t_{N_t}, \bm{\mu}_1)& \Phi_1(t_1, \bm{\mu}_2) & \cdots & \Phi_1(t_{N_t}, \bm{\mu}_{N_k}) \\ 
    \vdots & \vdots & \vdots & \vdots \\ 
    \Phi_{N_{h}}(t_1, \bm{\mu}_1) & \cdots & \Phi_{N_{h}}(t_{N_t}, \bm{\mu}_1) & \Phi_{N_{h}}(t_1, \bm{\mu}_2) & \cdots & \Phi_{N_{h}}(t_{N_t}, \bm{\mu}_{N_k})
    \end{bmatrix},
\end{equation*}
for $\Phi = \{\mathbf{u}, P, WSS \}$.
Commonly snapshot matrices are not square and denoting by $R \leq \text{min}(N_{h},N_s)$ the rank of $\mathcal{S}_{\Phi}$, the Singular Value Decomposition (SVD) allows to factorise $\mathcal{S}_{\Phi}$ as:
\begin{equation}
\mathcal{S}_{\Phi}=\mathcal{W} \mathcal{D} \mathcal{Z}^T,
\end{equation}
where $\mathcal{W}=\{ \mathbf{w}_1| \dots |\mathbf{w}_{N_{h}} \} \in \mathbb{R}^{N_{h} \times N_{h}}$ and $\mathcal{Z}=\{ \mathbf{z}_1| \dots |\mathbf{z}_{N_s}  \} \in \mathbb{R}^{ N_s \times N_s}$ are two orthogonal matrices composed of left singular vectors and right singular vectors respectively in columns, and $\mathcal{D} \in \mathbb{R}^{ N_{h} \times N_s}$ is a diagonal matrix with $R$ non-zero real singular values $\sigma_1 \geq \sigma_2 \geq \dots \geq \sigma_R > 0$.\\Our goal is to approximate the columns of $\mathcal{S}_{\Phi}$ by means of $L<R$ orthonormal vectors. The Schmidt-Eckart-Young theorem states that the POD basis of rank $L$  consists of the first $L$ left singular vectors of $\mathcal{S}_{\Phi}$, also named modes \cite{Eckart}. So we can introduce the matrix with the extrapolated modes as columns:
\begin{equation}
    \mathcal{V}=\{ \mathbf{w}_1| \dots |\mathbf{w}_L \} \in \mathbb{R}^{ N_{h}\times L}.
\end{equation}
It is well known \cite{Kunisch} that the POD basis of size $L$ is the solution to the minimization problem \cite{Negri}:
\begin{equation}
    \min_{\mathcal{V}} \Vert \mathcal{S}_{\Phi_h}- \mathcal{V}\mathcal{V}^T\mathcal{S}_{\Phi_h} \Vert \quad s.t. \quad \mathcal{V}^T\mathcal{V}=\mathcal{I},
    \label{min}
\end{equation}
where $\Vert \bullet \Vert$ is the Frobenius norm. Therefore, the reduced basis is the set of vectors that minimizes the distance between the snapshots and their projection onto the space spanned by the basis.
In addition, the error committed by approximating the columns of $\mathcal{S}_{\Phi}$ via the vectors of $\mathcal{V}$ is equal to the sum of the squares of the neglected singular values \cite{Negri}:
\begin{equation}
\sum_{i=L+1}^{\text{min}(N_{h},N_s)} \sigma_i^2.
     \label{err}
\end{equation}
So by controlling the size $L$, we can approximate the snapshot matrix $\mathcal{S}_{\Phi}$ with arbitrary accuracy. 
Since the error is strictly related to the magnitude of the singular values, a common choice is to set $L$ equal to the smallest integer $L$ such that:
\begin{equation}
    \frac{\sum_{i=1}^{L}\sigma_i}{\sum_{i=1}^{R}\sigma_i} \ge 1-\epsilon^2, 
    \label{energy}
\end{equation}
where $\epsilon$ is a user-provided tolerance and the left hand side of \eqref{energy} is the relative information content of the POD basis, namely the percentage of energy of the snapshots (or cumulative energy of the eigenvalues) retained by the first $L$ modes. 

Once the POD basis is available, the reduced solution $\Phi_{\text{rb}} (t_k, \mu_k)$ that approximates the full order solution $\Phi (t_k, \mu_k)$ is:
\begin{equation}
\Phi(t_k, \bm \mu_k) \approx \Phi_{\text{rb}} (t_k, \bm \mu_k) = 
\sum_{j=1}^{L} (\mathcal{V}^T \Phi (t_k, \bm \mu_k))_j \mathbf w_{j},
\end{equation}
where $(\mathcal{V}^T \Phi (t_k, \bm \mu_k))_j$ is the modal coefficient associated to the $j$-th mode. 

\subsection{Artificial neural network}
\label{sec:2.2}
An ANN is a computational model able to learn from observational data. It consists of neurons and a set of directed weighted synaptic connections among the neurons. It is an oriented graph, with the neurons as nodes and the synapses as oriented edges, whose weights are adjusted by means
of a training process to configure the network for a specific application (see \cite{Goodfellow,kriesel2007brief,calin2020deep}). 

Let us consider the neuron $j$. 
Three functions characterize completely the neuron $j$:
\begin{itemize}
    \item the propagation function $u_j$. It is used to transport values through the neurons of the ANN. We use the weighted sum: 
    $$u_j=\sum_{k=1}^m w_{s_k,j}y_{s_k} + b_j,$$ where $b_j$ is the bias, $y_{s_k}$ is the output related to the sending neuron $k$,  $w_{s_k,j}$ are proper weights and $m$ is the number of sending neurons linked with the neuron $j$. 
    Other choices are possible for $u$ \cite{Hesthaven}.
    \item the activation function $a_j$. It quantifies to which degree neuron $j$ is active. It is a function of the input $u_j$ and the bias $b_j$ chosen during the training process: 
    $$ a_j=f_{\text{act}}\left(\sum_{k=1}^m w_{s_k,j}y_{s_k}+b_j\right).$$
    Commonly the activation  functions  are  non-linear. Possible choices are sigmoid function, hyperbolic tangent, RELU, SoftMax. 
    The activation function is an hyperparameter to be tuned during the training stage in order to optimize the performance of the neural network. More details can be found in \cite{Sharma}.
    \item the output function $y_j$. It is related to the activation function $a_j$. Often it is the identity function, so that $a_j$ and $y_j$ coincide:
    \begin{equation*}
        y_j=f_{\text{out}}(a_j)=a_j.
    \end{equation*}
\end{itemize}
In this work, we will use a specific type of ANNs, the feedforward neural networks.

\subsubsection{The feedforward neural network}
\label{feed}
\begin{figure}
    \centering
	\includegraphics[width=0.8\linewidth]{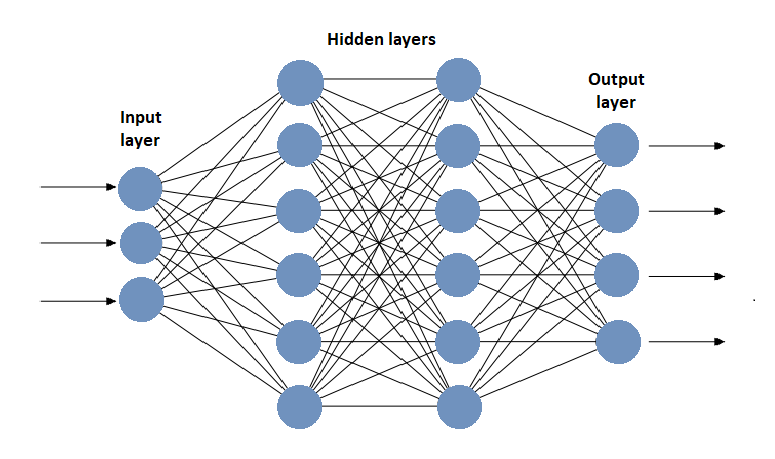}	
	\caption{Topology of the feedforward neural network.}
\label{topology}       
\end{figure}

In a feedforward neural network (Figure \ref{topology}), neurons are arranged into layers, so input nodes define one input layer and the same holds for the output layer and for the hidden layers. Neurons in a layer can only be linked with neurons in the next layer, towards the output layer (see \cite{Rosenblatt1958,minsky1969,fine2006feedforward}). We use fully  connected neural networks, so  each  node  in  layer  $l$  is  connected  to  all  nodes  in  layer  $l+1$   for  all  $l$. We highlight that the input layer of our network consists of the set of time/parameter instances $\{ (t_1, \bm{\mu}_1),\dots,(t_{N_t},\bm{\mu}_{N_k}) \}$, whilst the output one is given by the corresponding modal coefficients $\{(\mathcal{V}^T \Phi (t_k, \bm \mu_k))_1, \dots, (\mathcal{V}^T \Phi (t_k, \bm \mu_k))_L\}$. 


Layered  feedforward  networks  have  become  very  popular  firstly because they  have  been  found  in practice to  generalize  well.  Secondly,  a  training  algorithm  based on the so-called  backpropagation  can  often  find  a  good  set  of weights  (and  biases) in  a  reasonable  amount  of  iterations (or, equivalently, epochs). 
During the training procedure the weights of the connections in the network are repeatedly changed in order to minimize the difference between the actual output vector of the net $\Tilde{\bm{\pi}}$ and the required output vector $\bm{\pi}$. The  key  to  backpropagation  is  a  method  for  computing  the  gradient  of the  error with  respect  to  the  weights  for  a  given  input by  propagating  error  backwards  through  the  network \cite{rojas1996,Rumelhart}. A loss function is introduced to optimize the parameter values in a neural network model. This class of functions maps a set of parameter values for the network onto a scalar value that shows how well those parameters achieve the purpose the network is intended to do. \\ The loss function $\mathcal{L} = \mathcal{L}(\Tilde{\bm{\pi}},\bm{\pi})$ used in this work is the mean squared error (MSE), that is the most common choice for regression problems:
\begin{equation}
    \mathcal{L} = \sum_{i=1}^{N_s} \mathcal{L}_{i} = \sum_{i=1}^{N_s} \frac{1}{L}\sum_{j=1}^L (\pi_{i,j}-\Tilde{\pi}_{i,j})^2.
\end{equation}\\
To effectively compute the gradient of the loss function, the chain rule is used. 
The gradient of the loss function for a single weight $w_{s_k,j}^l$ and for the biases $b_j^l$ can be computed using the chain rule as:
\begin{equation}
\begin{split}
    & \frac{\partial \mathcal{L}}{ \partial w_{s_k,j}^l}=\frac{\partial \mathcal{L}}{\partial a_j^l}\frac{\partial a_j^l}{\partial u_j^l}\frac{\partial u_j^l}{\partial w_{s_k,j}^l} , \\
    & \frac{\partial \mathcal{L}}{ \partial b_j^l}=
   \frac{\partial \mathcal{L}}{\partial a_j^l}\frac{\partial a_j^l}{\partial u_j^l} \frac{\partial u_j^l}{\partial b_j^l}.
\end{split}
\end{equation}
Operatively, in the forward pass the values of the output layers from the inputs data are computed and the loss function is calculated. After each forward pass, backpropagation performs a backward pass to compute the gradient of the loss function while adjusting the model’s parameters as follow:
\begin{equation}
\begin{split}
    &  \bm{w}=\bm{w}-\eta \frac{\partial \mathcal{L}}{\partial \bm{w}}, \\
    & \bm{b} = \bm{b} -\eta  \frac{\partial \mathcal{L}}{\partial \bm{b}},
\end{split}
\end{equation}
where $\bm{w}$ and $\bm{b}$ are matrix representations of the weights and biases and $\eta$ is an hyperparameter tuned to minimize the loss, named learning rate.  The accuracy of the trained model is measured by the number of correct predictions over the total number of predictions with a tolerance of $10^{-3}$. \\ 
In addition to the hyperparameters already mentioned, also the number of hidden layers as well as the number of their neurons are decided during the learning process. Hence, given an initial amount of training samples, we train the network for increasing values of hidden neurons, stopping when overfitting of training data occurs, due to an excessive number of layers and neurons \cite{Goodfellow}.

For the creation and training of the neural networks, we employed the Python library PyTorch.

\subsubsection{Evaluation of the modal coefficients}

As already reported in the previous subsection, the neural network employed provides a reliable approximation of the following input-output relationship:
\begin{equation}
      (t_k, \bm \mu_k) 
     \mapsto \big[\mathcal{V}^T \Phi(t_k, \bm \mu_k)\big]_{j=1}^L \in \mathbb{R}^L.
\end{equation}
 Then during the online stage the solution for any new time instant $t_{new}$ and new parameter $\bm{\mu}_{\text{new}}$ can be simply computed as follows \cite{Chen,Hesthaven,pichi2021,shah2021,Siena}:
 \begin{equation}
 \Phi_{rb} (t_{\text{new}}, \bm \mu_{\text{new}}) = \sum_{j=1}^{L} {\pi}_j(t_{\text{new}}, \bm \mu_{\text{new}})\mathbf w_{j}. 
 \end{equation}

\section{Numerical results}
\label{sec:3}
In this section, we test the performance of our ROM approach. Two parametric cases are investigated:
\begin{itemize}
    \item \emph{Case 1}: we verify the functionality of the ROM method in a physical parametric setting by considering $f_i$ (see equation \eqref{flowrate}) as parameters. 
    \item \emph{Case 2}: we verify the functionality of the ROM method in a geometrical parametric setting by considering the stenosis severity as parameter. 
\end{itemize}
Trial and error process described in Section \ref{feed} is employed to optimize the hyperparameters of the neural networks. For both cases, the optimized neural network architecture consists of 3 hidden layers and Table \ref{param_nn_stenosis} shows the other hyperparameters which recorded the best performance. In particular, hyperparameters are tuned for \emph{Case 2}, however they show a good functionality also for \emph{Case 1}.  The neural network accuracy is about $93\%$ in all the cases. 
\begin{table}
\footnotesize
\centering
\caption{Optimized hyperparameters of the feed forward neural networks. }
\begin{tabular}{|c|c|c|c|c|}
\hline
 \textbf{Variable} & \textbf{Neurons per layer} & \textbf{Activation function} & \textbf{Number of epochs} & \textbf{Learning rate} \\
\hline
\textbf{p}  & 1300 & Tanh & 50.000 & 8.25e-6 \\
\cline{1-5}
\textbf{U} & 1300 & Sigmoid & 50.000 & 5.00e-5  \\
\cline{1-5}
\textbf{WSS} & 1300 & Tanh & 50.000 & 8.50e-6  \\
\hline
\end{tabular}
\label{param_nn_stenosis}
\end{table}

\subsection{Case 1}
We consider a 70\% stenosis. 
The following finite-dimensional set is used to train the ROM: 
$$f_i = \{0.66, 0.7, 0.8, 0.9, 1.1, 1.2, 1.33 \}, \qquad i=\text{LITA},\text{LMCA},$$ \\
whilst we set $f_i = 1$ as test point. Note that when $\bm \mu = f_{\text{LITA}}$, $f_{\text{LMCA}} = 1$ and viceversa. 

The average Reynolds number characterizing the dynamics of the problem can be evaluated taking into account both the fluid properties and geometrical features as: 
\begin{equation*}
    \overline{Re} = \frac{\overline{U}_i d_i}{\nu}=
    \begin{cases} 
        \simeq 
        87, & \mbox{if } i=\mbox{LMCA,} \\ 
        \simeq 
        161, & \mbox{if } i=\mbox{LITA,}
    \end{cases}
\end{equation*}
where $\overline{U}_i$ is the mean velocity at the inlet of the LMCA and of the LITA (with $f_i=1$), 
$d_i$ is the diameter of the LMCA/LITA and $\nu = 3.7\cdot 10^{-6}$ $m^2/s$. This result supports the employment of a laminar model.

100 full-order equally spaced time-dependent snapshots, one every $0.008$ s, are collected for each $f_i$ value, over a cardiac cycle $T=0.8s$. They are enough to generate a reliable reduced space as proved in \cite{Siena}. Then we get 1400 snapshots in all.

\begin{figure}
	\centering
	\subfloat[][Case 1:  $\bm \mu = f_{\text{LITA}}$.\label{cumulative_eig_s:a}]{\includegraphics[width=.45\textwidth]{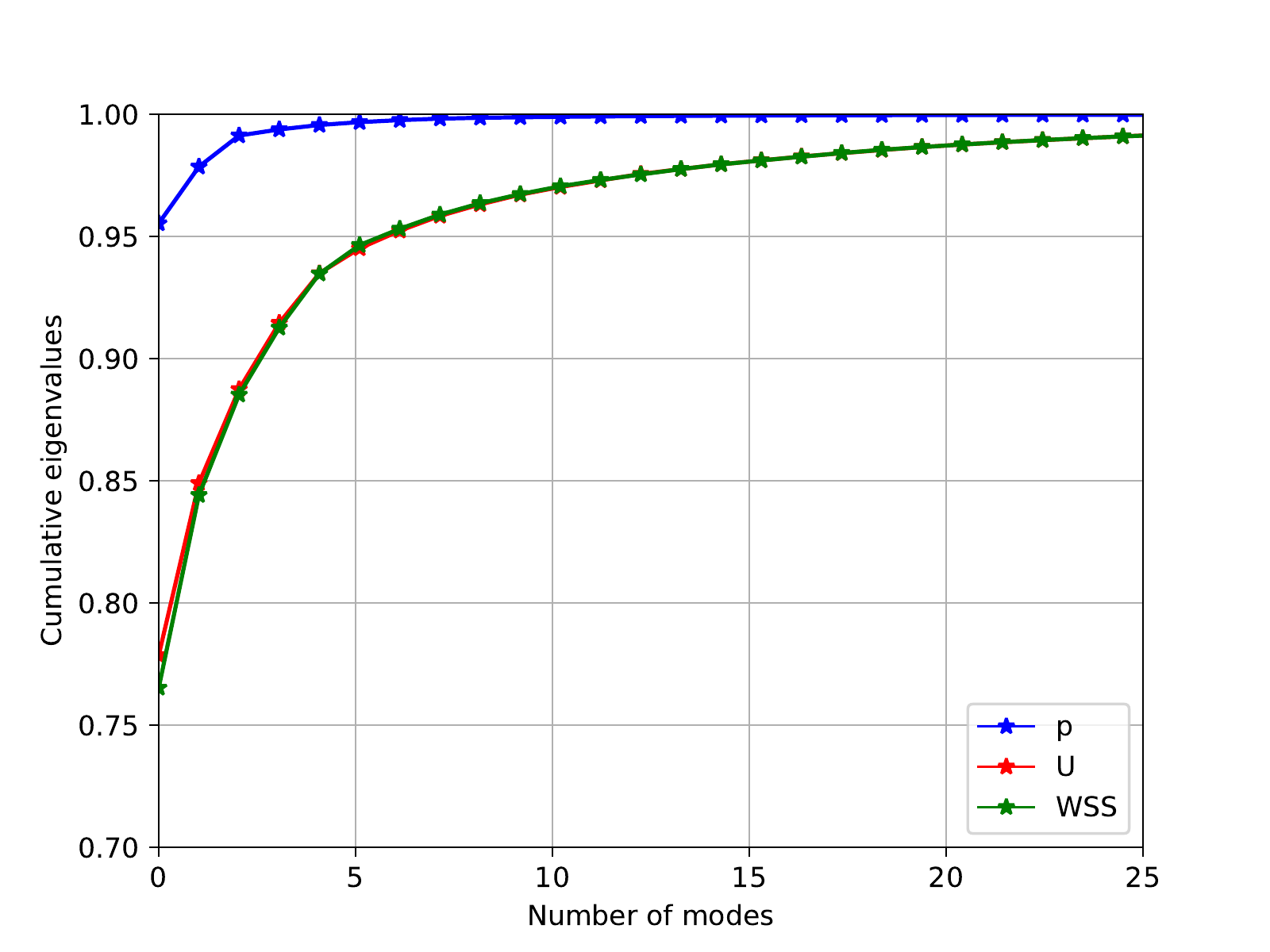}} 
	\subfloat[][Case 1: $\bm \mu = f_{\text{LMCA}}$.  \label{cumulative_eig_s:b}]{\includegraphics[width=.45\textwidth]{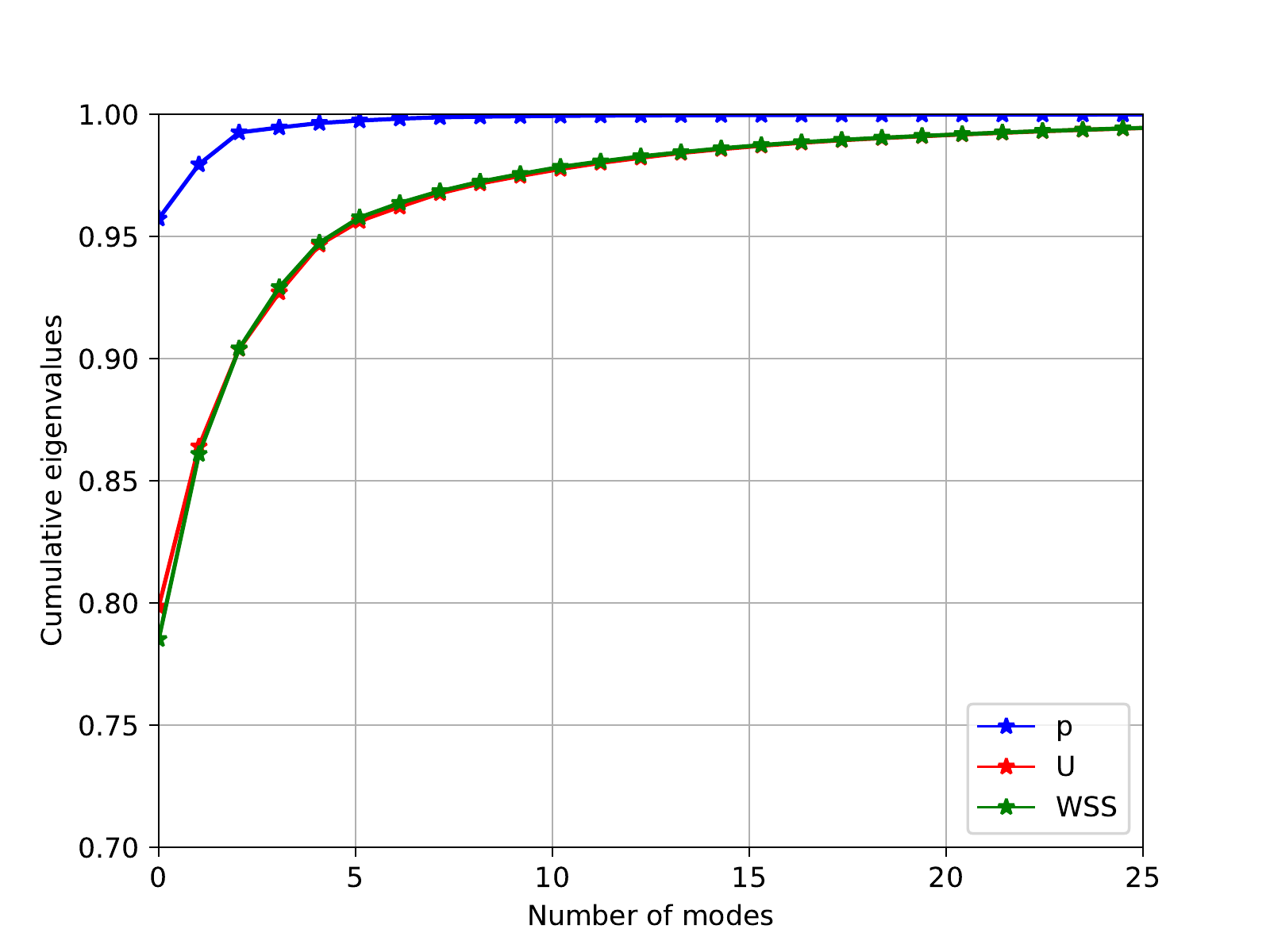}} \\
	\subfloat[][Case 2. \label{cumulative_eig_s:c}]{\includegraphics[width=.45\textwidth]{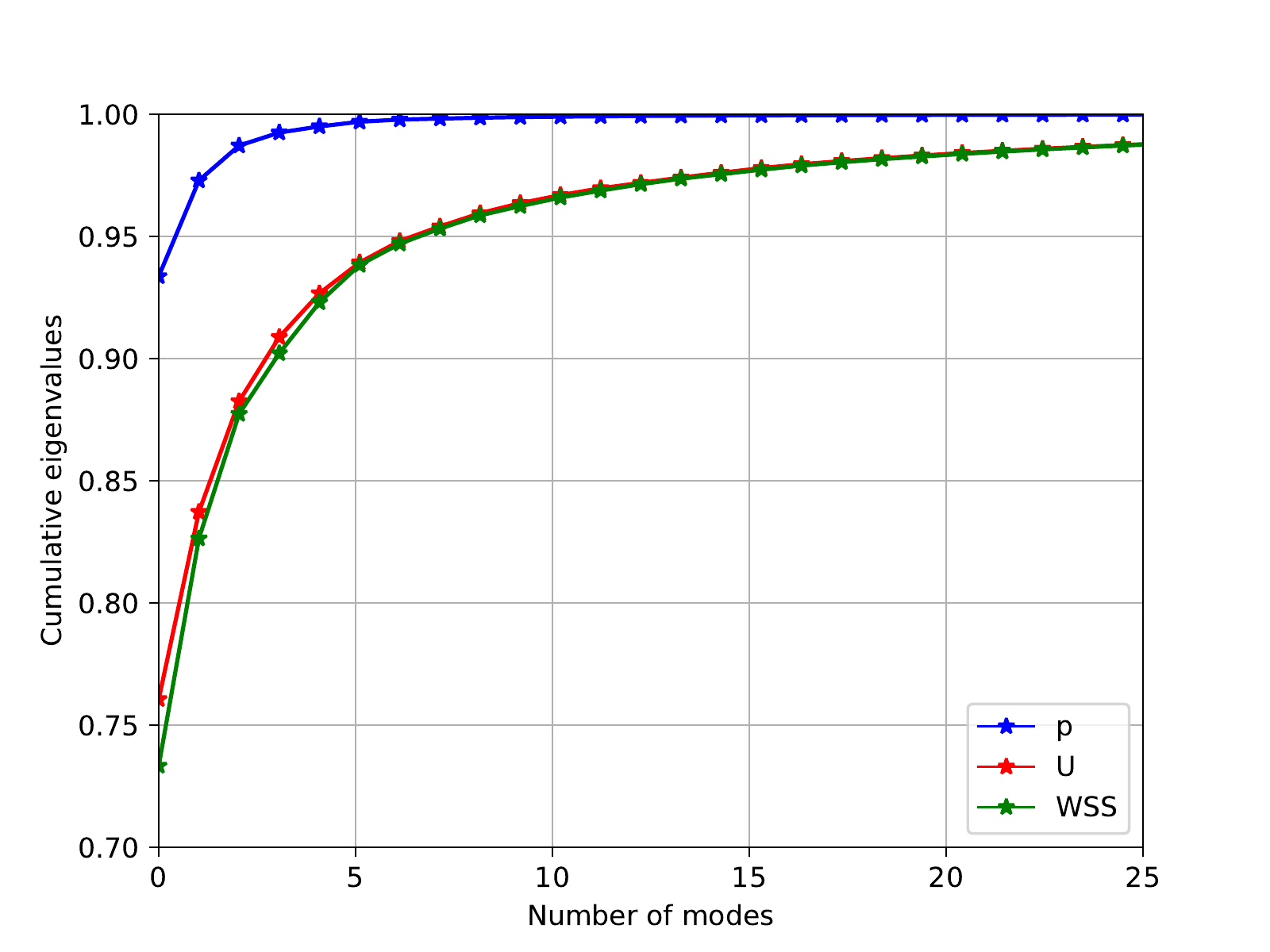}}
	\caption{Cumulative eigenvalues for pressure, velocity and WSS.}
	\label{cumulative_eig_s}
\end{figure}

\begin{figure}
	\centering
 	\subfloat[][2nd coefficient of  $P$.\label{func_f_lita:c}]{\includegraphics[width=.45\textwidth]{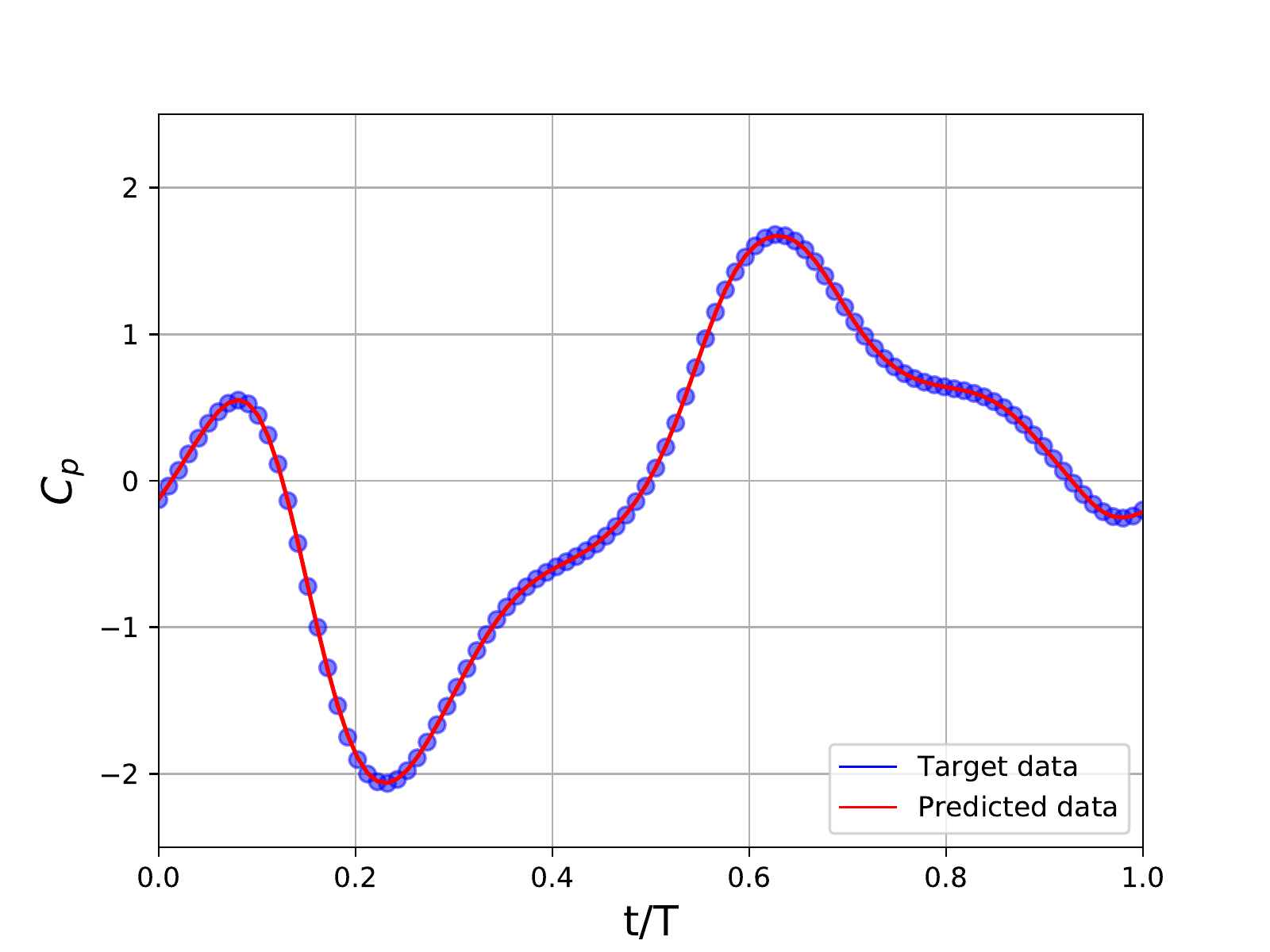}}
 	\subfloat[][6th coefficient of  $\mathbf{u}$.\label{func_f_lita:f}]{\includegraphics[width=.45\textwidth]{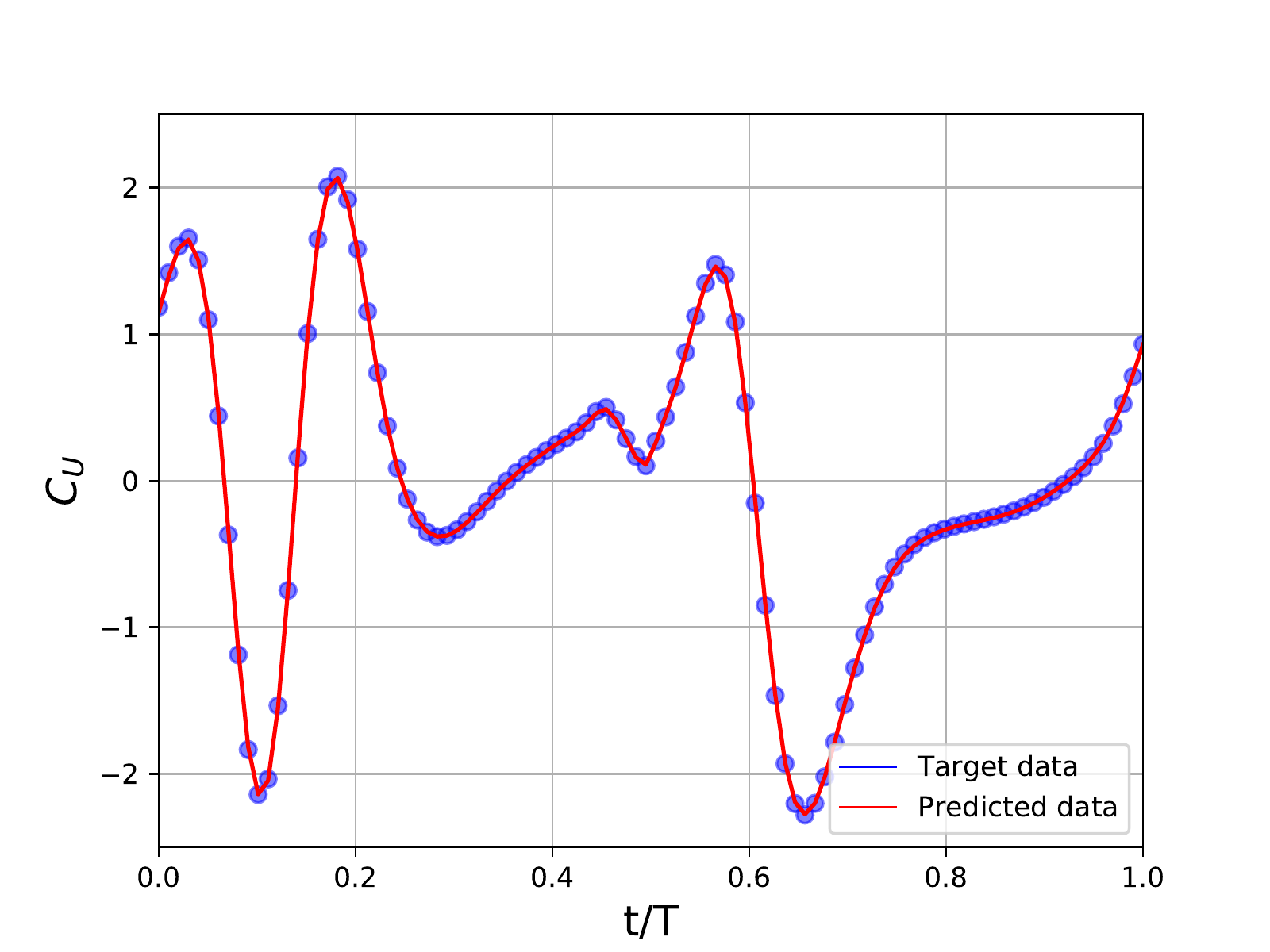}}\\
 	\subfloat[][2nd coefficient of  WSS.\label{func_f_lita:i}]{\includegraphics[width=.45\textwidth]{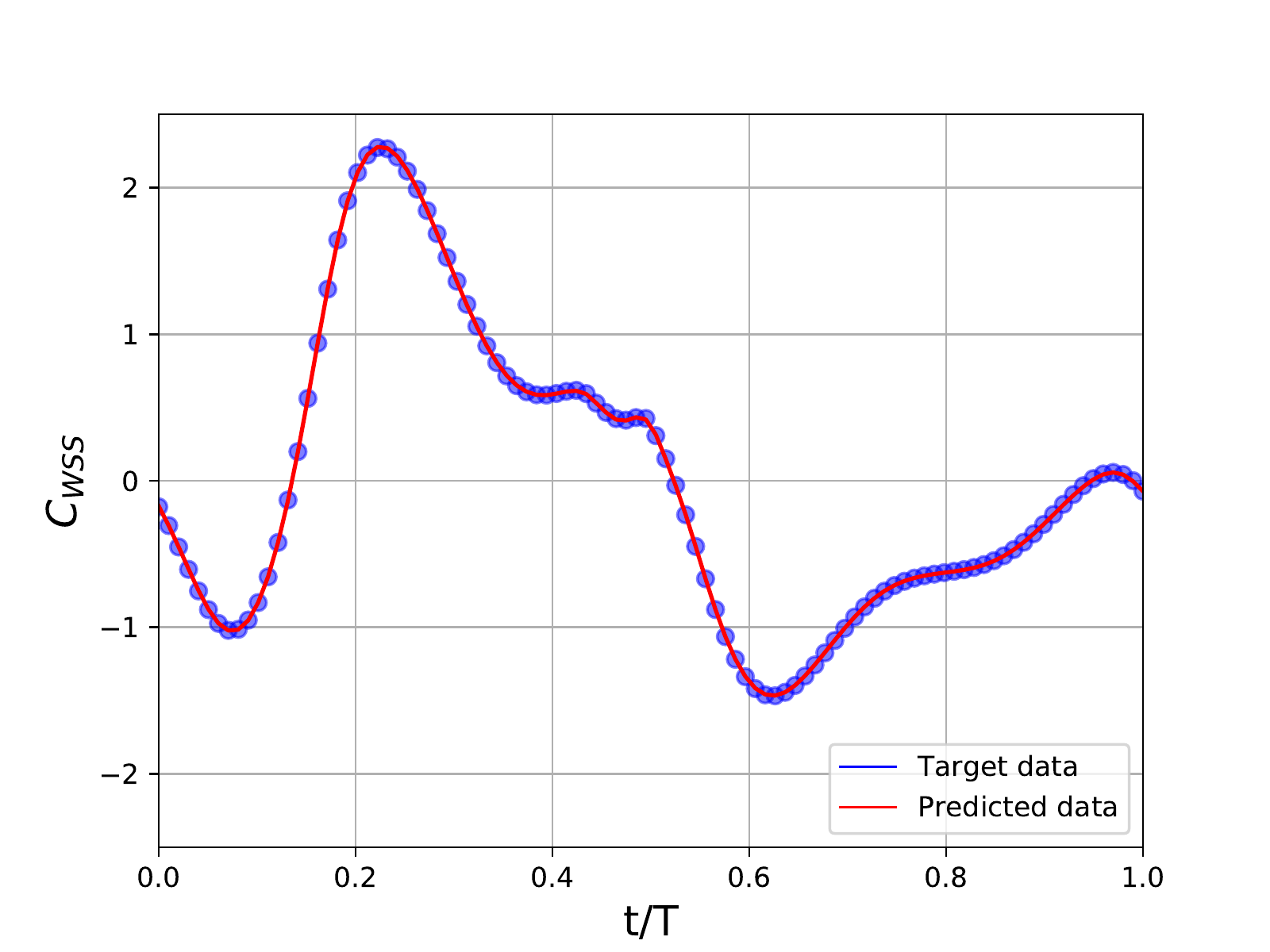}}
	\caption{Case 1: time evolution of some reduced coefficients including the prediction provided by NN (red line) and the FOM simulation (blue points) for $\bm \mu = f_{\text{LITA}}$.}
	\label{func_f_lita}
\end{figure}

\begin{figure}
	\centering
 	\subfloat[][3rd coefficient of  $P$.\label{func_f_lmca:c}]{\includegraphics[width=.45\textwidth]{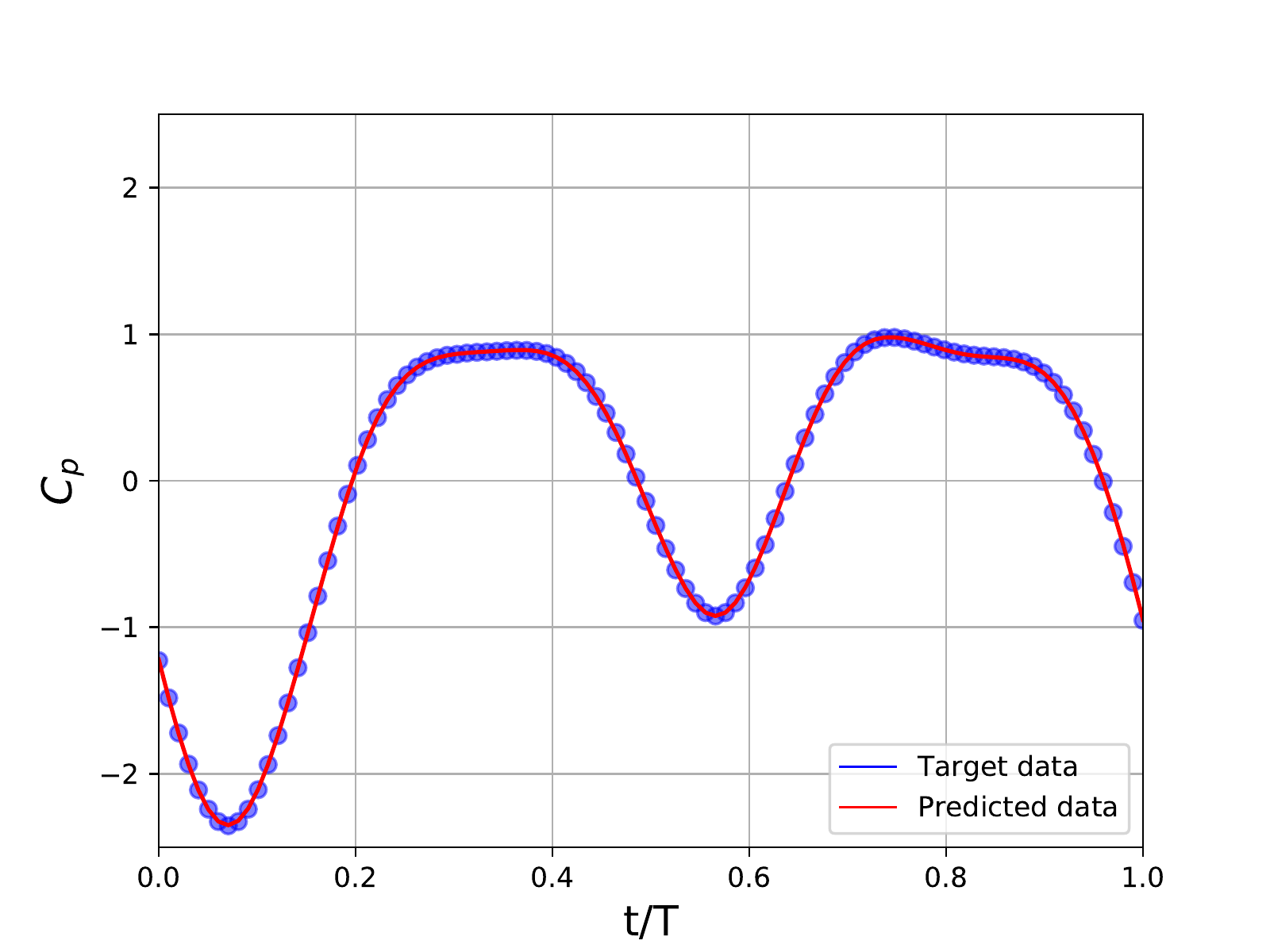}}
 	\subfloat[][8th coefficient of  $  \mathbf{u}$.\label{func_f_lmca:f}]{\includegraphics[width=.45\textwidth]{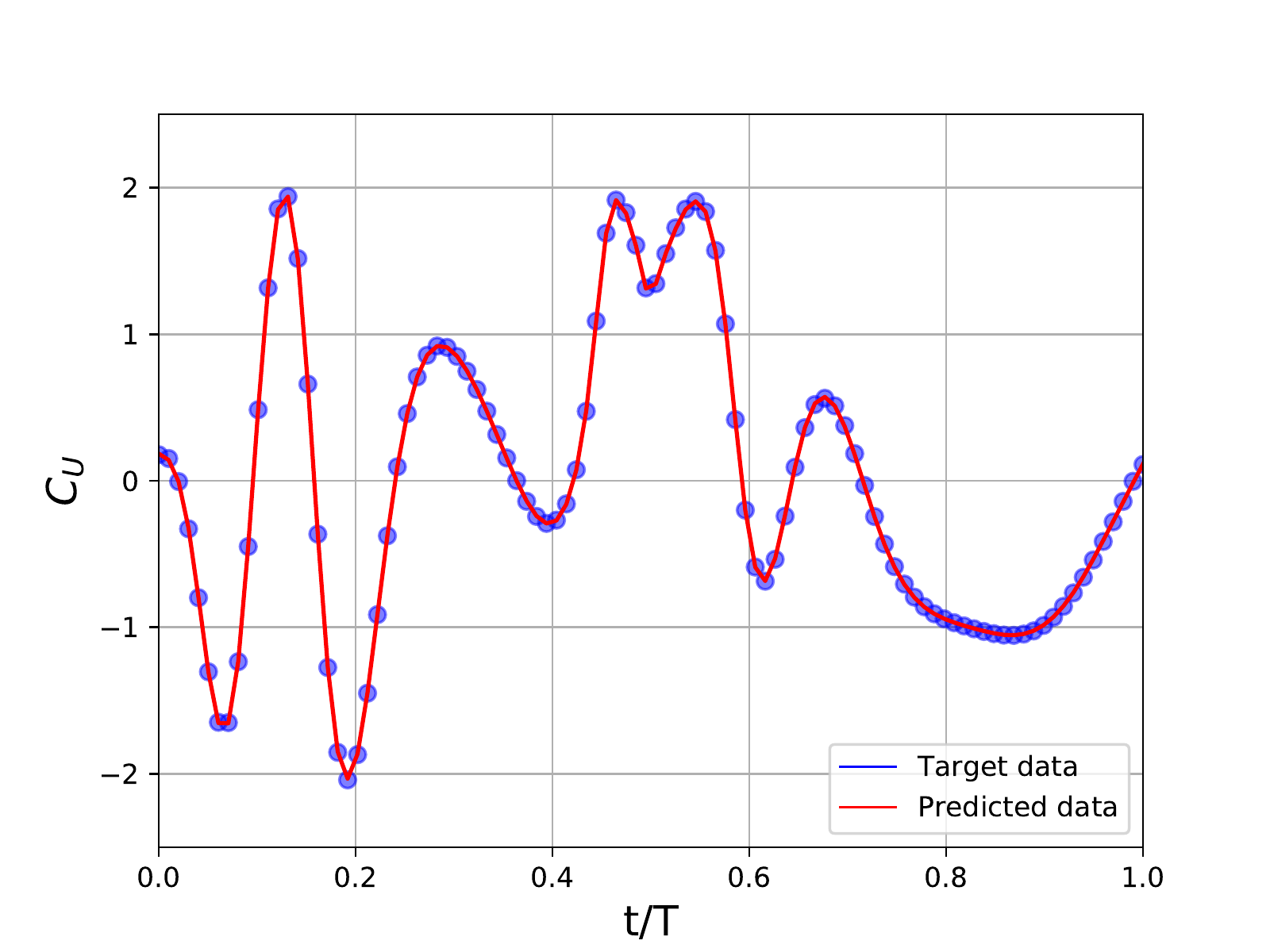}}\\
 	\subfloat[][6th coefficient of  WSS.\label{func_f_lmca:i}]{\includegraphics[width=.45\textwidth]{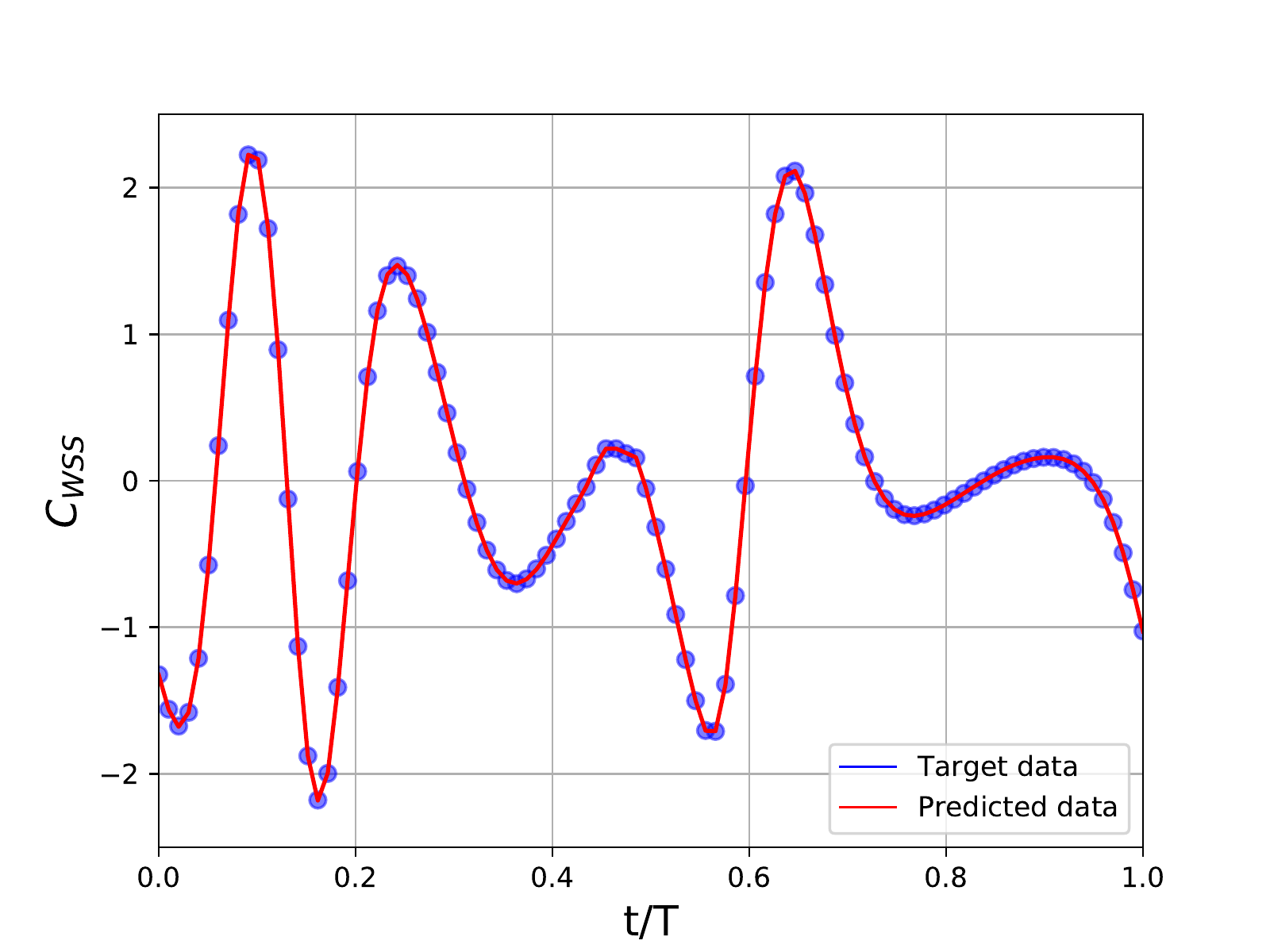}}
	\caption{Case 1: time evolution of some reduced coefficients including the prediction provided by NN (red line) and the FOM simulation (blue points) for $\bm \mu = f_{\text{LMCA}}$.}
	\label{func_f_lmca}
\end{figure}

Cumulative eigenvalues for pressure, velocity and WSS based on the first 50 most energetic POD
modes are shown in Figure \subref*{cumulative_eig_s:a} and \subref*{cumulative_eig_s:b}. In both cases, they exhibit a very similar trend for velocity and WSS whilst the pressure one grows faster. These results are expected because typically in hemodynamics applications the spatial variability in pressure is not that much compared to velocity (and WSS that is directly linked to the velocity): see, e.g., \cite{Infantino,Infantino1,Siena}. In addition, for each variable, one obtains similar trends for $\bm \mu = f_{\text{LITA}}$ and $\bm \mu = f_{\text{LMCA}}$. Even this result is not surprising by considering that $f_{\text{LITA}}$ and $f_{\text{LMCA}}$ play a very similar role and it is reasonable that they affect in the same way the dynamics of the system. \\
In  Figures \ref{func_f_lita} and \ref{func_f_lmca} 
the temporal evolution of the modal coefficients shows that the neural network is able to provide a prediction consistent with the FOM simulation. 

\begin{figure}
	\centering
	\subfloat[][$P$ error.\label{err_mod_lita:a}]{\includegraphics[width=.45\textwidth]{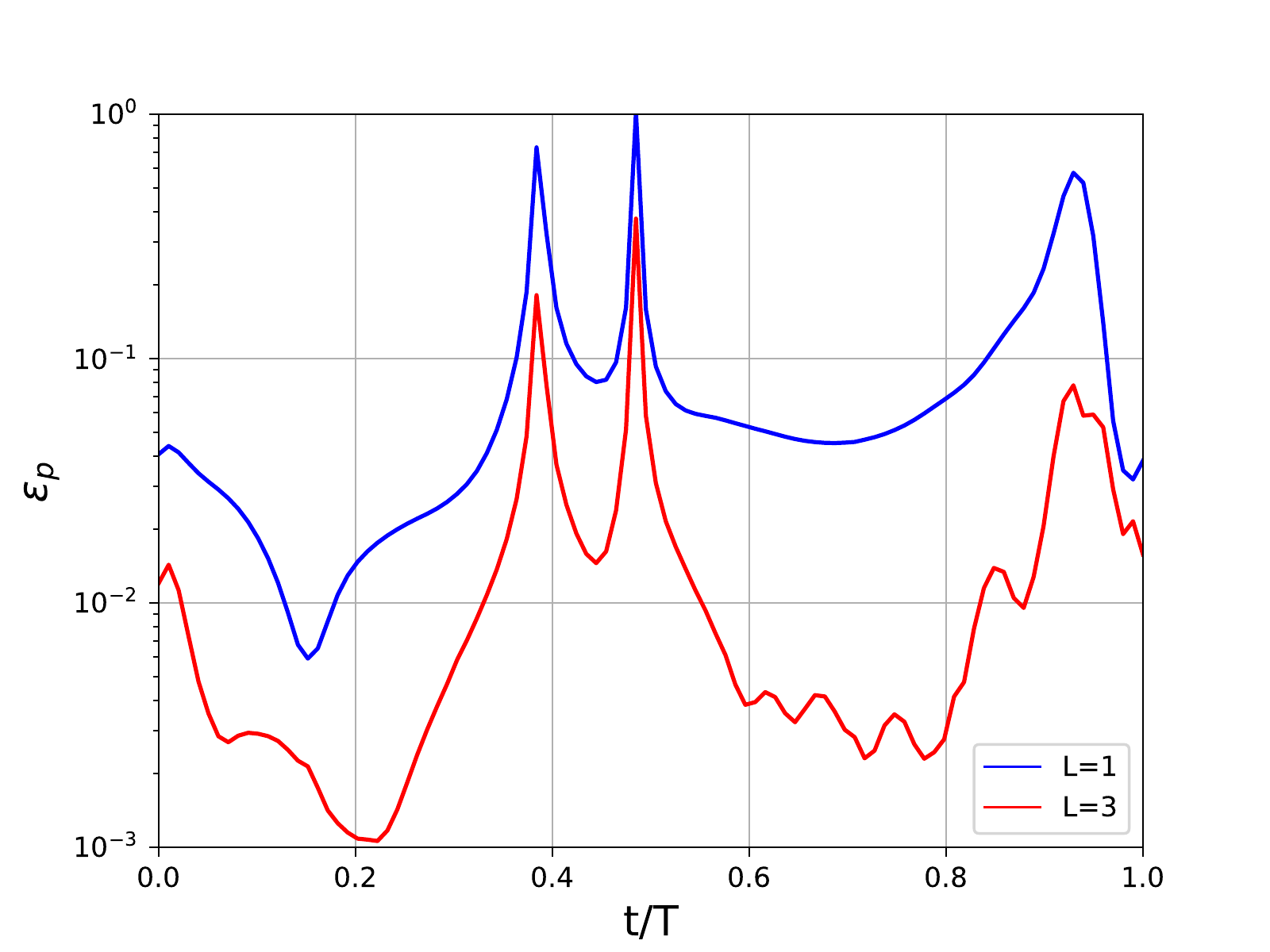}}
	\subfloat[][$\mathbf{u}$ error.\label{err_mod_lita:b}]{\includegraphics[width=.45\textwidth]{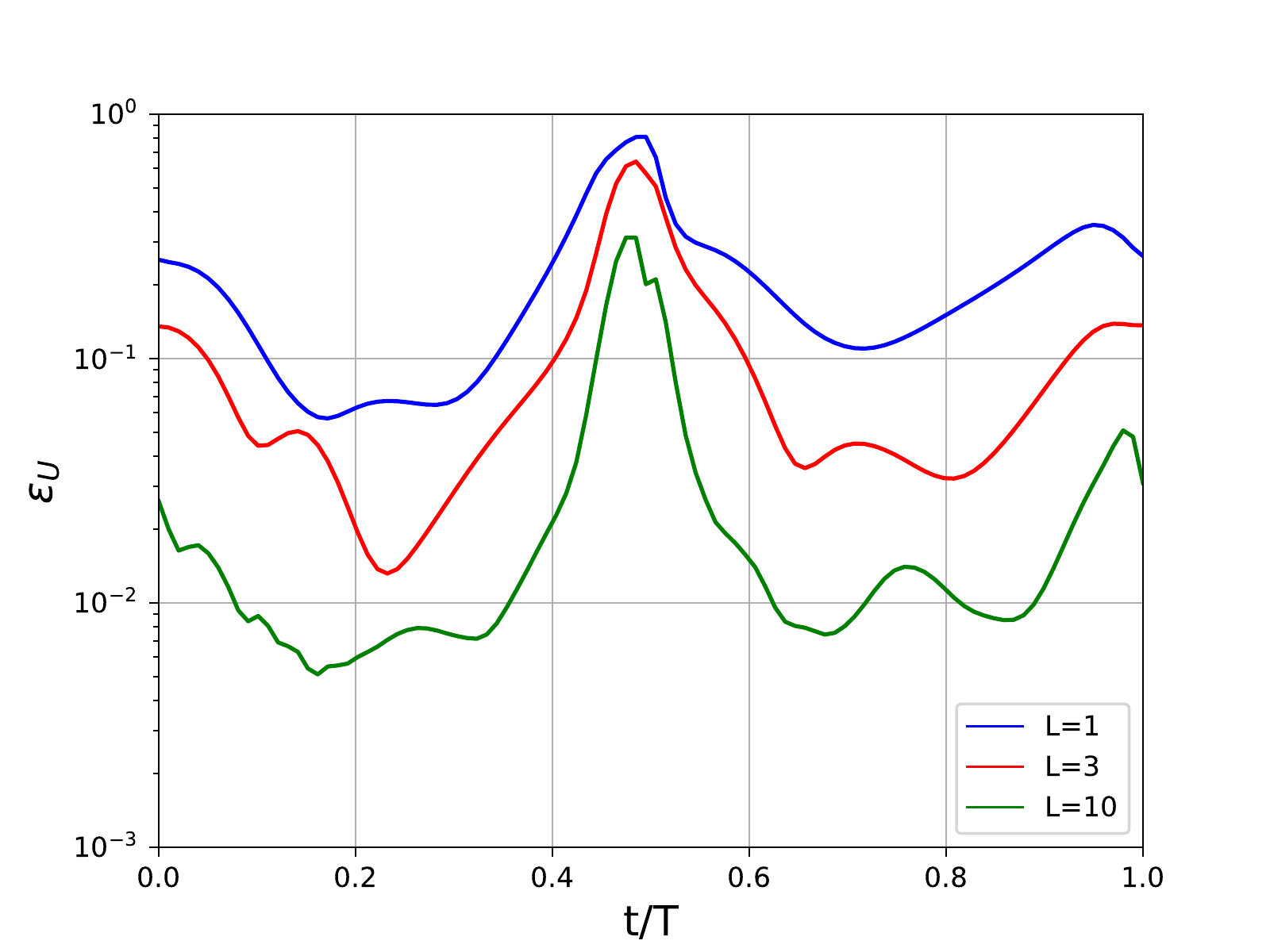}}\\
	\subfloat[][WSS  error.\label{err_mod_lita:c}]{\includegraphics[width=.45\textwidth]{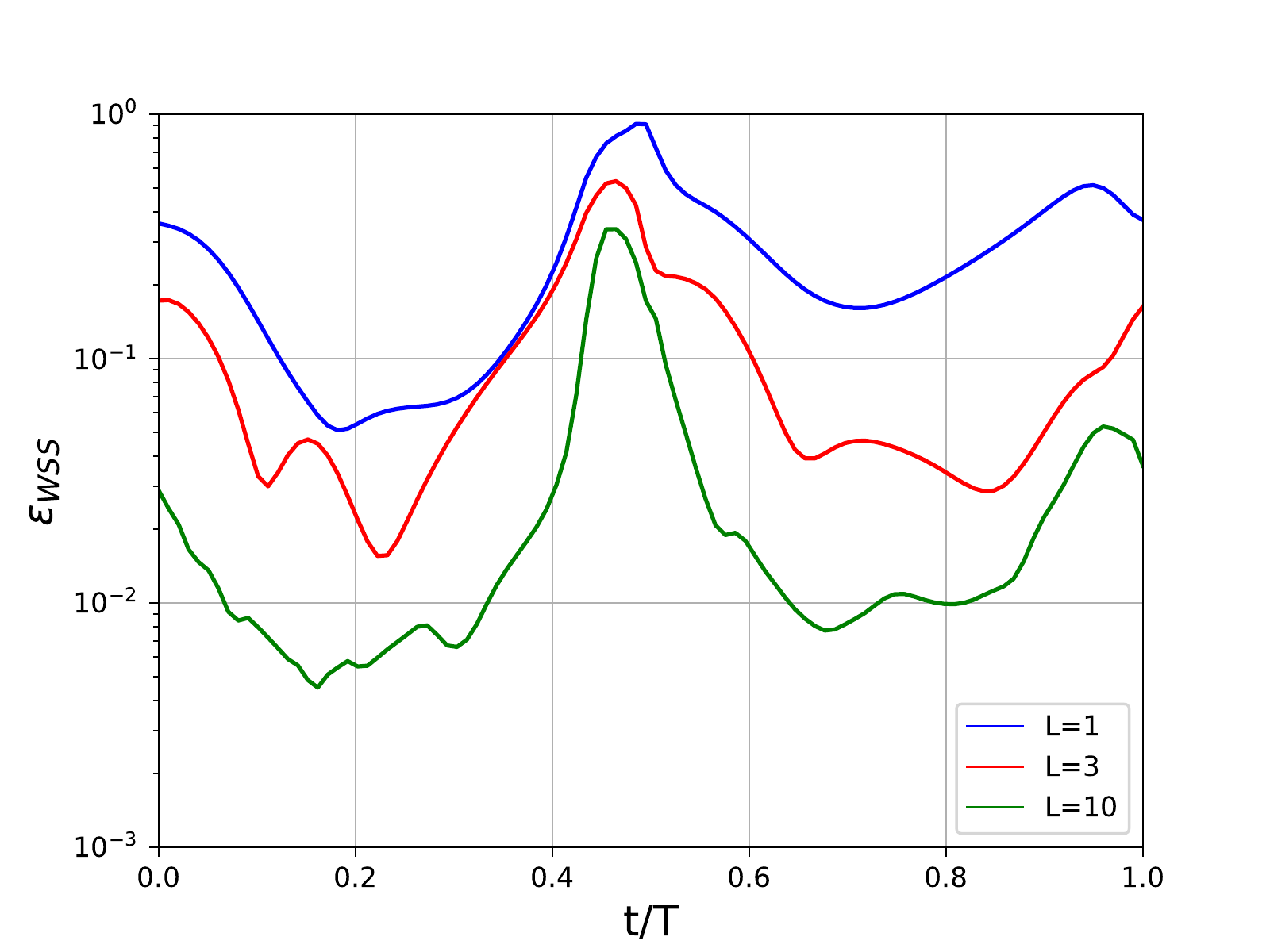}}
	\caption{Case 1: time evolution of the relative error for pressure, velocity and WSS at varying of the number of modes $L$ for $\bm \mu = f_{\text{LITA}}$.}
	\label{err_mod_lita}
\end{figure}

\begin{figure}
	\centering
	\subfloat[][$P$ error.\label{err_mod_lmca:a}]{\includegraphics[width=.45\textwidth]{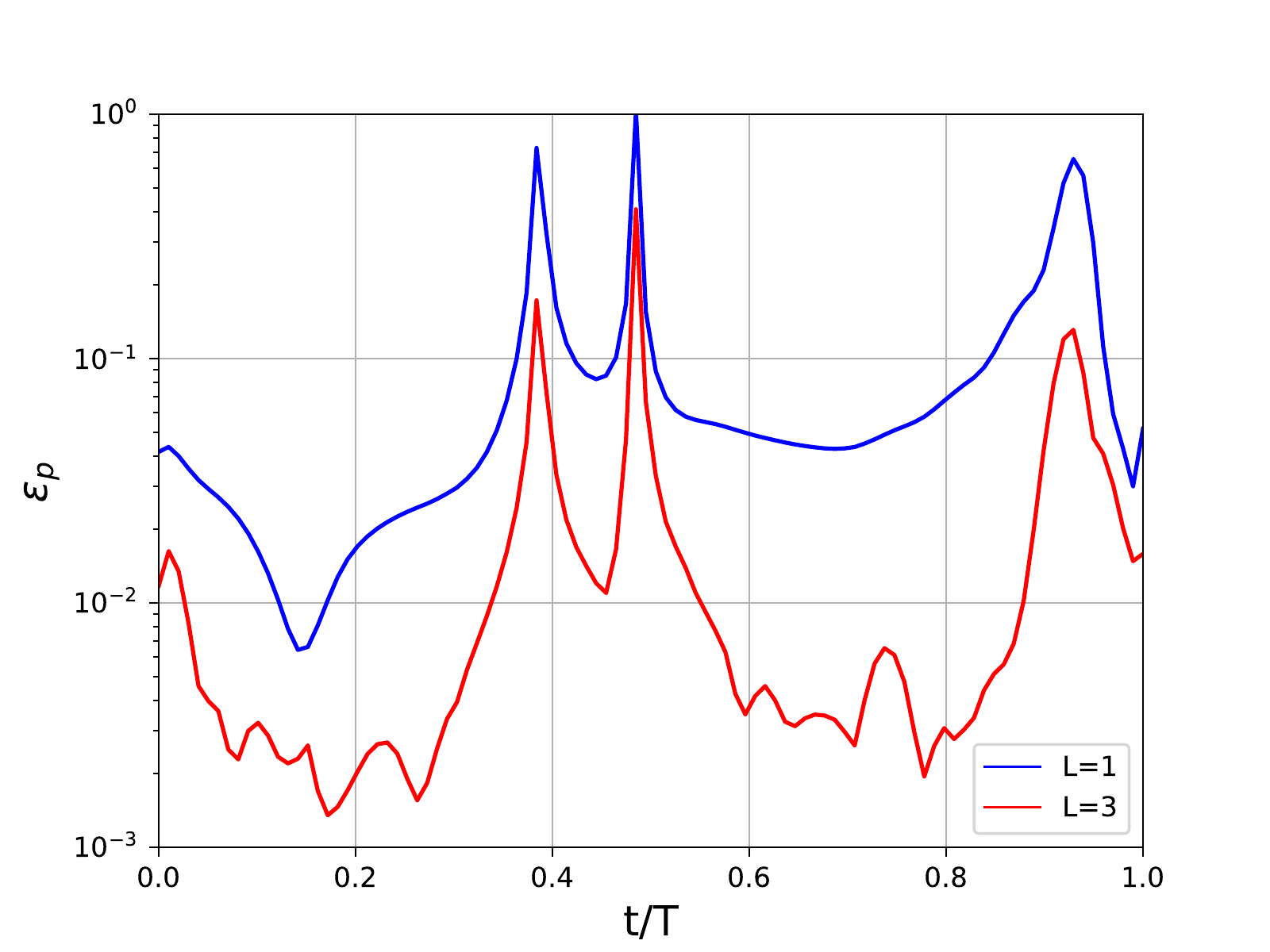}}
	\subfloat[][$\mathbf{u}$ error.\label{err_mod_lmca:b}]{\includegraphics[width=.45\textwidth]{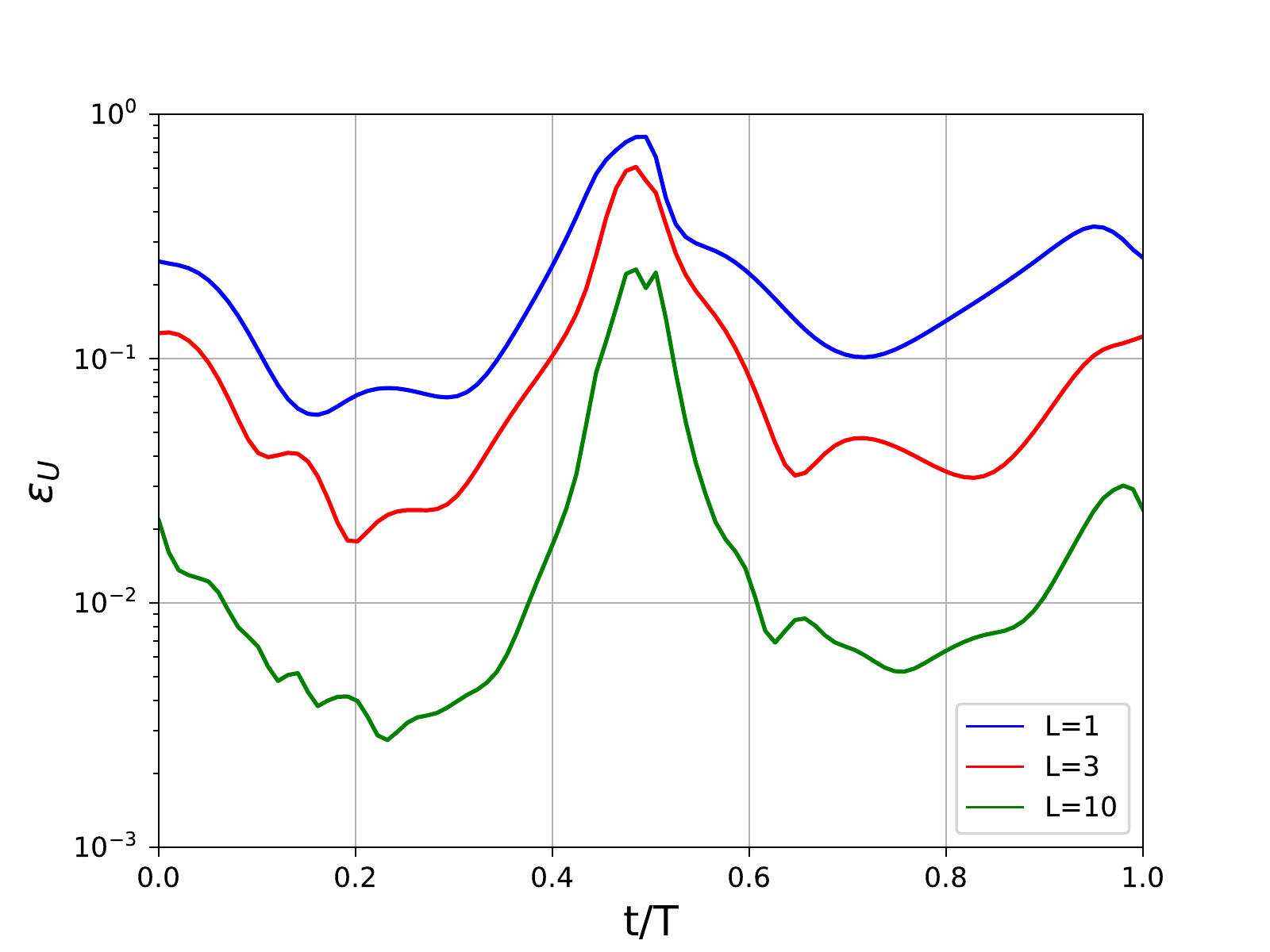}}\\
	\subfloat[][WSS  error.\label{err_mod_lmca:c}]{\includegraphics[width=.45\textwidth]{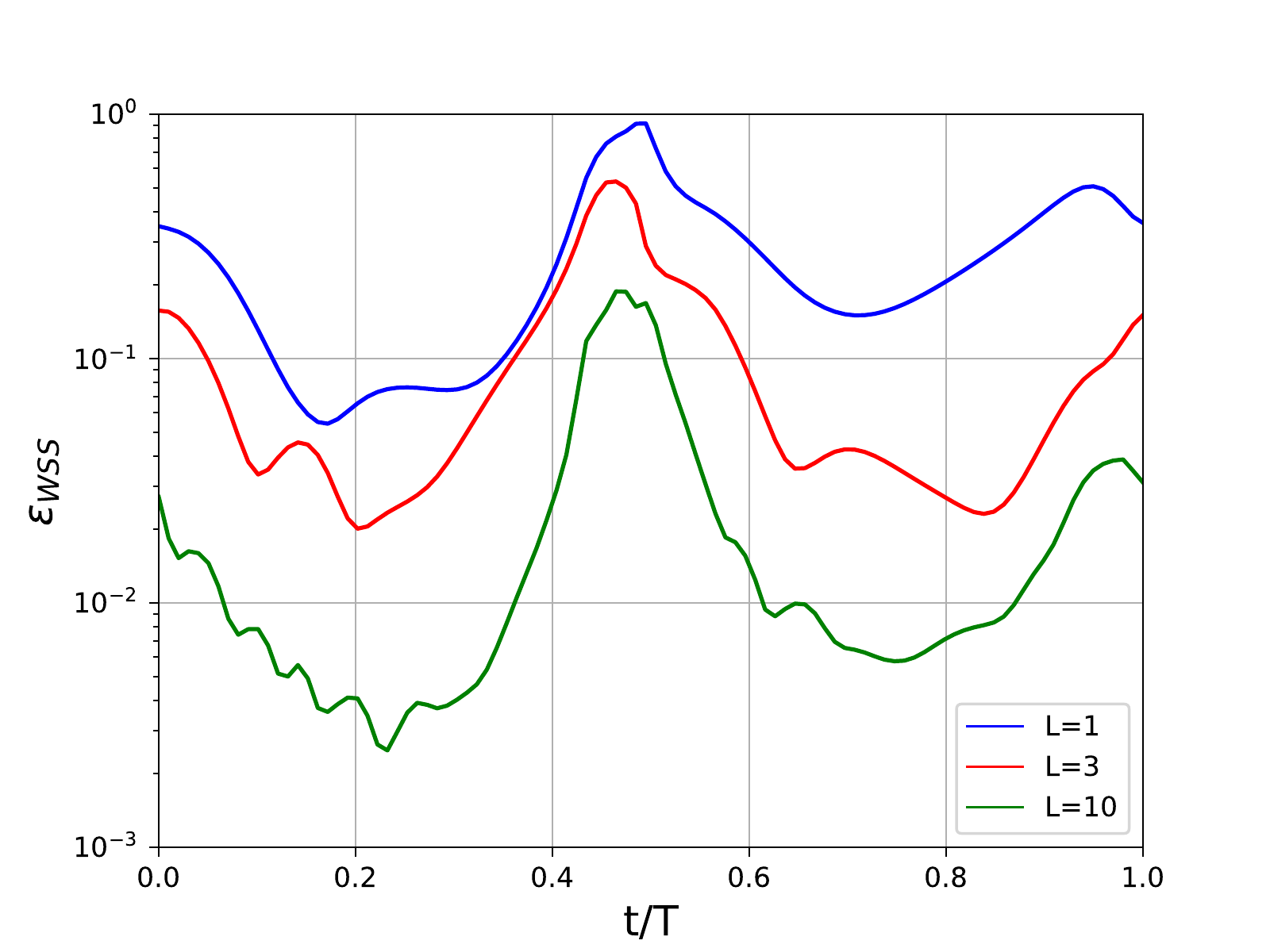}}
	\caption{Case 1: time evolution of the relative error for pressure, velocity and wall shear stress at varying of the number of modes $L$ for $\bm \mu = f_{\text{LMCA}}$.}
	\label{err_mod_lmca}
\end{figure}

Next, we carry out a convergence test with respect to the number of the modes. 
In Figures \ref{err_mod_lita} and \ref{err_mod_lmca}, all the variables show a monotonic convergence for the relative error $\varepsilon_i$: 
\begin{equation}
    \varepsilon_i = \frac{\left\lVert \Phi_i - \Phi_{\text{rb},i}\right\rVert}{\left\lVert \Phi_i \right\rVert},
    \label{erroree}
\end{equation}
with $i=p,U,WSS$, as the number of the modes is increased. Both for $\bm \mu = f_{\text{LITA}}$ and $\bm \mu = f_{\text{LMCA}}$, $L=3$ 
pressure modes (more than $99\%$ of the cumulative energy) are enough to obtain a time-averaged error of about $2\%$. 
On the other hand, for the velocity and WSS, we obtain an error of about $3\%$ 
using $L=10$ modes (more than $96\%$ of the cumulative energy). 
It should be noted that for the pressure we show only two cases ($L = 1,3$) 
because with more than 3 modes we obtain that the error is on average the same.

\begin{figure}
	\centering
	\subfloat[][\label{p_f_i:a}]{%
		\begin{overpic}[width=0.32\textwidth]{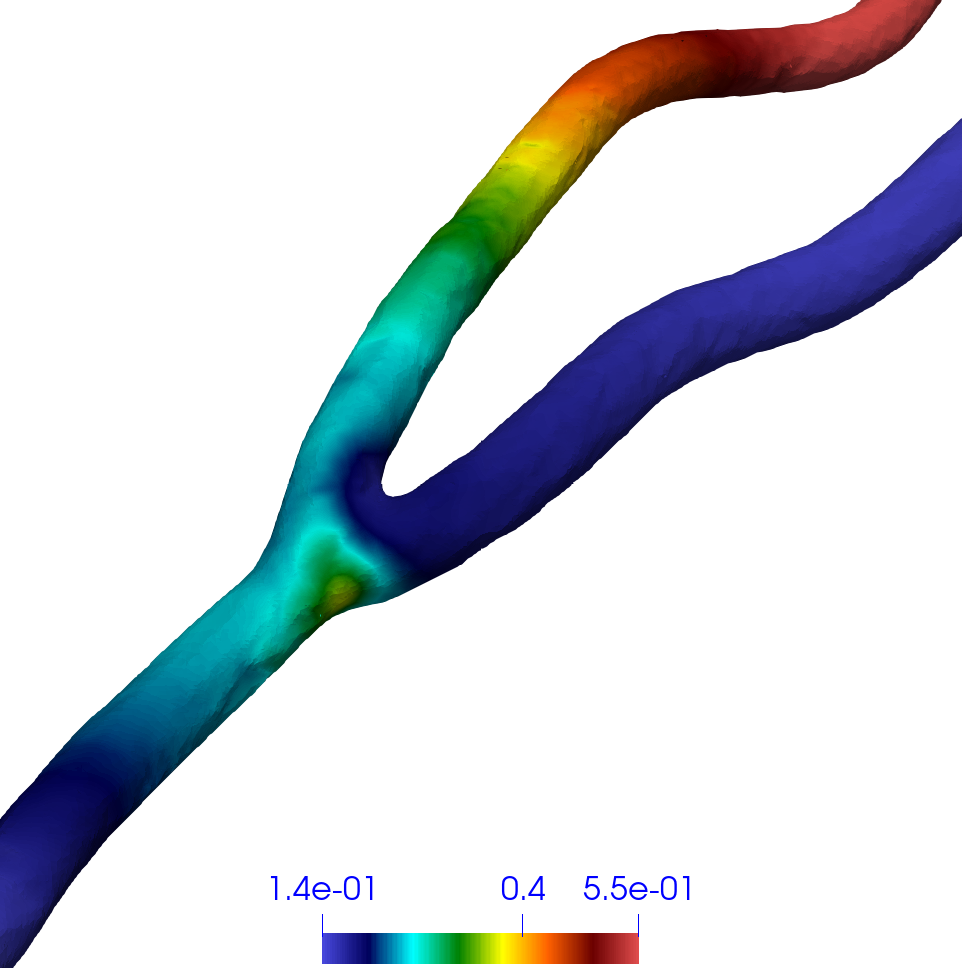}
	\put(18,105){ROM for $ \bm \mu = f_{\text{LMCA}}$}
	\end{overpic}}
	\subfloat[][\label{p_f_i:b}]{%
		\begin{overpic}[width=0.32\textwidth]{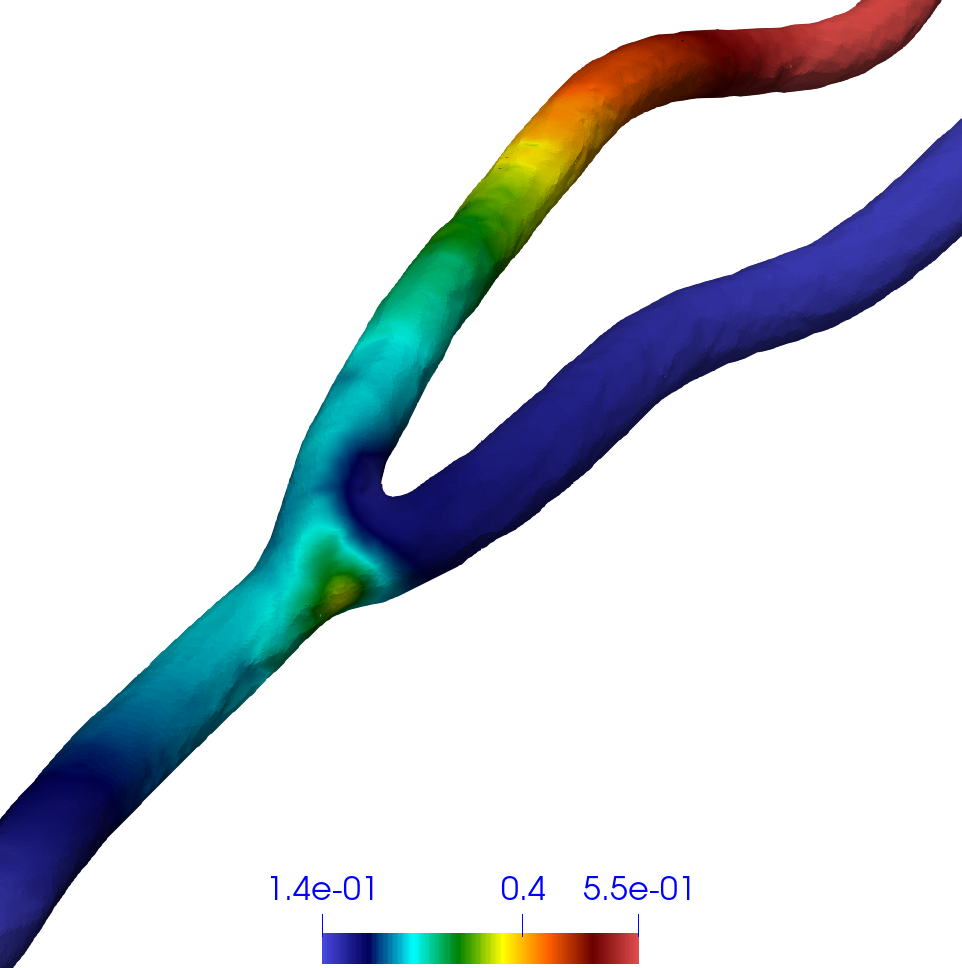}
	\put(18,105){ROM for $\bm \mu = f_{\text{LITA}}$}
	\end{overpic}}
	\subfloat[][\label{p_f_i:c}]{%
		\begin{overpic}[width=0.32\textwidth]{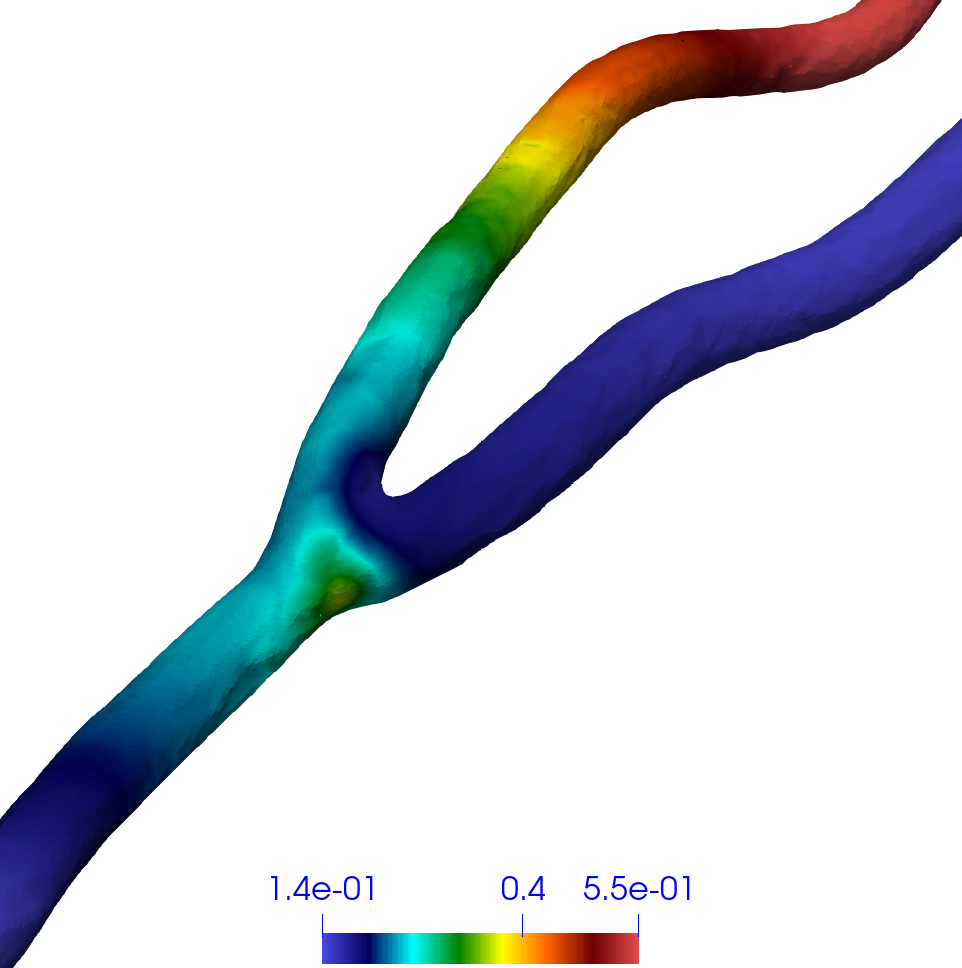}
	\put(42,105){FOM}
	\end{overpic}}
	\caption{Case 1: comparison between normalized pressure drop $P^* = P/P_{\text{max}}$ 
		in the anastomosis region computed by the FOM and by the ROM at $t/T = 0.8$ for the test point $f_{\text{LMCA}}=f_{\text{LITA}}=1$.}
    \label{p_f_i}
\end{figure}

\begin{figure}
	\centering
	\subfloat[][\label{p_f_i:c}]{%
		\begin{overpic}[width=0.4\textwidth]{p_FOM_f_1}
	\put(30,105){FOM for $ f_{\text{LITA}} = 1$}
	\end{overpic}}
	\subfloat[][\label{p_f_i:d}]{%
		\begin{overpic}[width=0.4\textwidth]{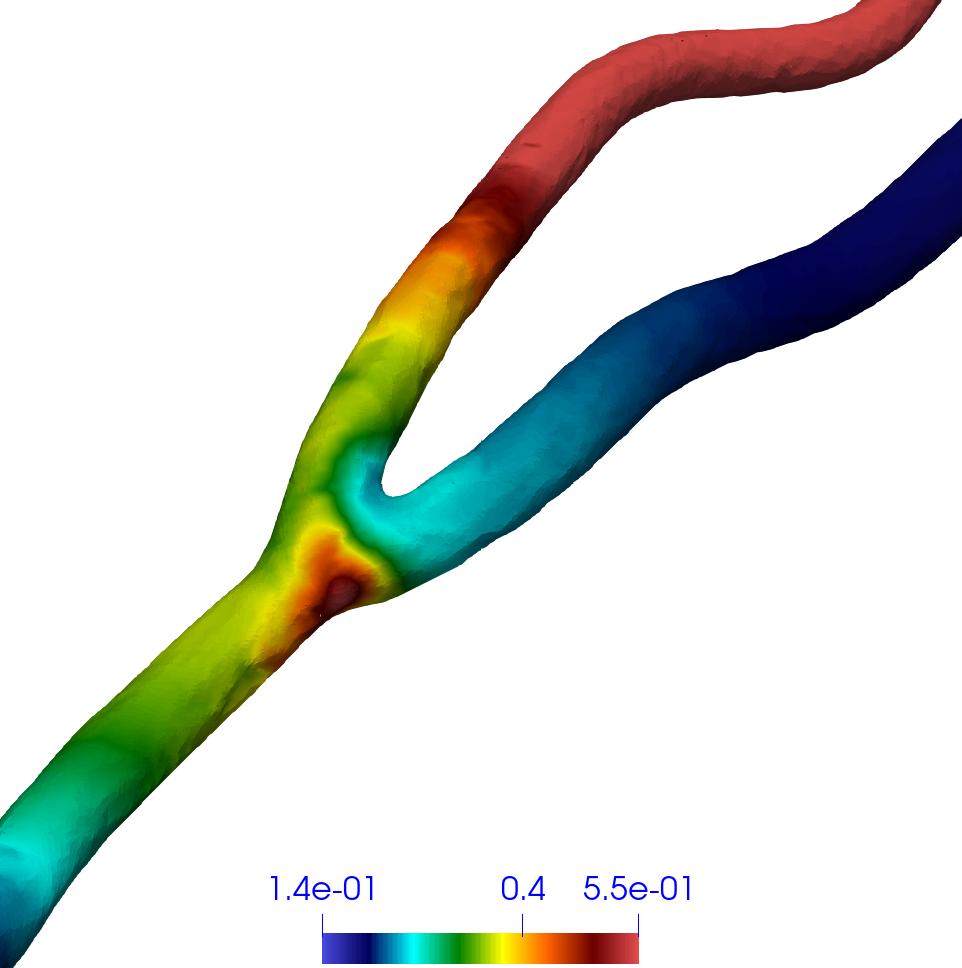}
	\put(30,105){FOM for $f_{\text{LITA}} = 1.33$}
	\end{overpic}}
	\caption{Case 1: comparison between normalized pressure drop $P^* = P/P_{\text{max}}$ 
		in the anastomosis region at $t/T = 0.8$ related to two different values of $f_{\text{LITA}}$. We set $f_{\text{LMCA}} = 1$.} 
    \label{p_f_ii}
\end{figure}

\begin{figure}
	\centering
	\subfloat[][\label{wss_f_i_sten:a}]{%
		\begin{overpic}[width=0.32\textwidth]{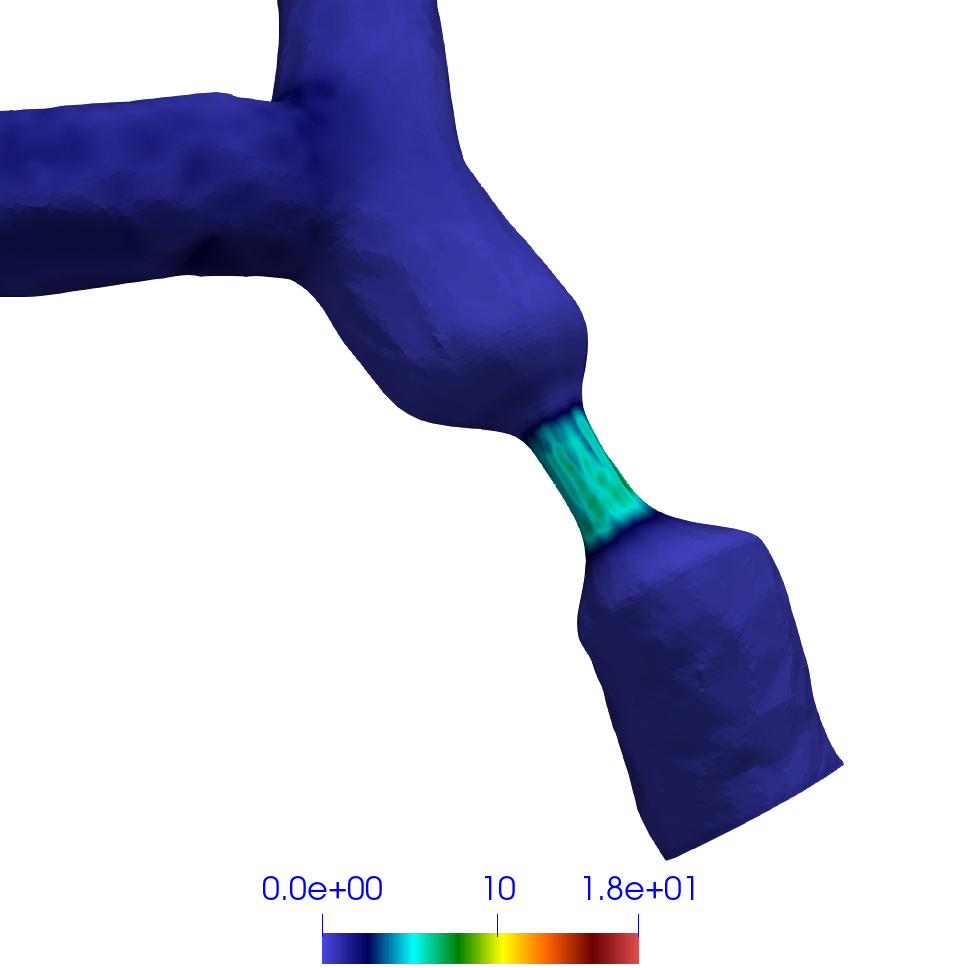}
	\put(13,105){ROM for $\bm \mu = f_{\text{LMCA}}$}
	\end{overpic}}
	\subfloat[][\label{wss_f_i_sten:b}]{%
		\begin{overpic}[width=0.32\textwidth]{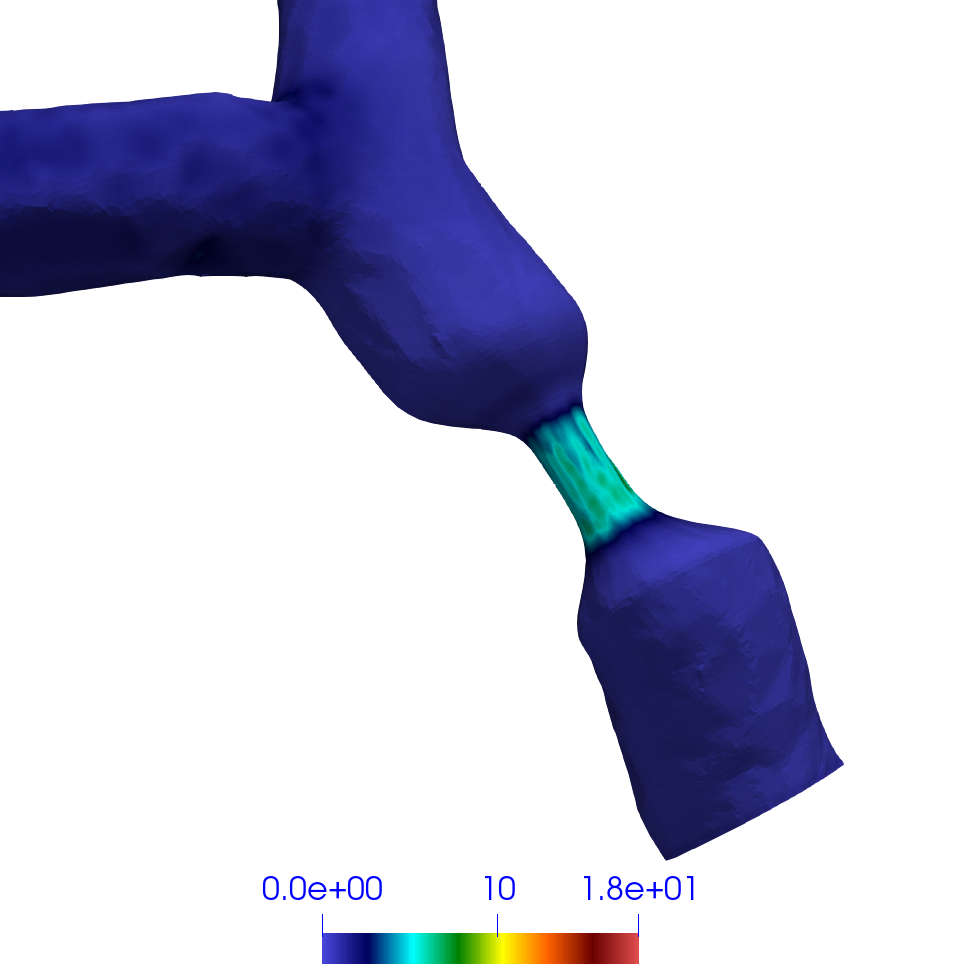}
	\put(13,105){ROM for $\bm \mu = f_{\text{LITA}}$}
	\end{overpic}}
	\subfloat[][\label{wss_f_i_sten:c}]{%
		\begin{overpic}[width=0.32\textwidth]{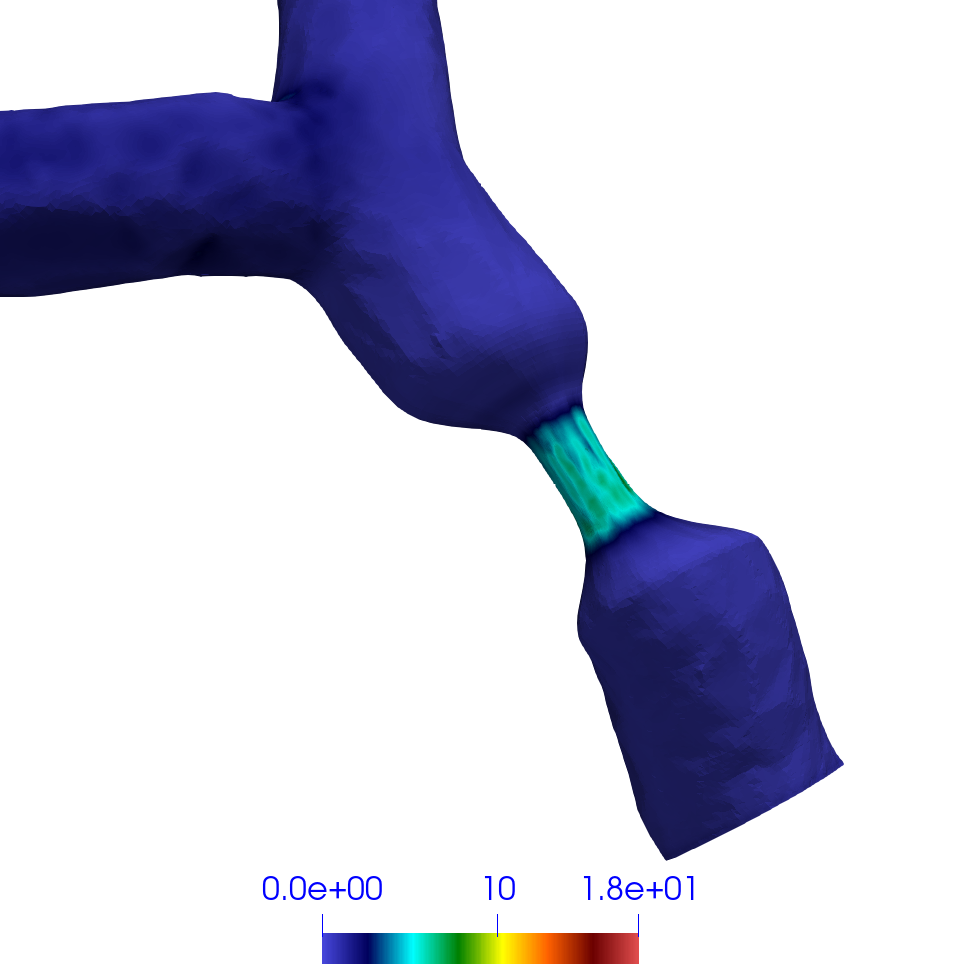}
	\put(35,105){FOM}
	\end{overpic}}\\
	\caption{Case 1: comparison between WSS (Pa) 
		in the stenosis region computed by FOM and ROM at $t/T = 0.8$ for the test point $f_{\text{LITA}} = f_{\text{LMCA}} = 1$.}
    \label{wss_f_i_sten}
\end{figure}

\begin{figure}
	\centering
	\subfloat[][\label{wss_f_i_sten_new:c}]{%
		\begin{overpic}[width=0.32\textwidth]{WSS_sten_FOM_f_1}
	\put(13,105){FOM for $f_{\text{LMCA}}=1$}
	\end{overpic}}
	\subfloat[][\label{wss_f_i_sten_new:d}]{%
		\begin{overpic}[width=0.32\textwidth]{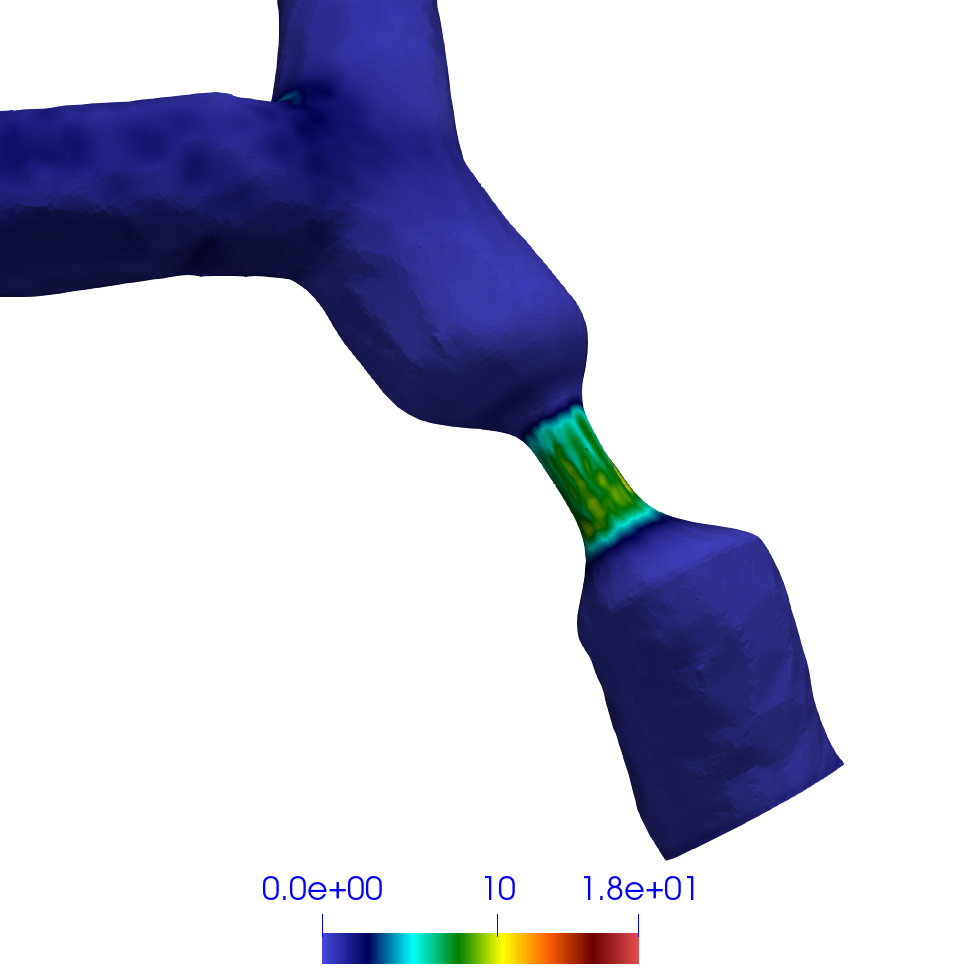}
	\put(13,105){FOM for $f_{\text{LMCA}}=1.33$}
	\end{overpic}}
	\caption{Case 1: comparison between WSS (Pa) 
		in the stenosis region at $t/T = 0.8$ related to two different values of $f_{\text{LMCA}}$. We set  $f_{\text{LITA}} = 1$.}
    \label{wss_f_i_sten_new}
\end{figure}

\begin{figure}
	\centering
	\subfloat[][\label{wss_f_i:a}]{%
		\begin{overpic}[width=0.32\textwidth]{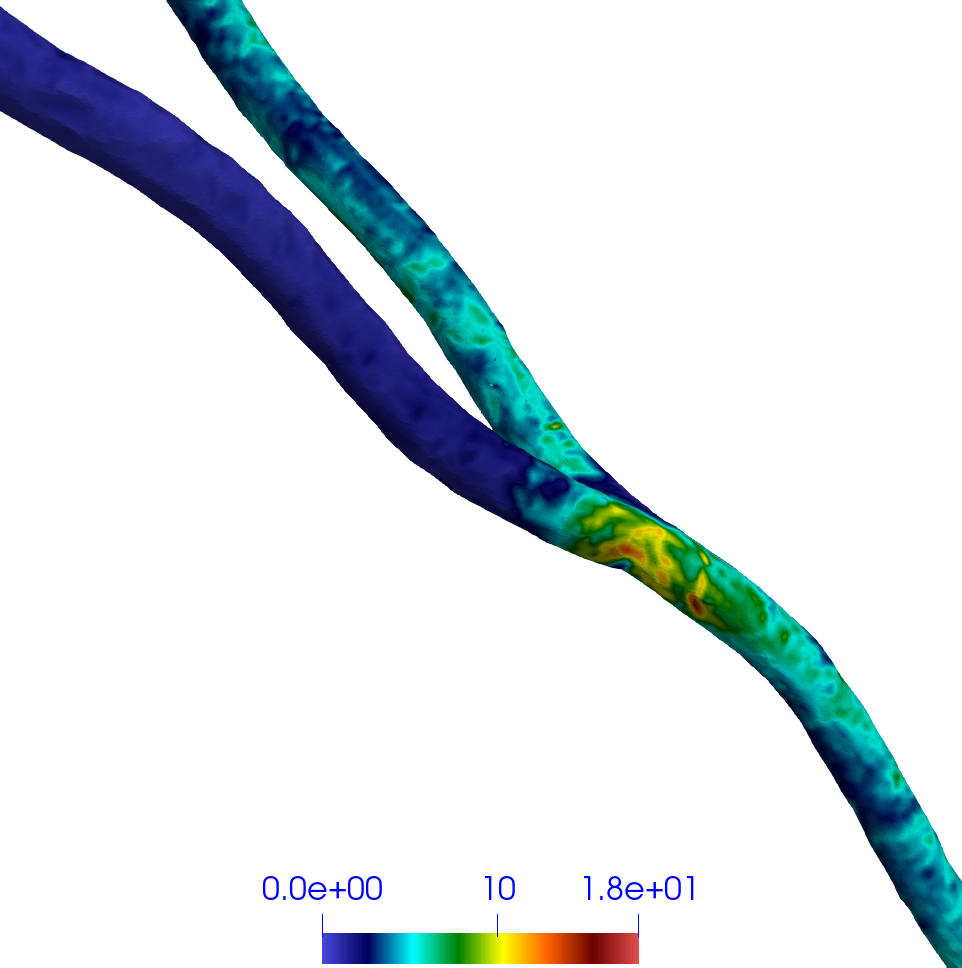}
	\put(13,105){ROM for $\bm \mu = f_{\text{LMCA}}$}
	\end{overpic}}
	\subfloat[][\label{wss_f_i:b}]{%
		\begin{overpic}[width=0.32\textwidth]{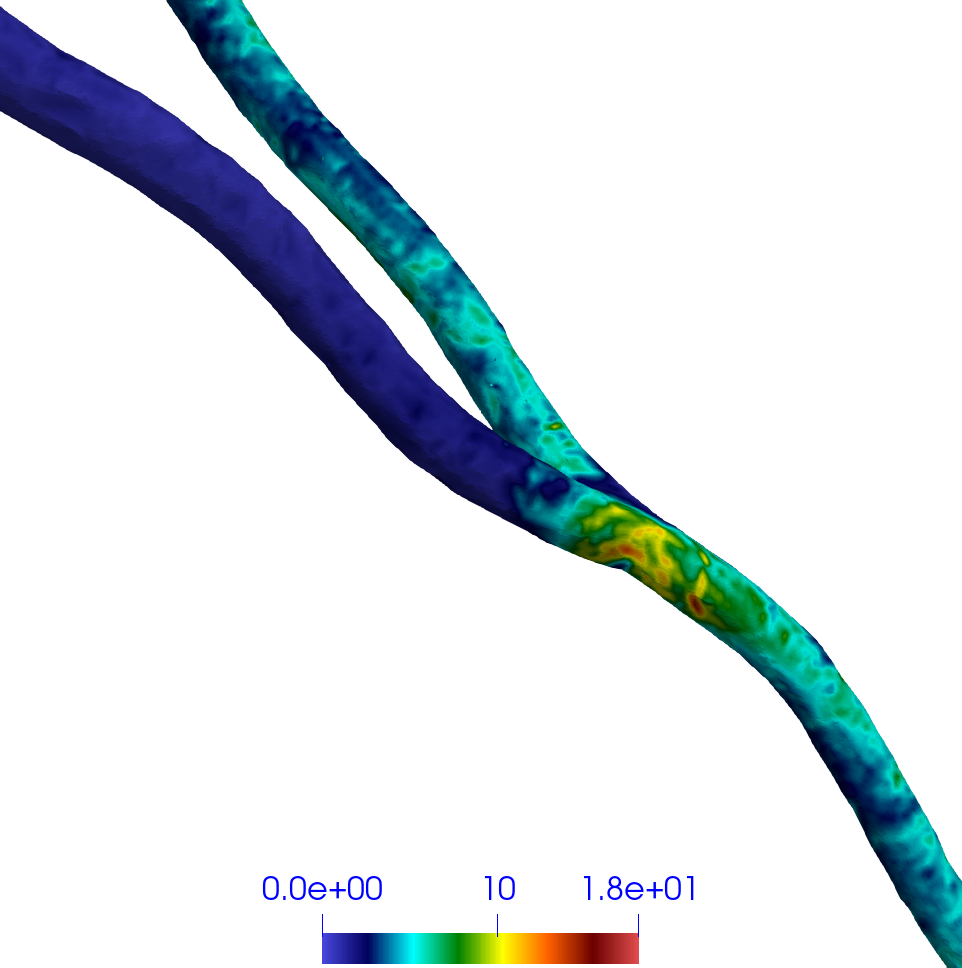}
	\put(13,105){ROM for $\bm \mu = f_{\text{LITA}}$}
	\end{overpic}}
	\subfloat[][\label{wss_f_i:c}]{%
		\begin{overpic}[width=0.32\textwidth]{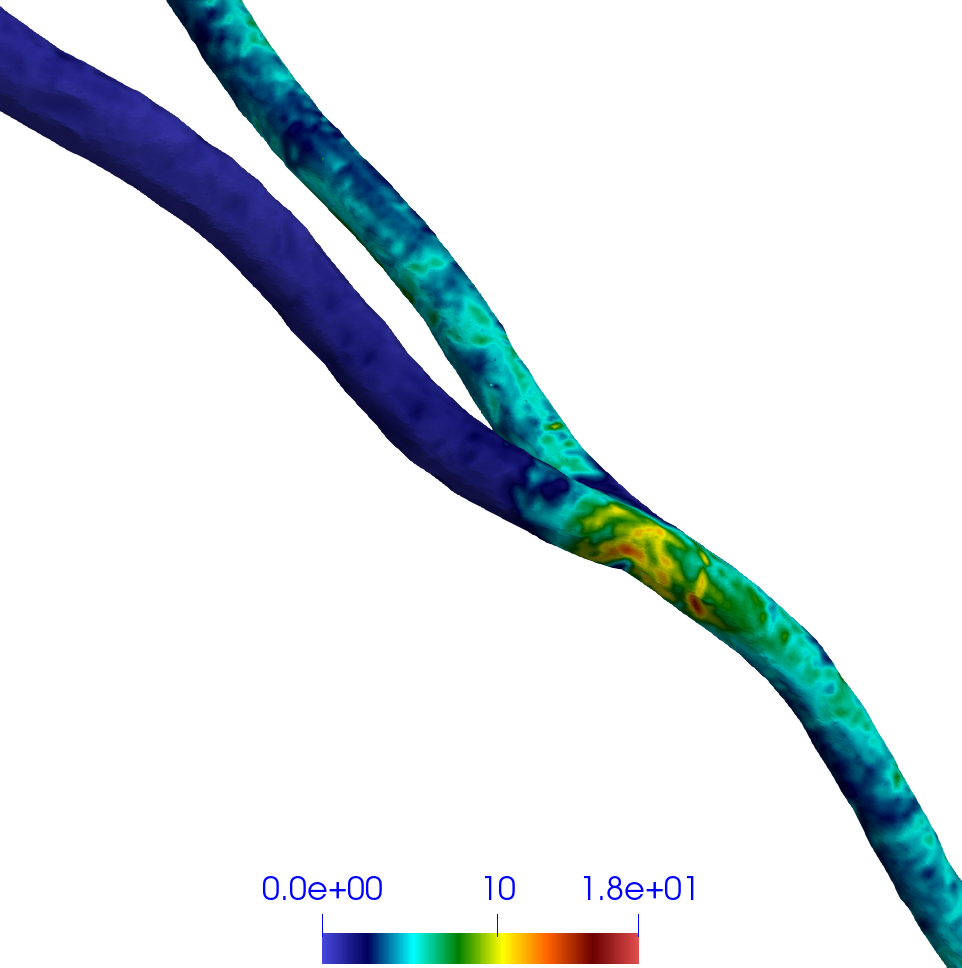}
	\put(40,105){FOM }
	\end{overpic}}\\
	\caption{Case 1: comparison between WSS (Pa) 
		in the anastomosis region computed by the FOM and by the ROM at $t/T = 0.8$ for the test point $f_{\text{LMCA}}=f_{\text{LITA}}=1$. 
		}
    \label{wss_f_i}
\end{figure}

\begin{figure}
	\centering
	\subfloat[][\label{wss_f_i_new:c}]{%
		\begin{overpic}[width=0.32\textwidth]{WSS_FOM_f_1}
	\put(13,105){FOM for $f_{\text{LITA}}=1$}
	\end{overpic}}
	\subfloat[][\label{wss_f_i_new:d}]{%
		\begin{overpic}[width=0.32\textwidth]{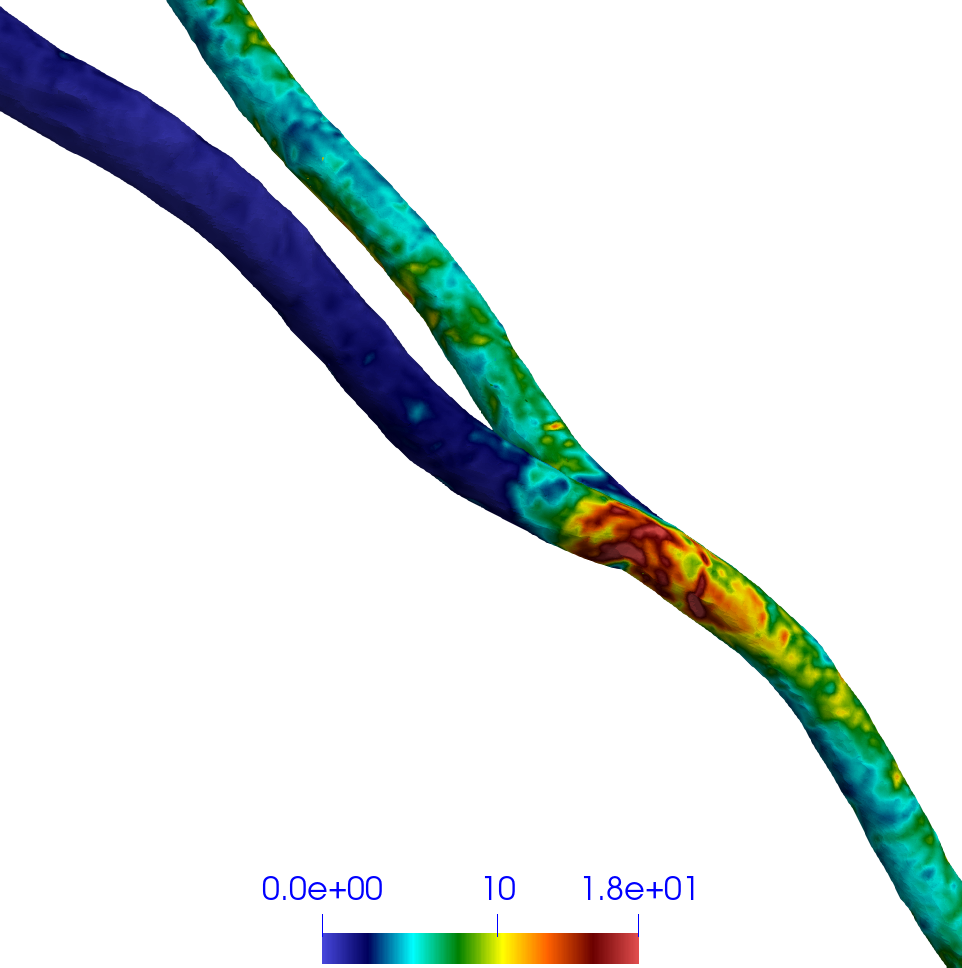} 
	\put(13,105){FOM for $f_{\text{LITA}}=1.33$}
	\end{overpic}}
	\caption{Case 1: comparison between WSS (Pa) 
		in the anastomosis region at $t/T = 0.8$ related to two different values of $f_{\text{LITA}}$. We set $f_{\text{LMCA}}=1$. 
		}
    \label{wss_f_i_new}
\end{figure}

\begin{figure}
	\centering
	\subfloat[][\label{u_f_i_graft:a}]{%
		\begin{overpic}[width=0.32\textwidth]{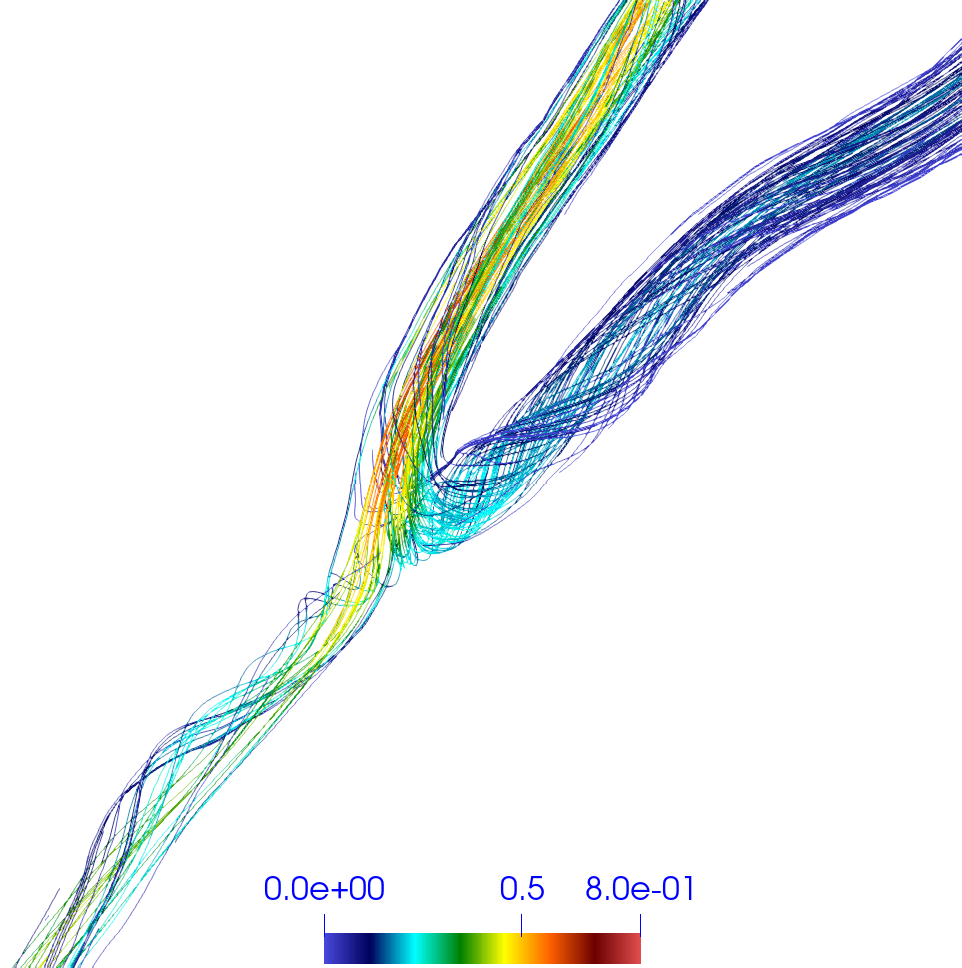}
	\put(18,105){ROM for $\bm \mu = f_{\text{LMCA}}$}
	\end{overpic}}
	\subfloat[][\label{u_f_i_graft:b}]{%
		\begin{overpic}[width=0.32\textwidth]{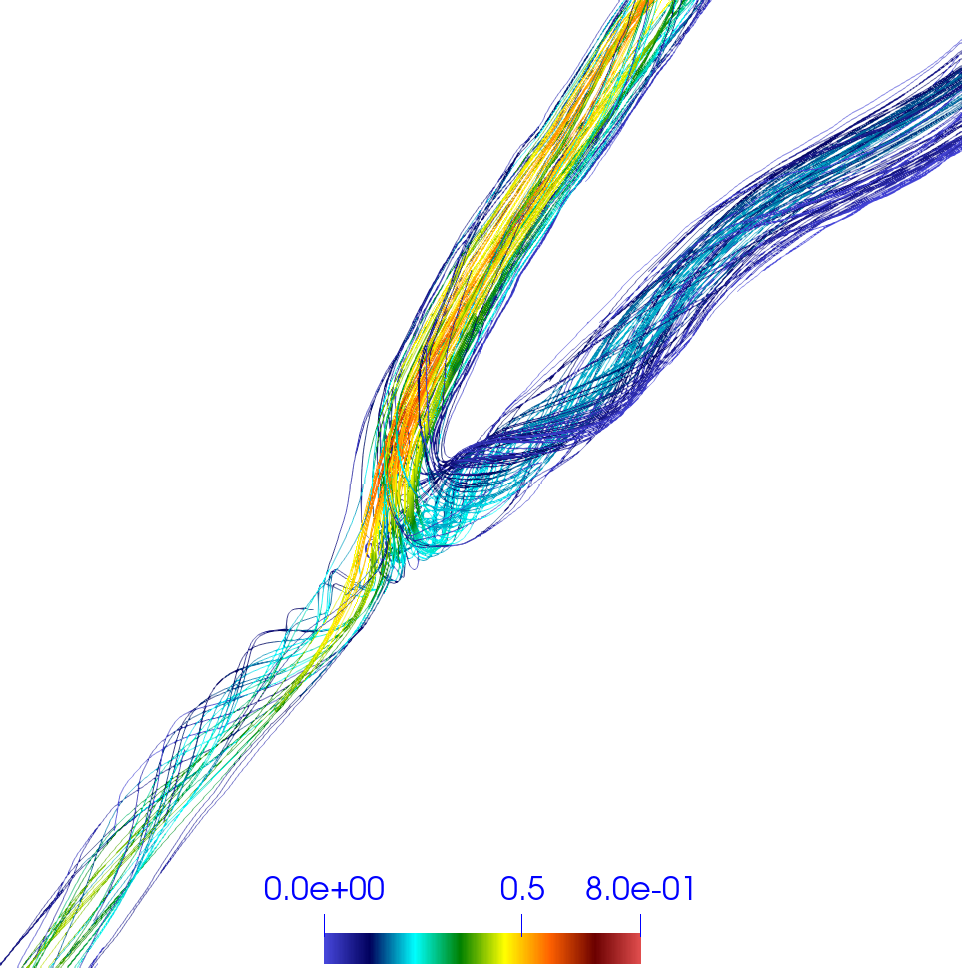}
	\put(18,105){ROM for $\bm \mu = f_{\text{LITA}}$}
	\end{overpic}}
	\subfloat[][\label{u_f_i_graft:c}]{%
		\begin{overpic}[width=0.32\textwidth]{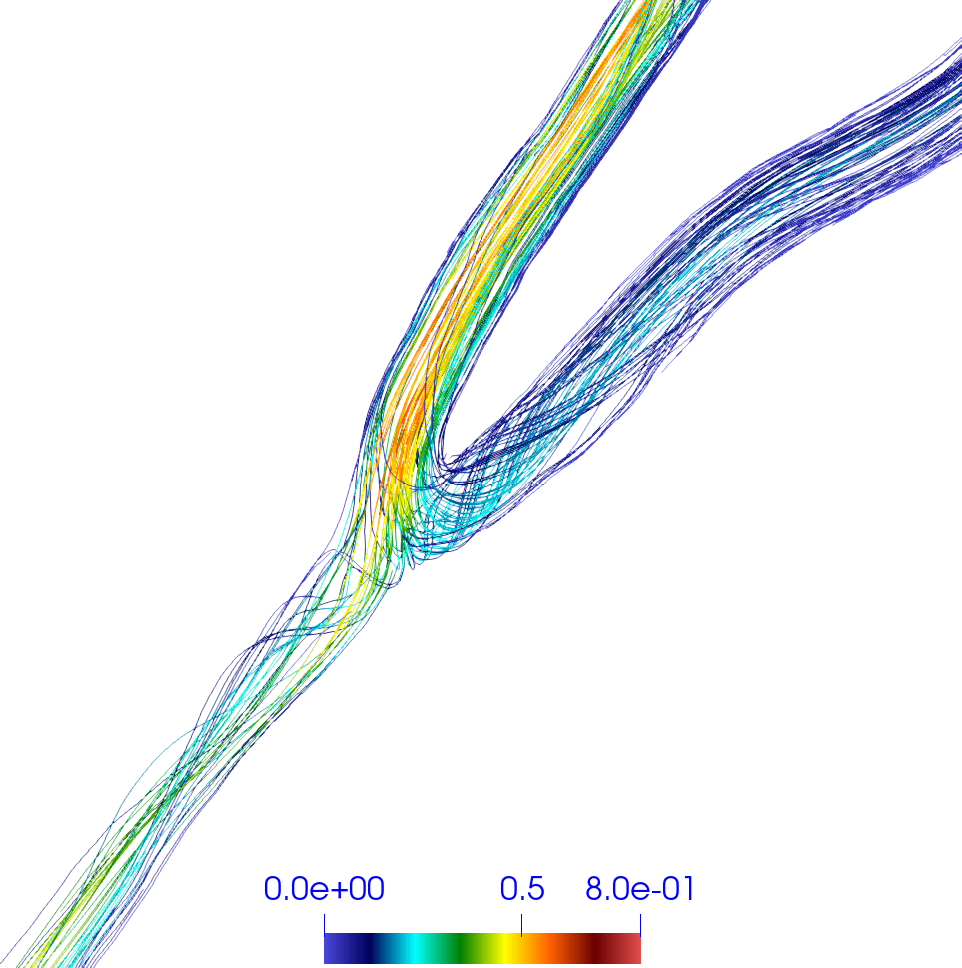}
	\put(42,105){FOM }
	\end{overpic}}
	\caption{Case 1: comparison between velocity (m/s) streamlines 
		in the anastomosis region computed by the FOM and by the ROM at $t/T = 0.8$ for the test point $f_{\text{LMCA}}=f_{\text{LITA}}=1$. 
		}
    \label{u_f_i_graft}
\end{figure}

\begin{figure}
	\centering
	
	\subfloat[][\label{u_f_i_graft_new:c}]{%
		\begin{overpic}[width=0.32\textwidth]{U_FOM_f_1_graft}
	\put(18,105){FOM for $f_{\text{LITA}}=1$}
	\end{overpic}}
	\subfloat[][\label{u_f_i_graft_new:d}]{%
		\begin{overpic}[width=0.32\textwidth]{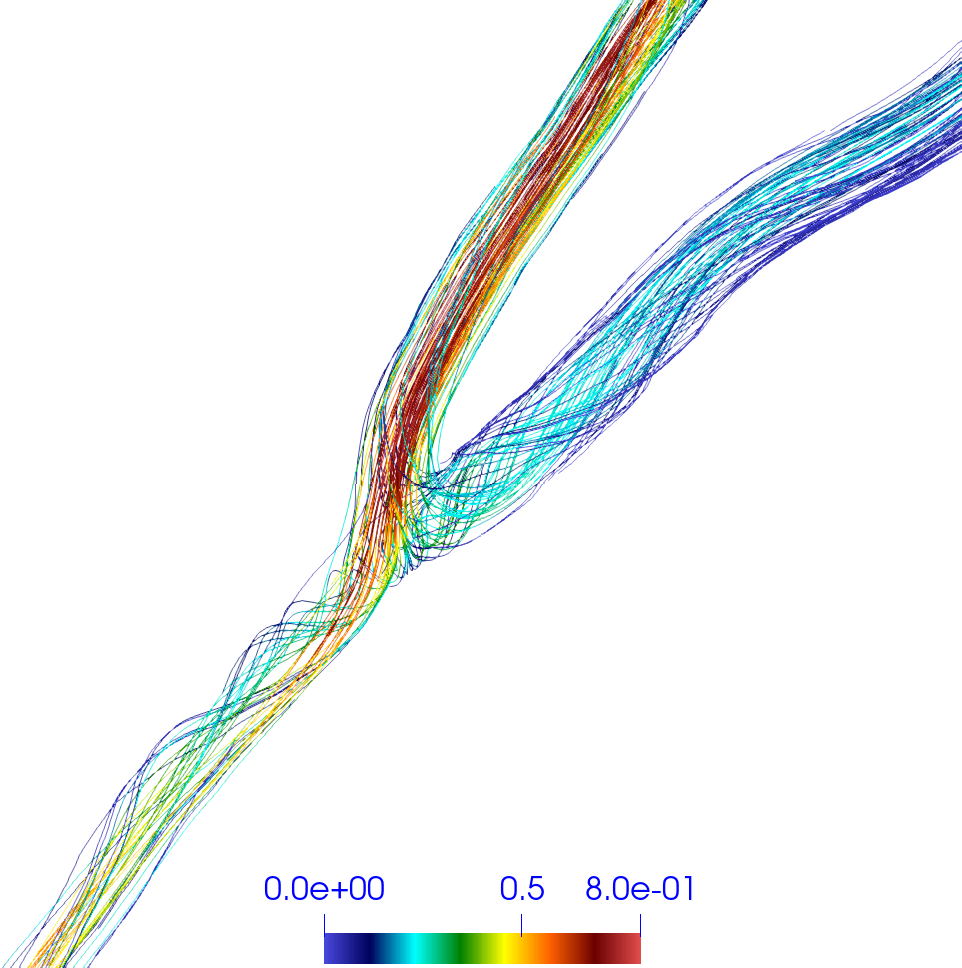}
	\put(18,105){FOM for $f_{\text{LITA}}=1.33$}
	\end{overpic}}
	\caption{Case 1: comparison between velocity (m/s) streamlines 
		in the anastomosis region at $t/T = 0.8$ for the test case $f_{\text{LMCA}}=f_{\text{LITA}}=1$ related to two different values of $f_{\text{LITA}}$. We set $f_{\text{LMCA}}=1$.
		}
    \label{u_f_i_graft_new}
\end{figure}

Qualitative comparisons between FOM and ROM at $t/T=0.8$ are displayed in Figures \ref{p_f_i}, \ref{wss_f_i_sten}, \ref{wss_f_i} and \ref{u_f_i_graft}. Moreover, in Figures \ref{p_f_ii}, \ref{wss_f_i_sten_new}, \ref{wss_f_i_new} and \ref{u_f_i_graft_new} we report some further full order solutions at the aim to provide some physical insights. We display the stenosis and the anastomosis region, which are those of major interest, because they are modifications with respect to the healthy configuration.   
In Figure \ref{p_f_i} a good ROM prediction for the pressure drop $P^*=P/P_{\text{max}}$ is found across the anastomosis. 
It is an important quantity because is related to the intimal thickening of the blood vessels and therefore it represents a significant indicator for heart diseases \cite{Loth}. As expected, if an higher flow is imposed on the LITA, the pressure increases across the anastomosis (Figure \ref{p_f_ii}).

In Figure \ref{wss_f_i_sten} and \ref{wss_f_i}, we can appreciate a good performance of our ROM in the reconstruction of WSS. We also observe that a region of locally high WSS is found across the stenosis and the anastomosis. It can represent a significant indication for the restenosis process. In addition, Figure \ref{wss_f_i_sten_new} shows that as the inlet flow rate of the LMCA is increased, also the WSS magnitude in the stenosis rises. In addition, as consequence of increased LITA inlet flow, in Figure \ref{wss_f_i_new} a significant growth of the WSS magnitude in the region of the anastomosis can be observed. 

Figure \ref{u_f_i_graft} shows the functionality of the ROM framework for the velocity.  Furthermore, we can observe that the high velocity regions coincide with the high WSS ones. 
In addition, as LITA inlet flow is increased (Figure \ref{u_f_i_graft_new}), the velocity is higher both on the LITA and the LAD, which indicate on the whole a good functionality of the bypass. 

\subsection{Case 2}%
We set $f_{\text{LMCA}} = 1.12 $ and $f_{\text{LITA}}=0.82$ adapted from \cite{Keegan,Ishida,Verim}. To train the ROM, we consider a uniform sample distribution of the stenosis severy ranging between $50\%$ to $75\%$ with step $5\%$, except $70\%$ which is considered as test point. This results in 500 snapshots. 

Cumulative eigenvalues based on the first 120 
most energetic POD modes are shown in Figure \subref*{cumulative_eig_s:c}. It can be seen that, with respect to \emph{Case 1} (Figures \subref*{cumulative_eig_s:a} and \subref*{cumulative_eig_s:b}), they increase slightly slower.

\begin{figure}
	\centering
 	\subfloat[][3rd coefficient of  P.\label{coeff:c}]{\includegraphics[width=.45\textwidth]{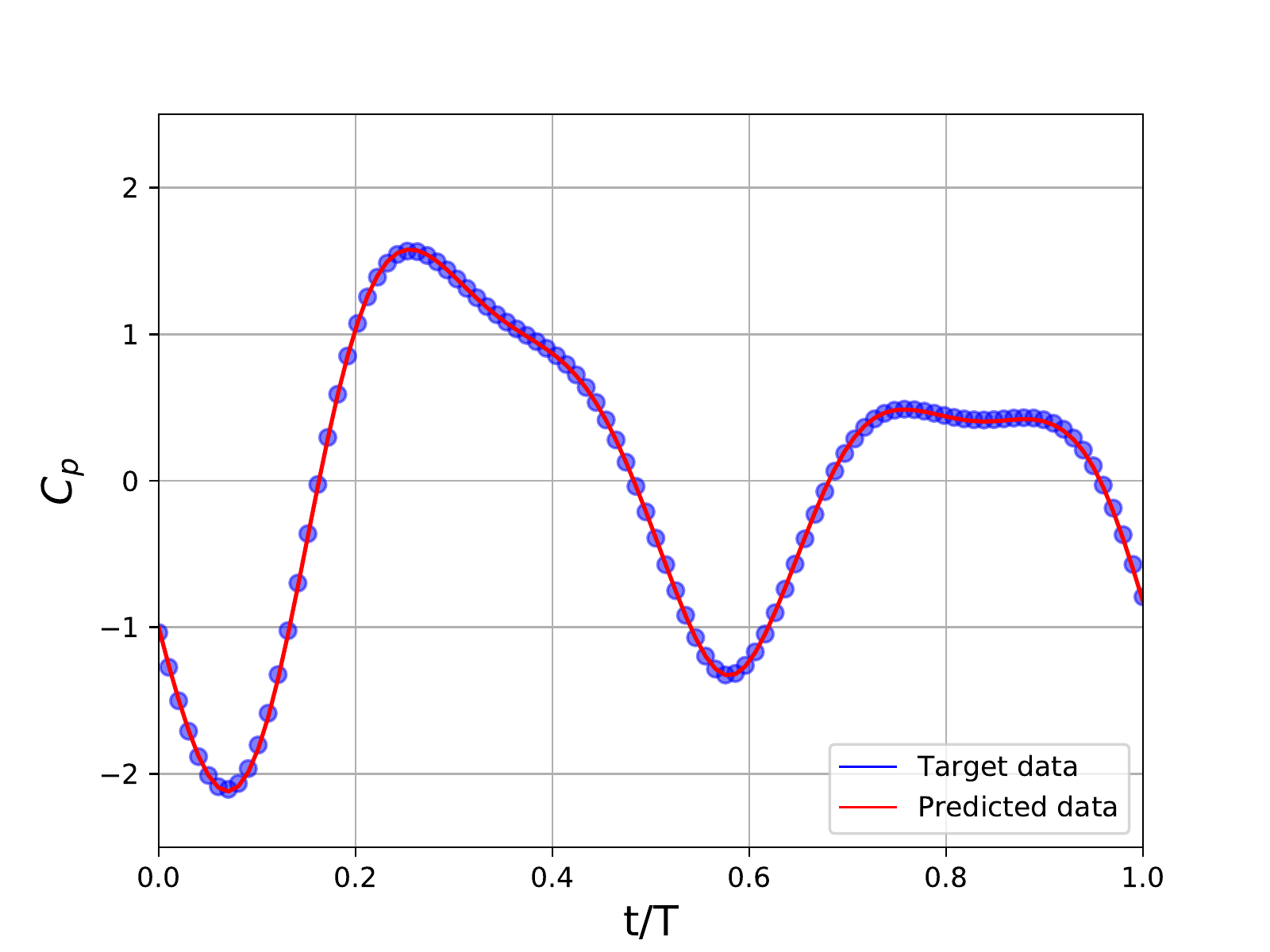}}
	\subfloat[][1st coefficient of U.\label{coeff:f}]{\includegraphics[width=.45\textwidth]{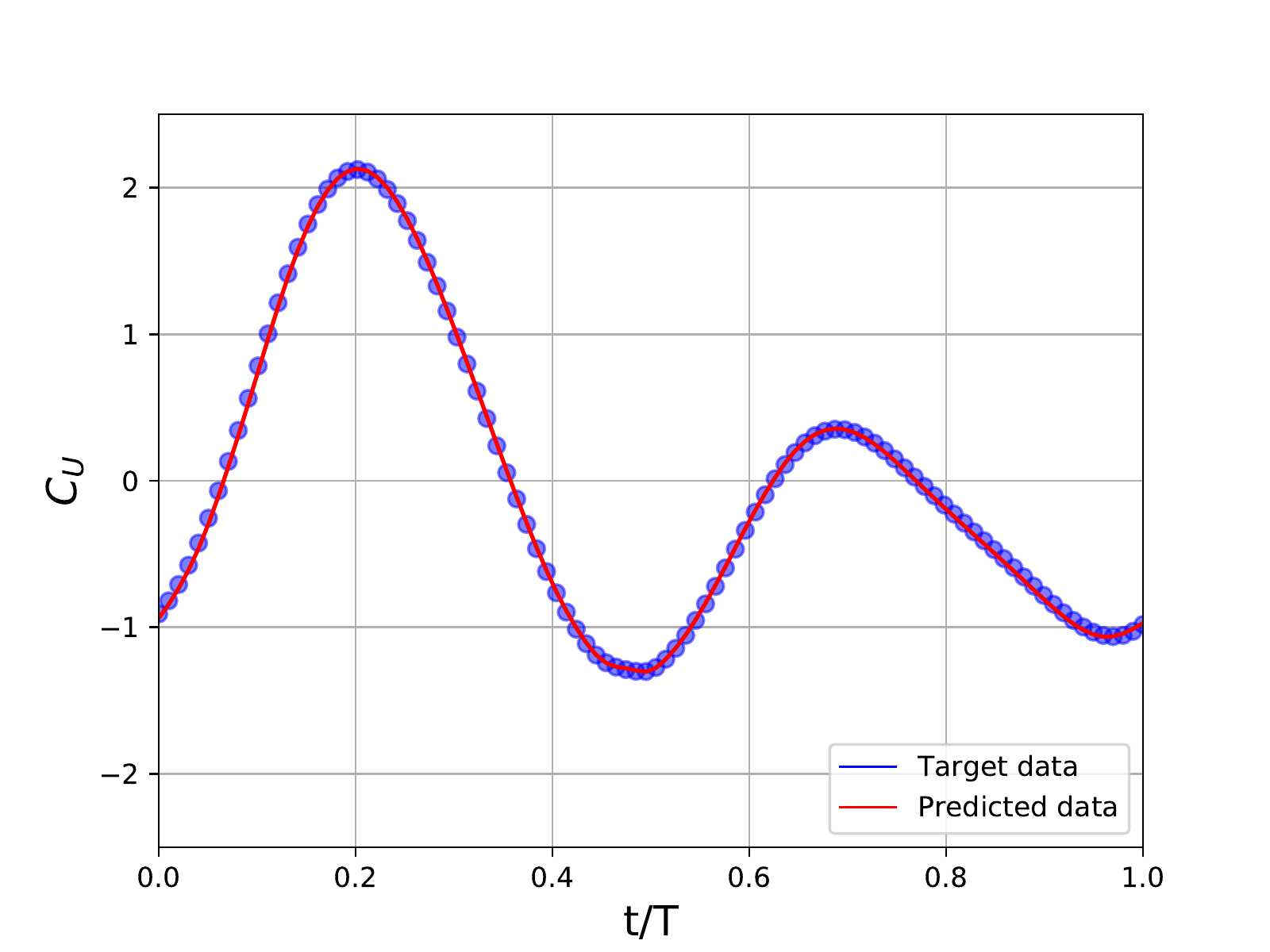}}\\
	\subfloat[][2nd coefficient of WSS.\label{coeff:i}]{\includegraphics[width=.45\textwidth]{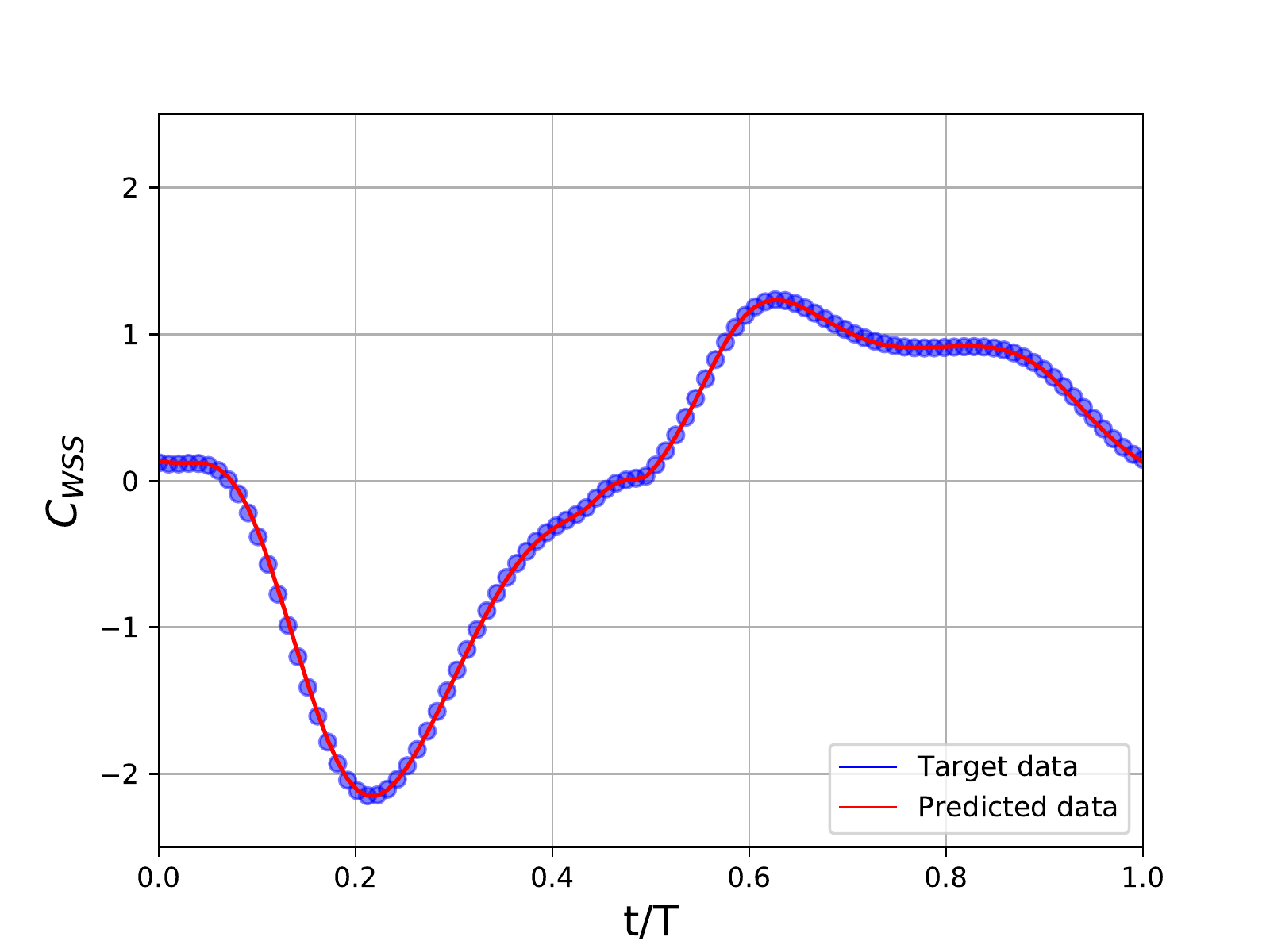}}
	\caption{
	Case 2: time evolution of some reduced coefficients including the prediction
provided by NN (red line) and the FOM simulation (blue points)
	}
	\label{coeff}
\end{figure}

The time evolution of some reduced coefficients are displayed in Figures \ref{coeff}. Even in this case, we can observe
that there is consistency between the neural newtork prediction and the FOM solution. 

\begin{figure}
	\centering
	\subfloat[][$P$ error.\label{err_mod:a}]{\includegraphics[width=.45\textwidth]{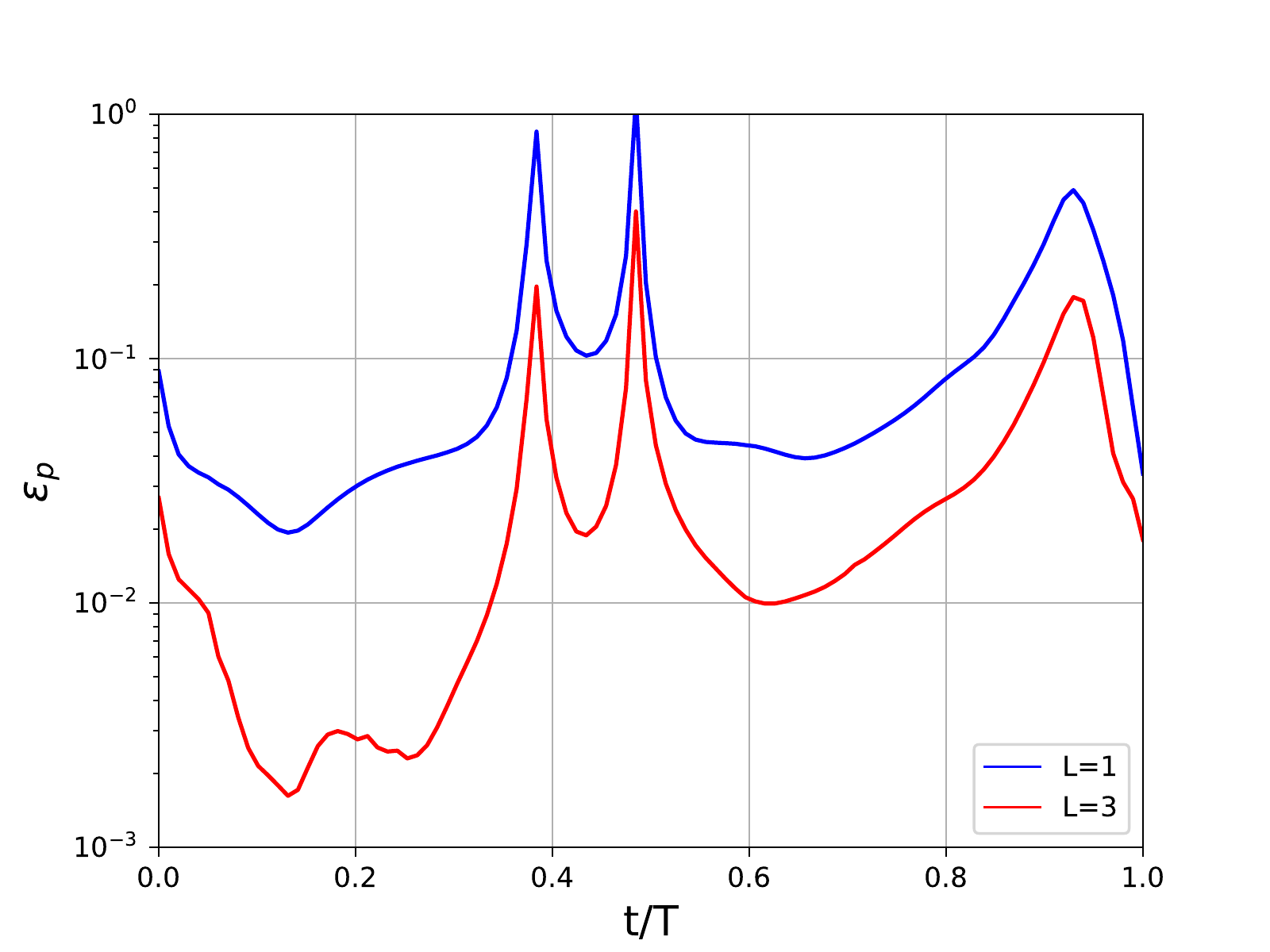}}
	\subfloat[][$\mathbf{u}$ error.\label{err_mod:b}]{\includegraphics[width=.45\textwidth]{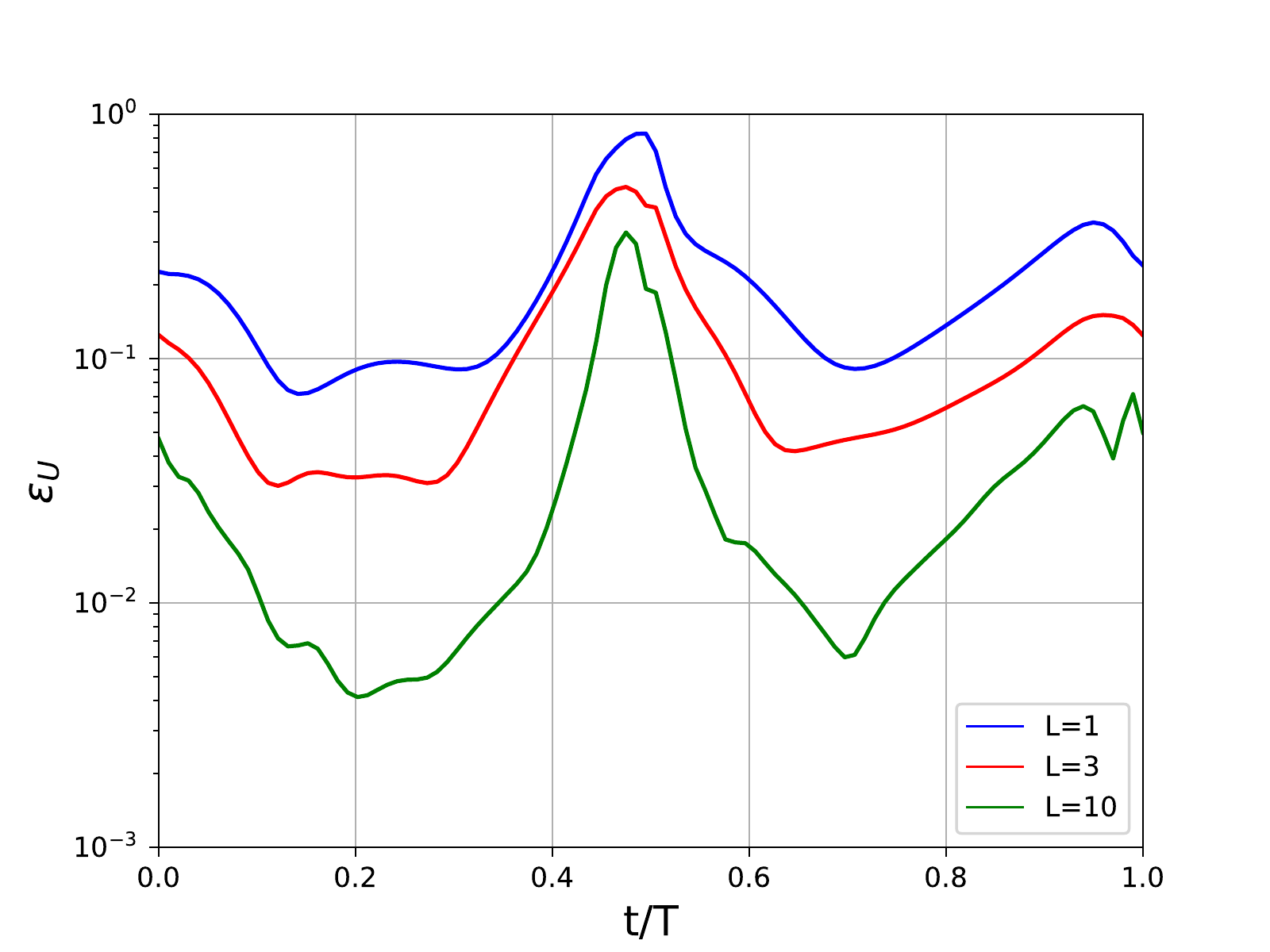}}\\
	\subfloat[][WSS  error.\label{err_mod:c}]{\includegraphics[width=.45\textwidth]{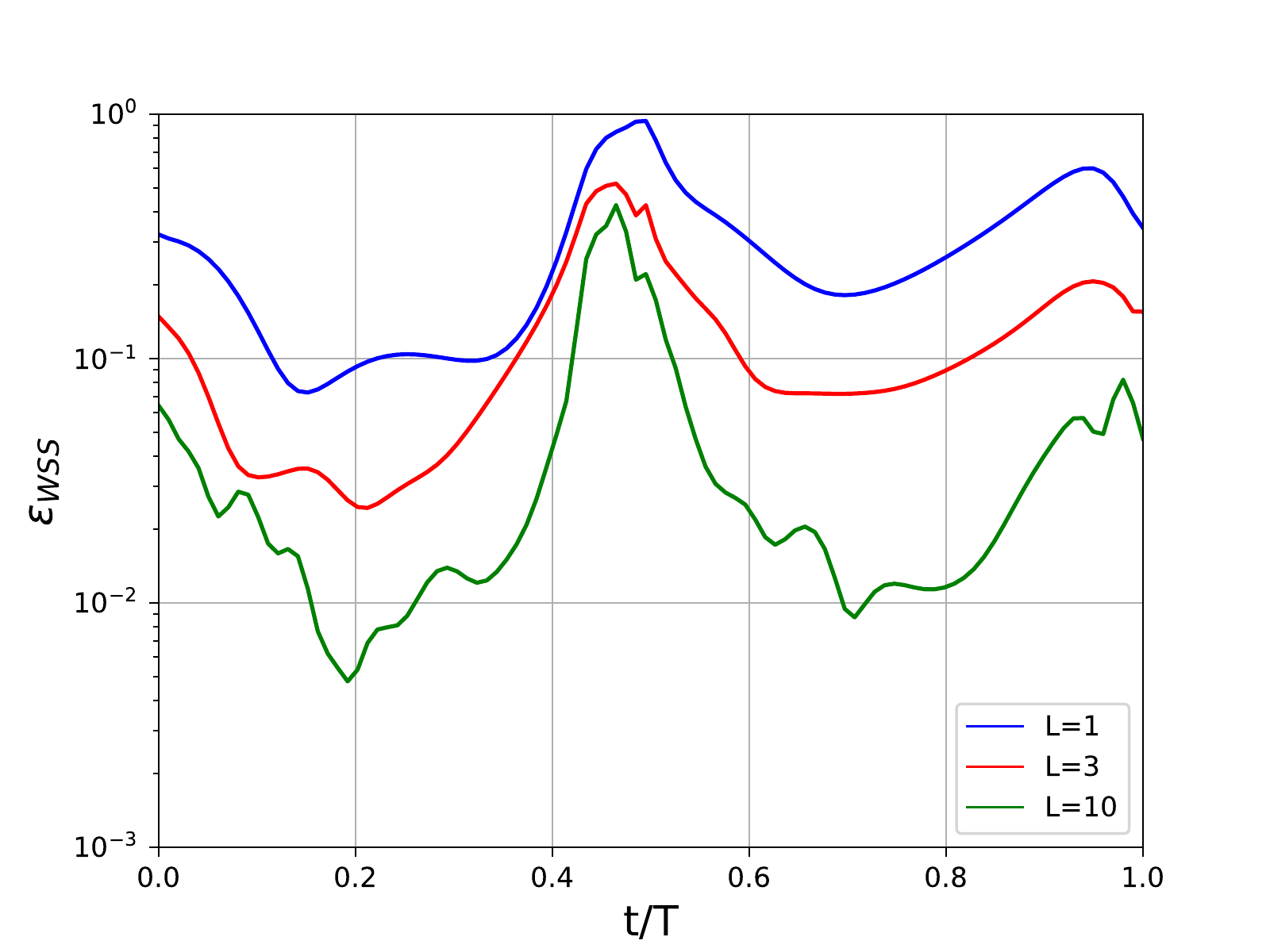}}
	\caption{
	Case 2: time evolution of the relative error for pressure, velocity and wall shear stress at varying of the number of modes $L$.} 
	\label{err_mod}
\end{figure}

A convergence test at varying of the number of the modes is reported in Figure~\ref{err_mod}. We use the same number of modes employed for \emph{Case 1}. Even in this case all the variables show a monotonic convergence for the relative error $\varepsilon_i$ (equation~\eqref{erroree}) as the number of the modes is increased. A time-averaged error of about $3.2\%$ is obtained for the pressure with $L = 3$ modes (corresponding to the 99\% of the cumulative energy) and of about $3.8\%$ and $4.9\%$ for the velocity and WSS, respectively, using $L = 10$ modes (more than 96\% of the cumulative energy).


\begin{figure}
	\centering
	\subfloat[][\label{p_stenosi:a}]{%
		\begin{overpic}[width=0.32\textwidth]{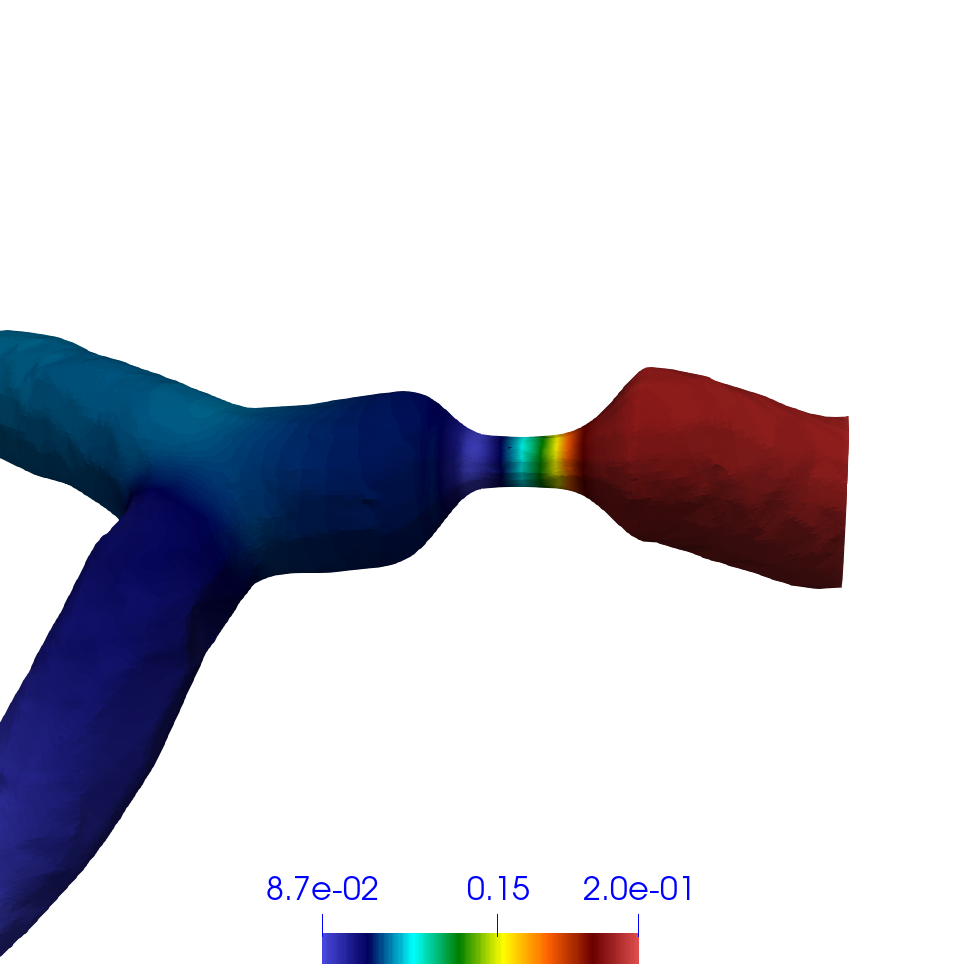}
			\put(35,80){ROM}
	\end{overpic}}
	\subfloat[][\label{p_stenosi:b}]{%
		\begin{overpic}[width=0.32\textwidth]{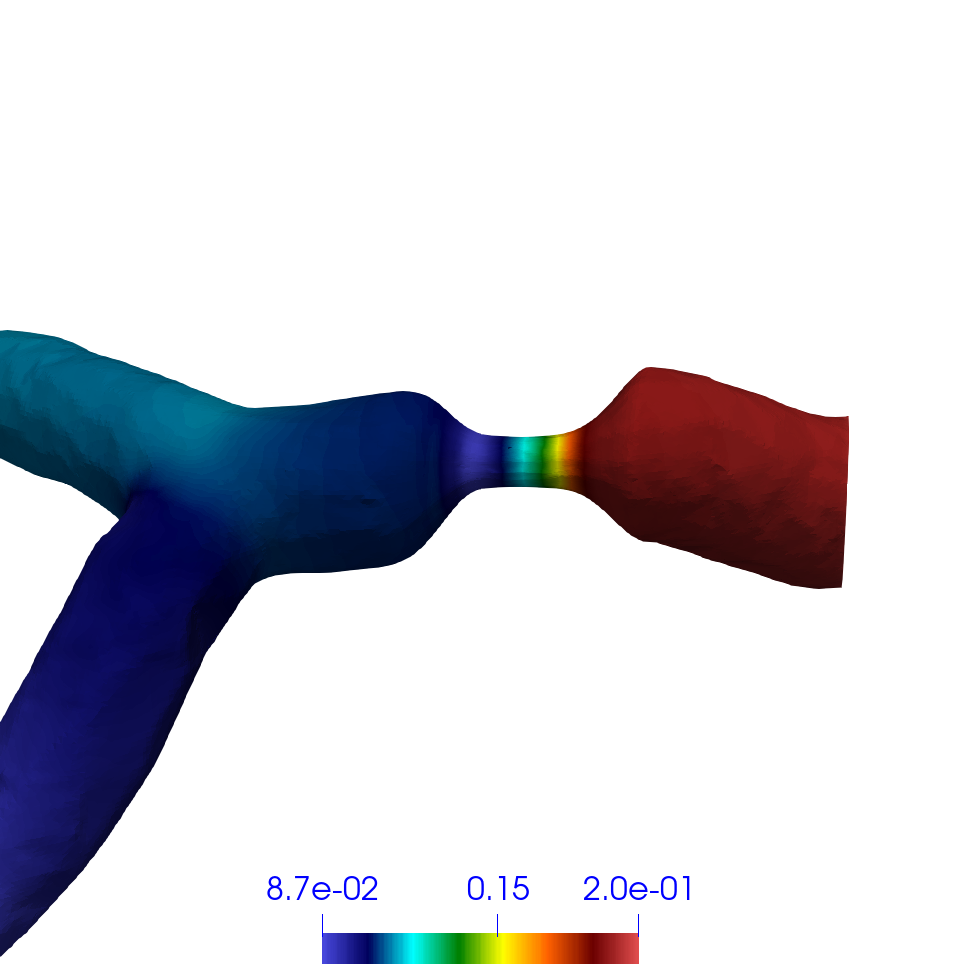}
			\put(35,80){FOM}
	\end{overpic}}

	\caption{
	Case 2: comparison between normalized pressure drop $P^* = P/P_{\text{max}}$ 
	in the stenosis region computed by the FOM and by the ROM at $t/T = 0.8$ for the test point (70\% stenosis).}
	\label{p_stenosi}
\end{figure}

\begin{figure}
	\centering
	\subfloat[][\label{p_stenosi_new:b}]{%
		\begin{overpic}[width=0.32\textwidth]{p_FOM_sten_70}
			\put(18,80){FOM for $\bm \mu = 70\%$}
	\end{overpic}}
 	\subfloat[][\label{p_stenosi_new:c}]{%
 		\begin{overpic}[width=0.32\textwidth]{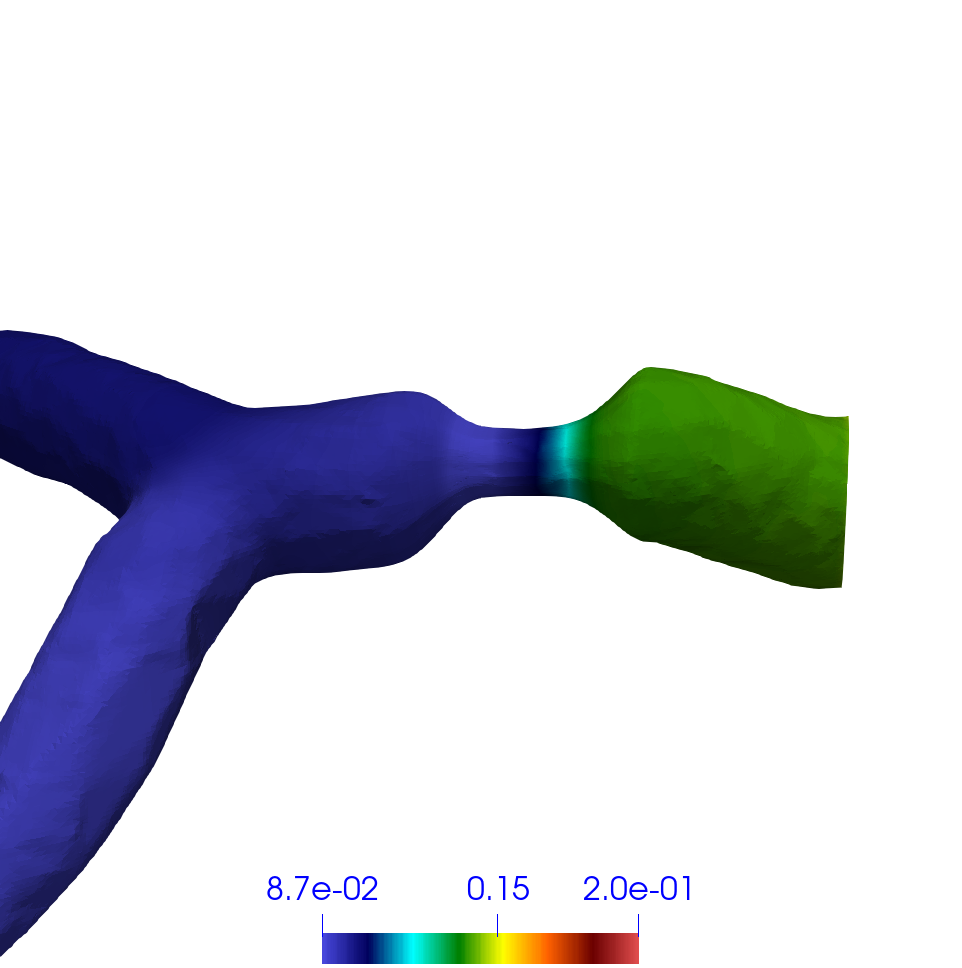}
 			\put(18,80){FOM for $\bm \mu = 60\%$}
 	\end{overpic}}
 	\subfloat[][\label{p_stenosi_new:d}]{%
 		\begin{overpic}[width=0.32\textwidth]{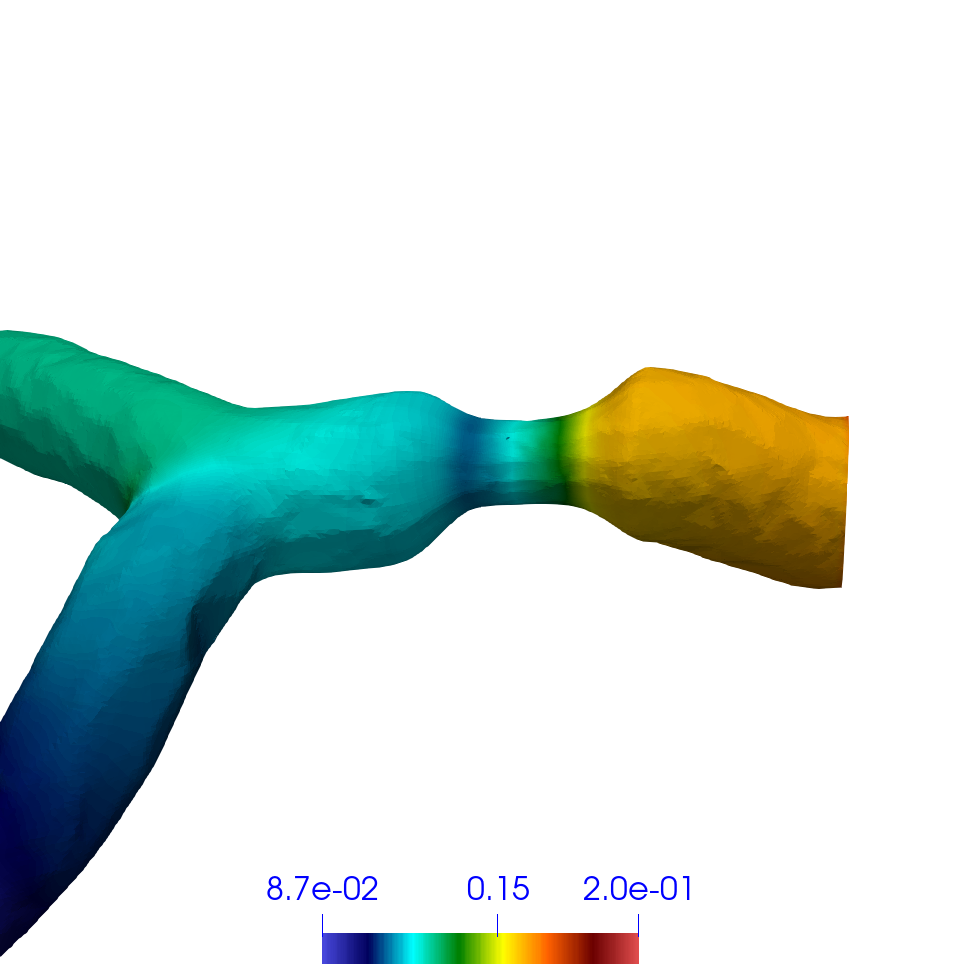}
 			\put(18,80){FOM for $\bm \mu = 50\%$}
 	\end{overpic}}

	\caption{
	Case 2: comparison between normalized pressure drop $P^* = P/P_{\text{max}}$ 
	in the stenosis region at $t/T = 0.8$ related to three different values of stenosis degree.}
	\label{p_stenosi_new}
\end{figure}

We conclude by showing a qualitative comparison between FOM and ROM simulations at $t/T = 0.8$ and giving some physical insights on the patterns showed by the variables at hand. 
As one can see from Figure \ref{p_stenosi}, our ROM is able to provide a good reconstruction of the normalized pressure drop across the stenosis. 
As expected, the pressure drop decreases with the stenosis severity (Figure \ref{p_stenosi_new}).

\begin{figure}
	\centering
	\subfloat[][\label{wss_stenosi:a}]{%
		\begin{overpic}[width=0.32\textwidth]{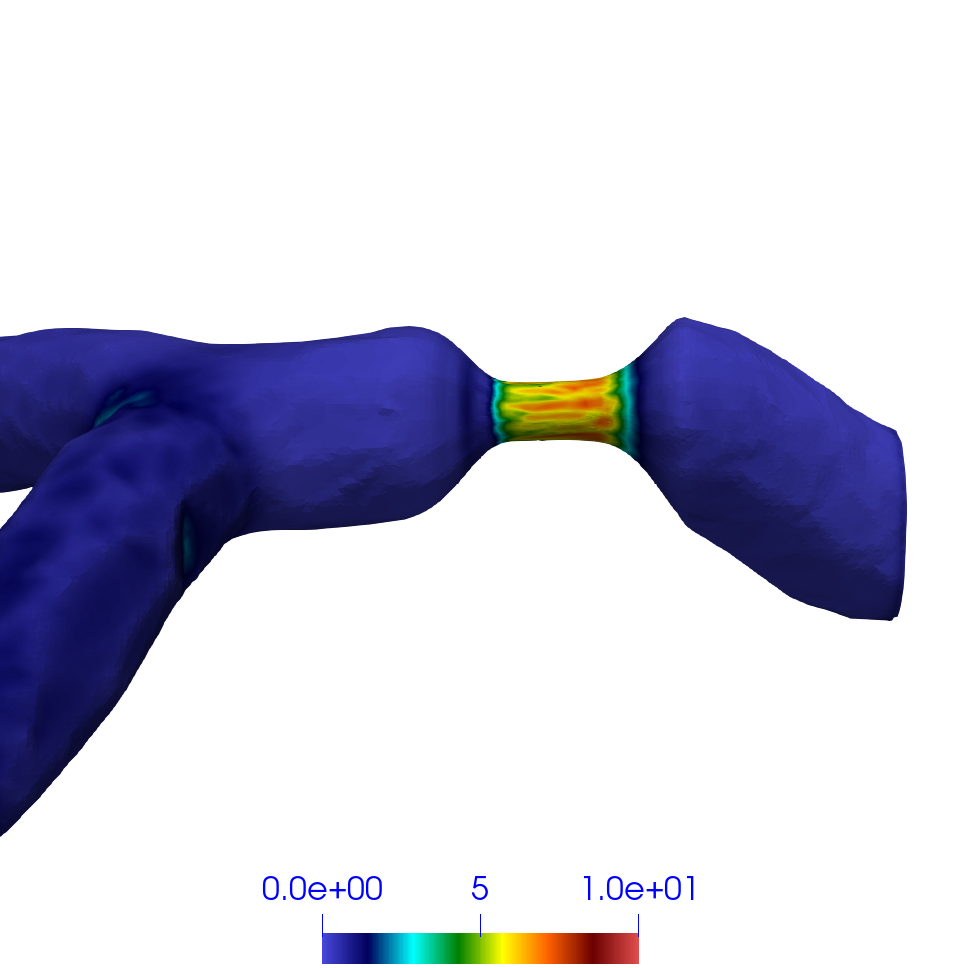}
			\put(35,80){ROM}
	\end{overpic}}
	\subfloat[][\label{wss_stenosi:b}]{%
		\begin{overpic}[width=0.32\textwidth]{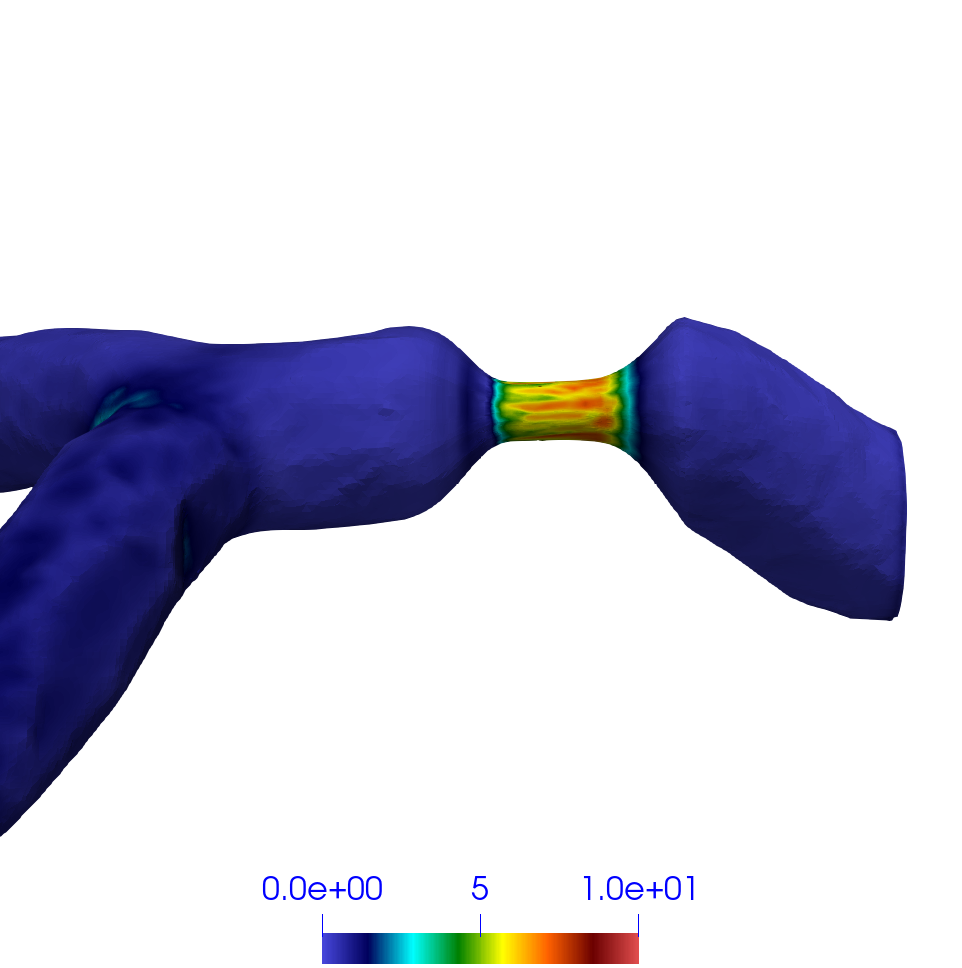}
			\put(35,80){FOM }
	\end{overpic}}

	\caption{
    Case 2: comparison between WSS (Pa) distribution
	in the stenosis region computed by the FOM and by the ROM at $t/T = 0.8$ for the test point ($70\%$ stenosis).}
	\label{wss_stenosi}
\end{figure}

\begin{figure}
	\centering
	\subfloat[][\label{wss_stenosi_new:b}]{%
		\begin{overpic}[width=0.32\textwidth]{WSS_FOM_sten_70}
			\put(18,80){FOM  for $\bm \mu = 70\%$}
	\end{overpic}}
	\subfloat[][\label{wss_stenosi_new:c}]{%
		\begin{overpic}[width=0.32\textwidth]{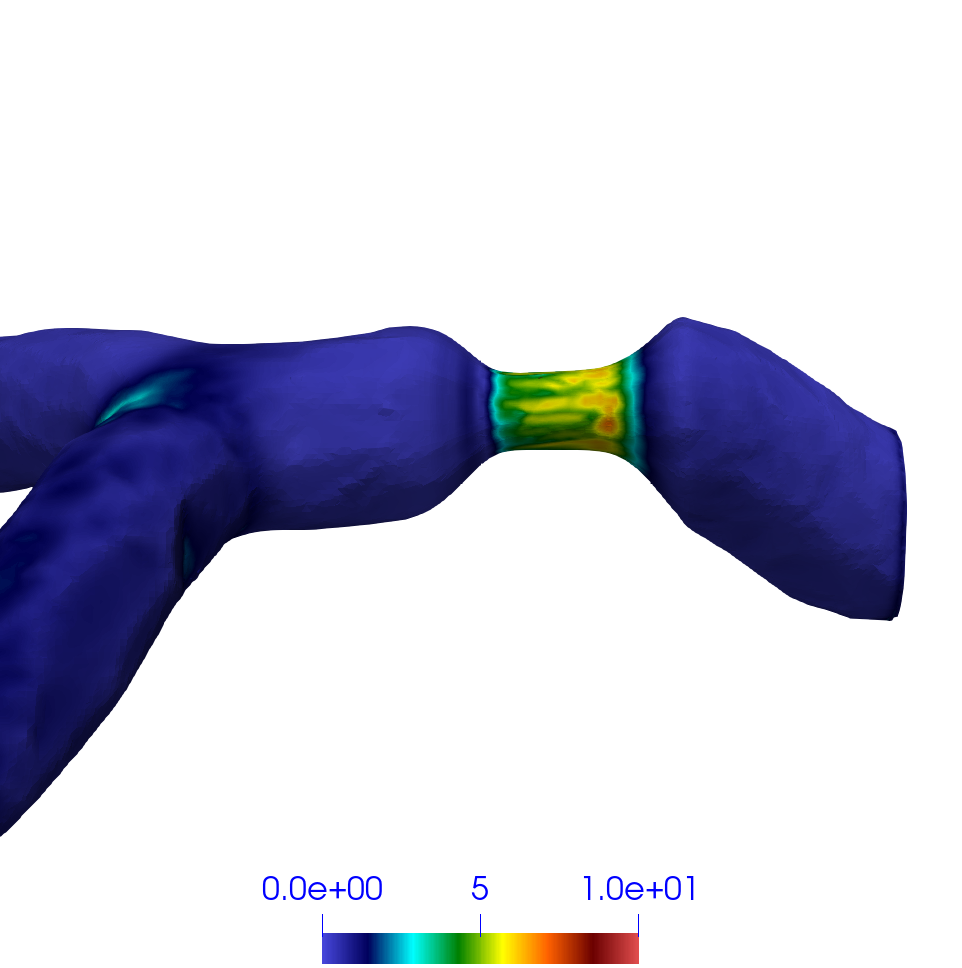}
			\put(18,80){FOM for $\bm \mu = 60\%$}
	\end{overpic}}
	\subfloat[][\label{wss_stenosi_new:d}]{%
		\begin{overpic}[width=0.32\textwidth]{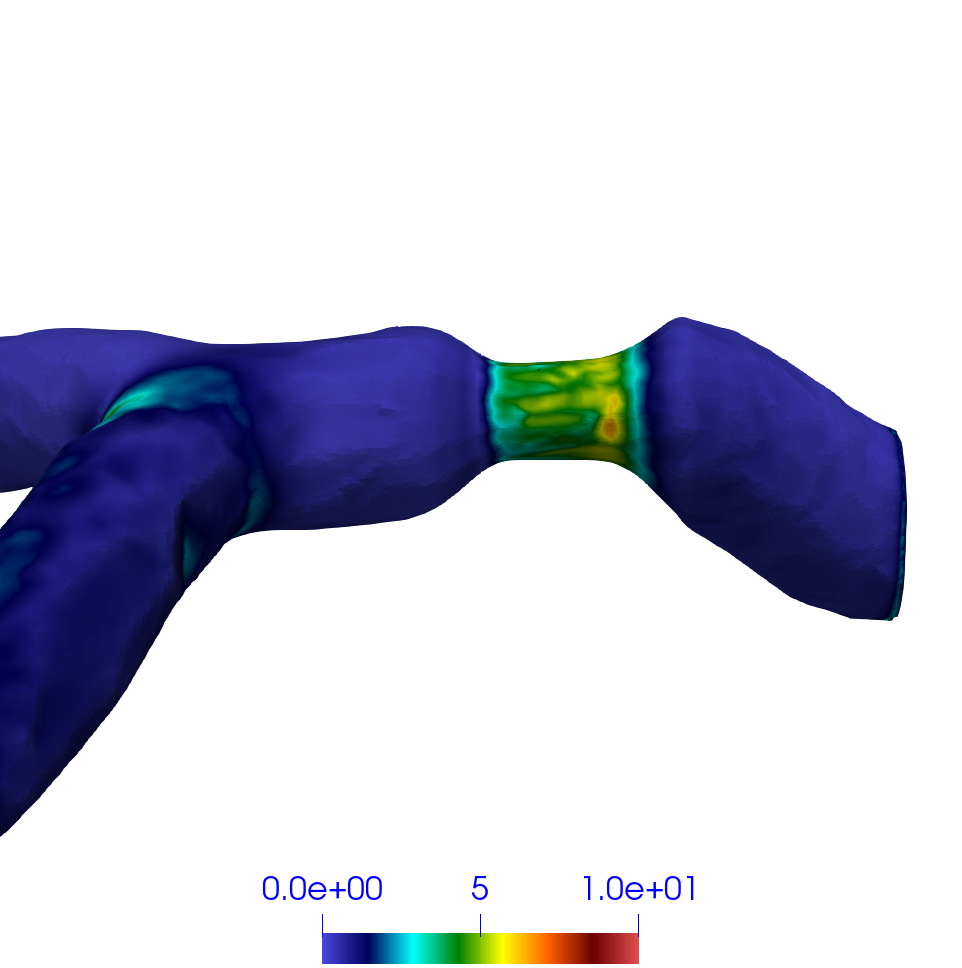}
			\put(18,80){FOM for $\bm \mu = 50\%$}
	\end{overpic}}

	\caption{
	Case 2: comparison between WSS (Pa) distribution
	in the stenosis region at $t/T = 0.8$ related to three different values of stenosis degree.}
	\label{wss_stenosi_new}
\end{figure}

\begin{figure}
	\centering
	\subfloat[][\label{wss_stenosi_graft:a}]{%
		\begin{overpic}[width=0.32\textwidth]{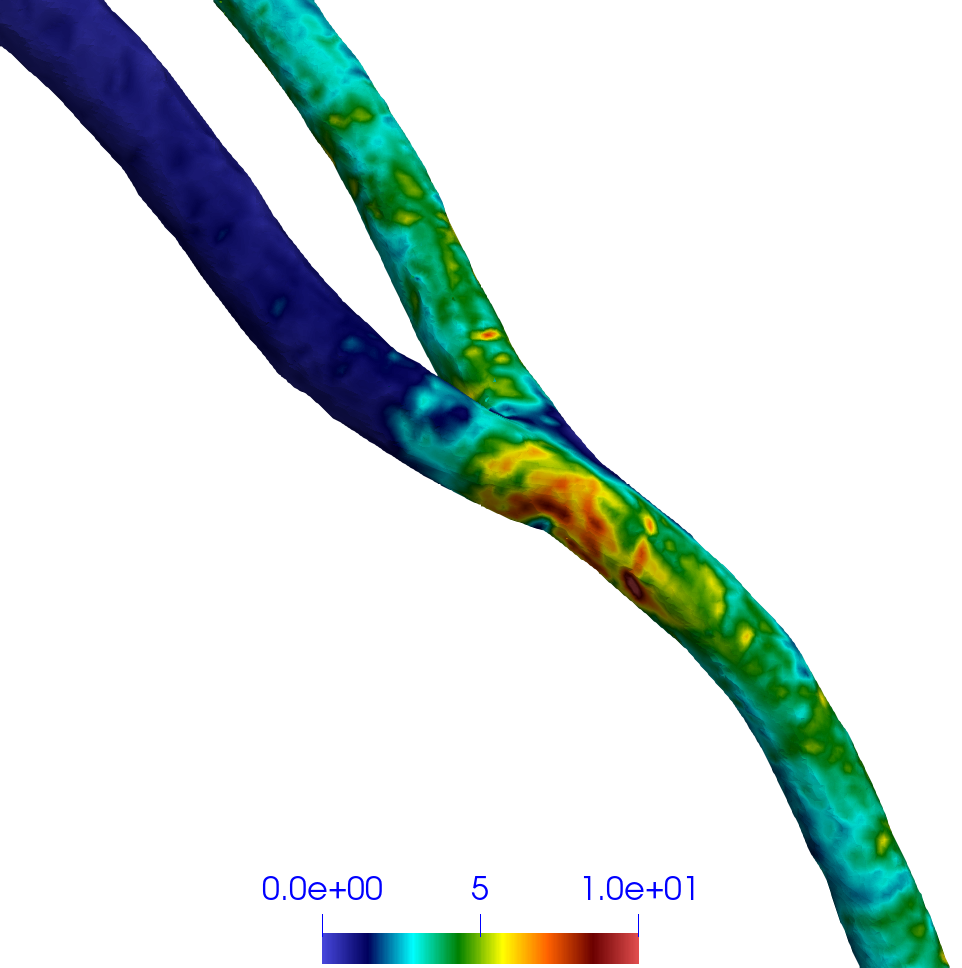}
			\put(35,105){ROM}
	\end{overpic}}
	\subfloat[][\label{wss_stenosi_graft:b}]{%
		\begin{overpic}[width=0.32\textwidth]{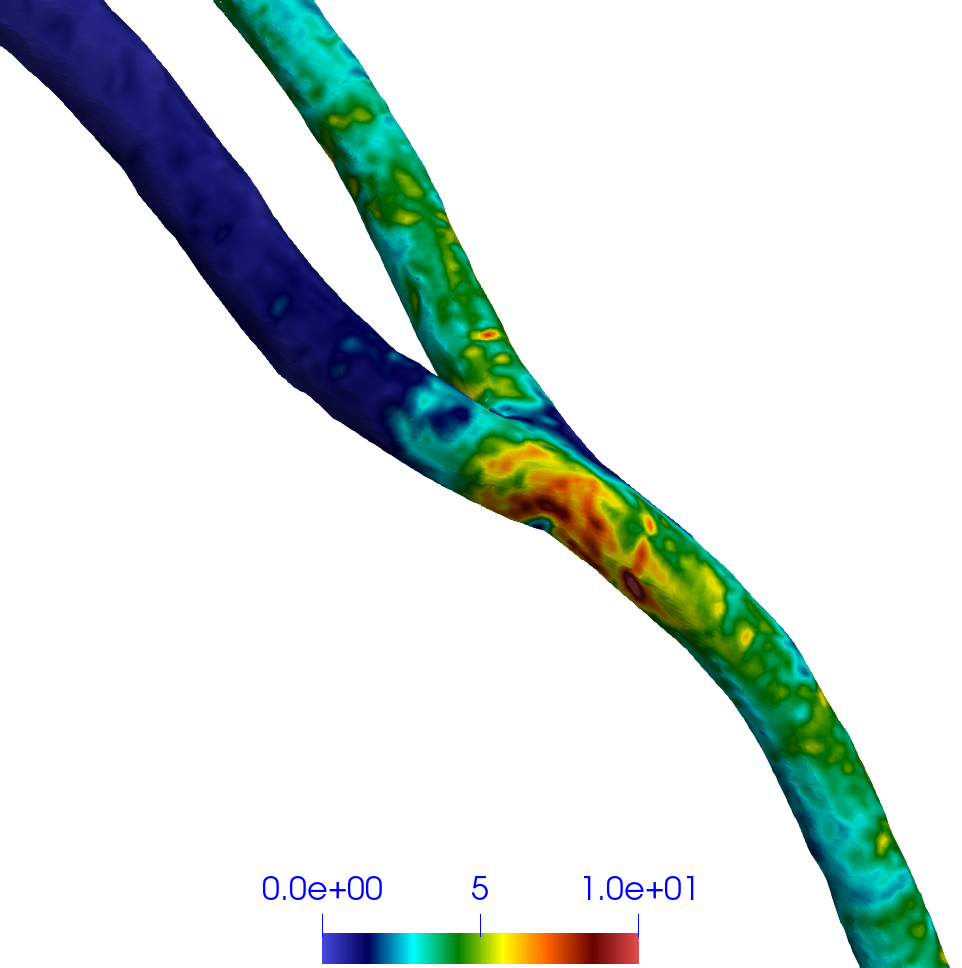}
			\put(35,105){FOM }
	\end{overpic}}

	\caption{
	Case 2: comparison between WSS (Pa) distribution
	in the anastomosis region computed by the FOM and by the ROM at $t/T = 0.8$ for the test point ($70\%$ stenosis).
	}
	\label{wss_stenosi_graft}
\end{figure}

\begin{figure}
	\centering
	\subfloat[][\label{wss_stenosi_graft_new:b}]{%
		\begin{overpic}[width=0.32\textwidth]{WSS_FOM_sten_70_graft}
			\put(18,105){FOM for $\bm \mu  = 70\%$ }
	\end{overpic}}
	\subfloat[][\label{wss_stenosi_graft_new:c}]{%
		\begin{overpic}[width=0.32\textwidth]{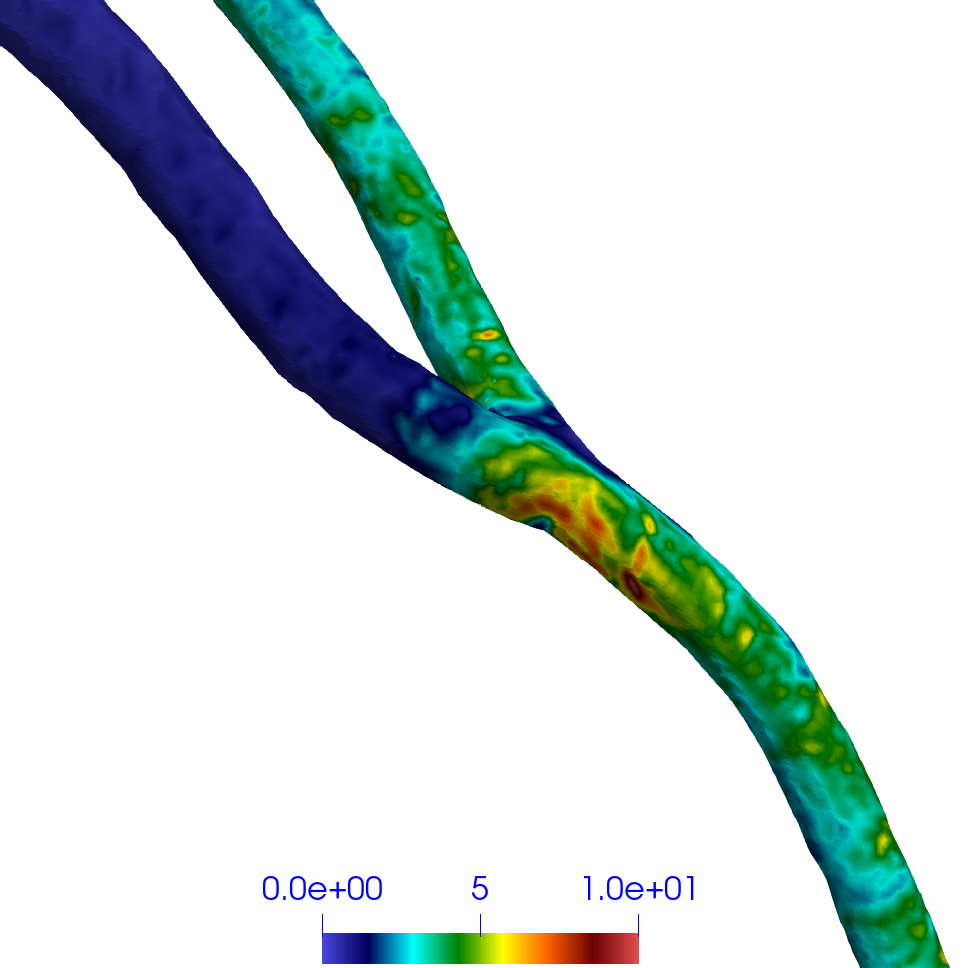}
			\put(18,105){FOM for $\bm \mu  = 60\%$}
	\end{overpic}}
	\subfloat[][\label{wss_stenosi_graft_new:d}]{%
		\begin{overpic}[width=0.32\textwidth]{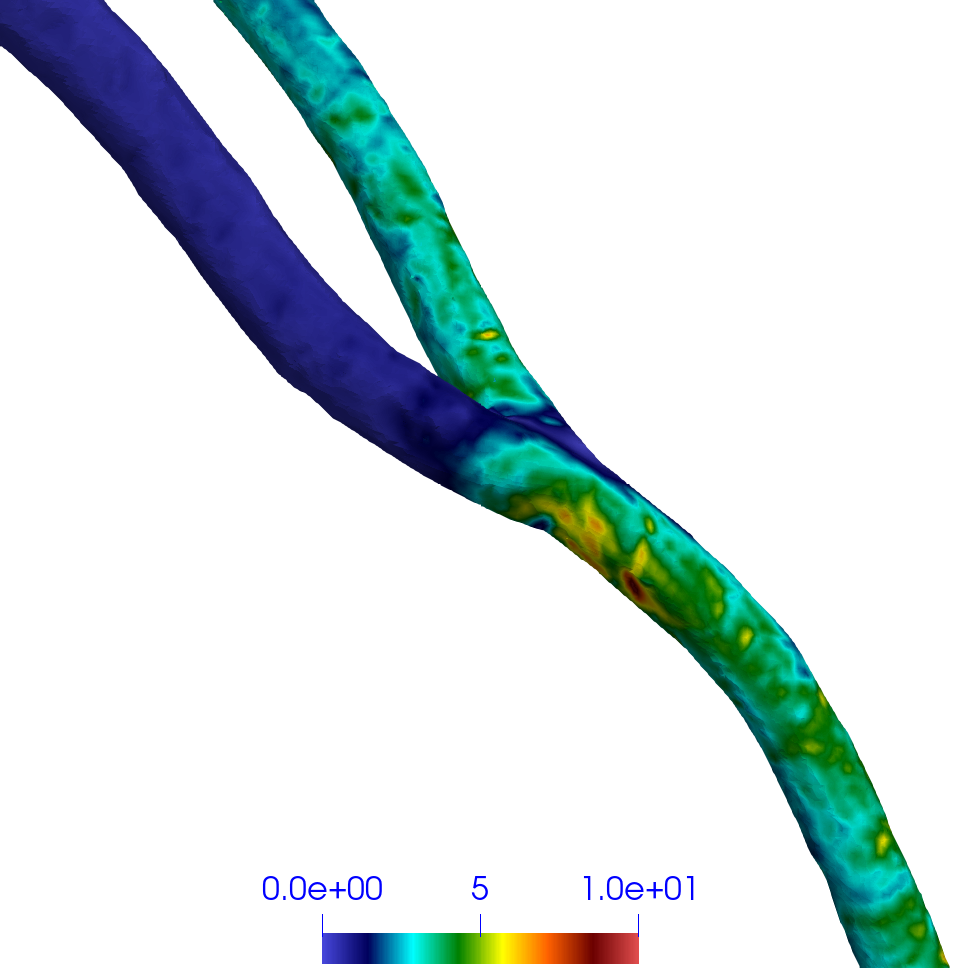}
			\put(18,105){FOM for $\bm \mu  = 50\%$}
	\end{overpic}}

	\caption{
	Case 2: comparison between WSS (Pa) distribution
	in the anastomosis region at $t/T = 0.8$ related to three different values of stenosis degree.}
	\label{wss_stenosi_graft_new}
\end{figure}

From Figures \ref{wss_stenosi} and \ref{wss_stenosi_graft} we can observe that 
FOM and ROM solutions are very similar for WSS.  
Furthermore, we observe that the WSS magnitude rises as the severity of the stenosis increases both in the stenosis (Figure \ref{wss_stenosi_new}) and the anastomosis (Figure \ref{wss_stenosi_graft_new}) regions. This could be due to the higher flow rate enforced on the LITA to compensate the lack of blood in the LMCA.


\begin{figure}
	\centering
	\subfloat[][\label{u_stenosi_graft:a}]{%
		\begin{overpic}[width=0.32\textwidth]{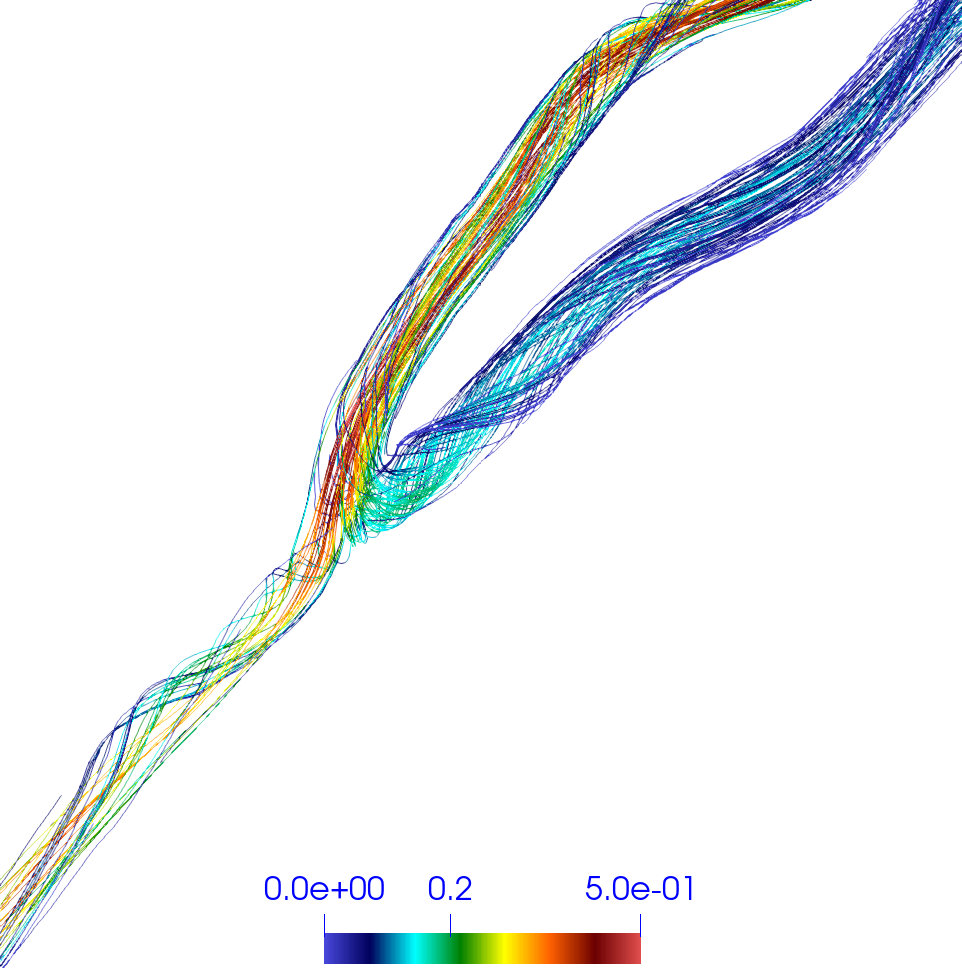}
			\put(50,108){ROM}
	\end{overpic}}
	\subfloat[][\label{u_stenosi_graft:b}]{%
		\begin{overpic}[width=0.32\textwidth]{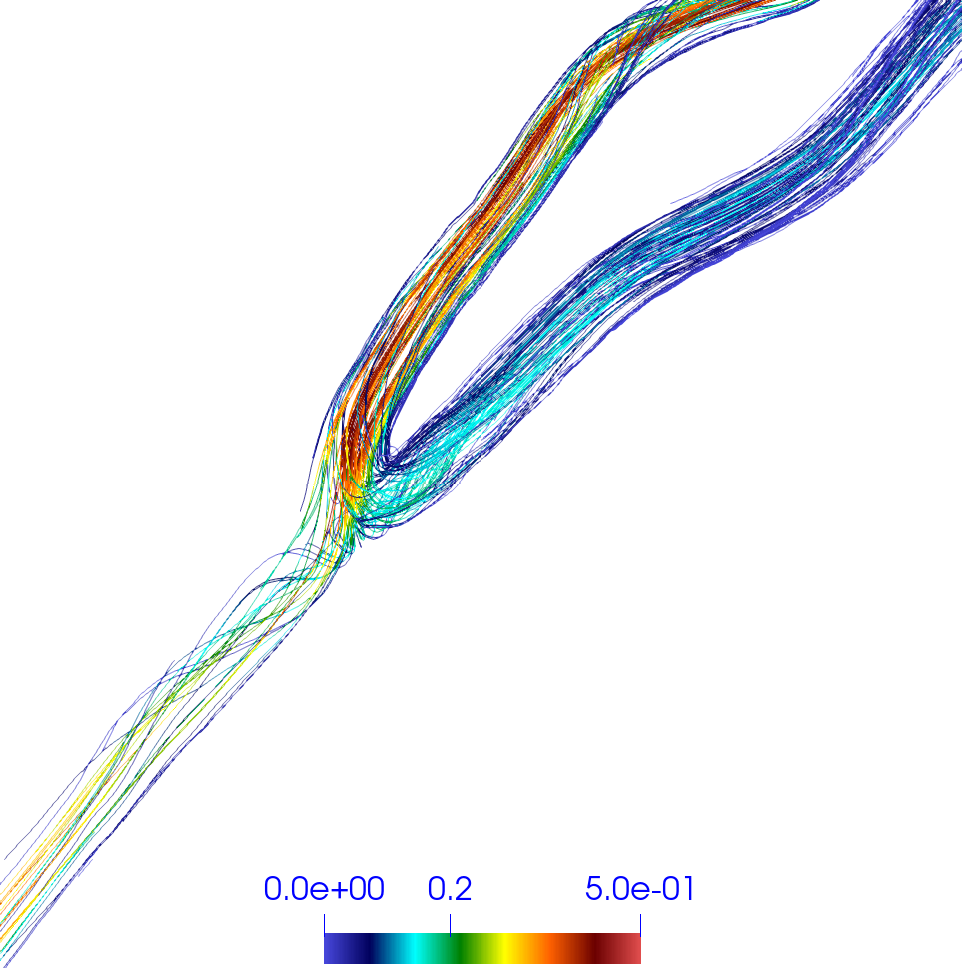}
			\put(50,108){FOM }
	\end{overpic}}

	\caption{
	Case 2: comparison between velocity (m/s) streamilines
	in the anastomosis region computed by the FOM and by the ROM at $t/T = 0.8$ for the test point ($70\%$ stenosis).}
	\label{u_stenosi_graft}
\end{figure}

\begin{figure}
	\centering
	\subfloat[][\label{u_stenosi_graft:b}]{%
	\begin{overpic}[width=0.32\textwidth]{U_FOM_70_graft}
			\put(25,108){FOM  for $\bm \mu = 70\%$}
	\end{overpic}}
 	\subfloat[][\label{u_stenosi_graft_new:c}]{%
 		\begin{overpic}[width=0.32\textwidth]{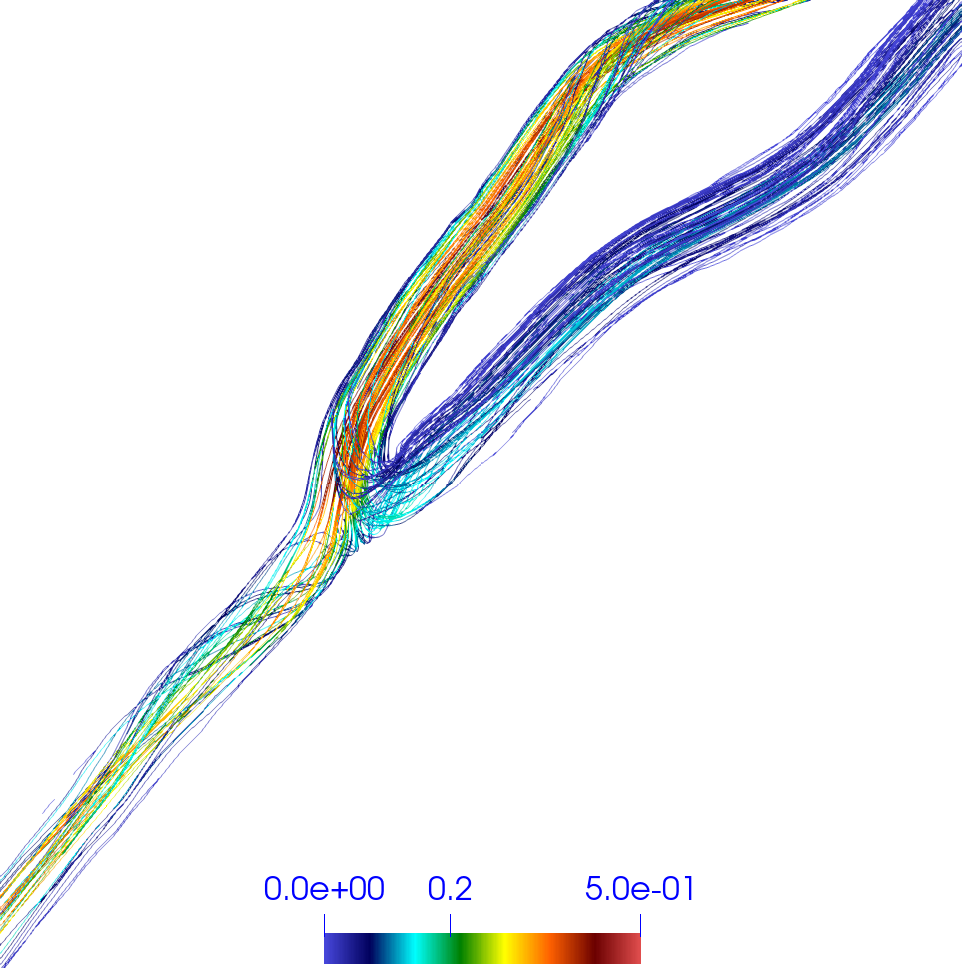}
 			\put(25,108){FOM for $\bm \mu = 60\%$}
 	\end{overpic}}
 	\subfloat[][\label{u_stenosi_graft_new:d}]{%
		\begin{overpic}[width=0.32\textwidth]{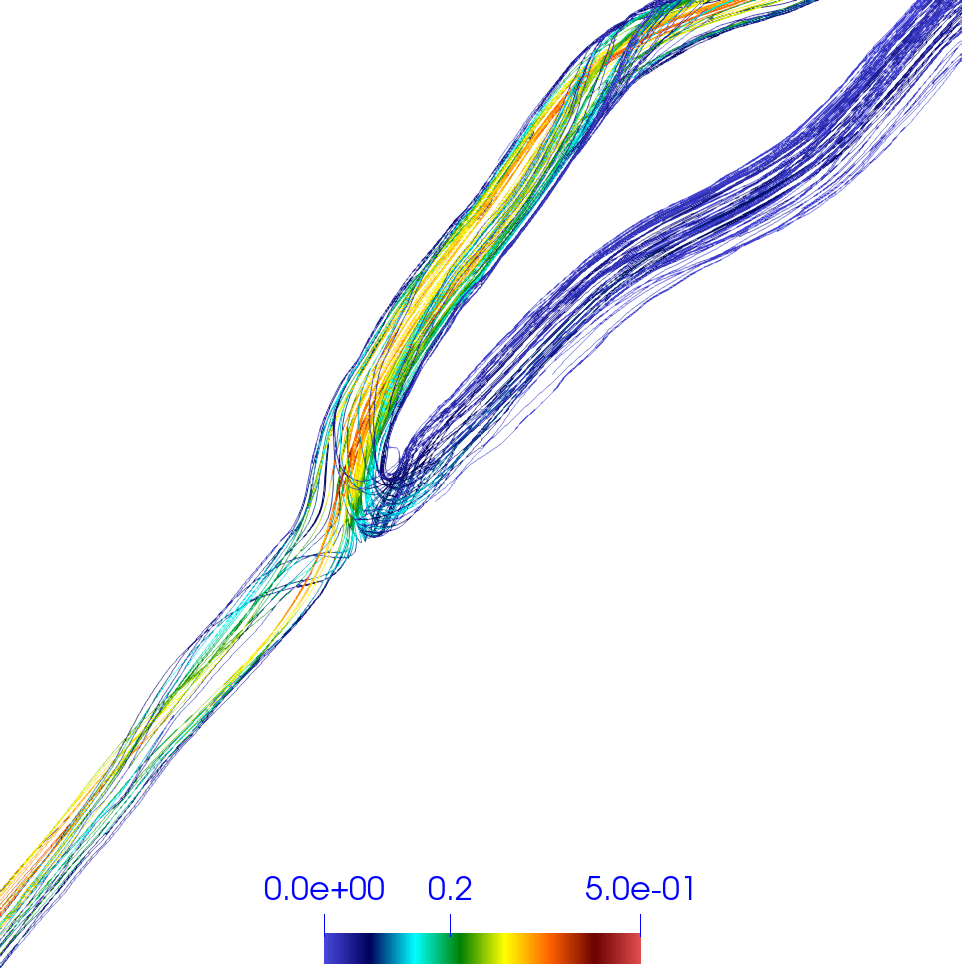}
 			\put(25,108){FOM  for $\bm \mu = 50\%$}
 	\end{overpic}}

	\caption{
    Case 2: comparison between velocity (m/s) streamlines
	in the anastomosis region at $t/T = 0.8$ related to three different values of stenosis degree.}
	\label{u_stenosi_graft_new}
\end{figure}

In Figure \ref{u_stenosi_graft} the FOM and ROM streamlines for the velocity field are depicted. We can appreciate a good matching between the two solutions. In Figure \ref{u_stenosi_graft_new} we observe that, as for the WSS, the velocity increases with the stenosis. 
The velocity is higher in the LITA because it supplies blood to the entire vessels network, oxygenating LAD and, going up this vessel, LCx too.

\subsection{Computational cost}
We ran the FOM simulations in parallel using 20 processor cores. The simulations are run on the
SISSA HPC cluster Ulysses (200 TFLOPS, 2TB RAM, 7000 cores). Each FOM simulation takes roughly 41 h in terms of wall time, or 820 h in terms of total CPU time (i.e., wall time multiplied by the number of cores). On the other hand, the ROM has been run on an Intel(R) Core(TM) i5-8265U CPU @ 1.60GHz 8GB RAM by using one processor core only. Our ROM approach takes less then 10 s for the computation of the reduced coefficients for each variable. So we obtain a speed-up of at least $10^{5}$.
   In Table \ref{time_stenosi} we can find a more detailed description of the estimation of the time required for the online  phase,  for  the  SVD  analysis and  for  the  training  of  the  ANNs for each variable. Notice that for the WSS we obtain the highest speed-up because it is related to only the boundary of the domain. 
\begin{table}
\centering
\caption{Time required for offline/online stages.}
\begin{tabular}{|c|c|c|c|c|c|c|}
\hline
 & $t_{\text{online}}[s]$  & $t_{\text{SVD}}[s]$ & $t_{\text{training}}[s]$ & Speedup & $t_{\text{FOM}}[s]$  \\
\hline
P  & 5.43       & 2371.64 & 13182.88 & 5.42e5 & 
\\
\cline{1-5}
$\mathbf{u}$ & 9.42       & 2827.82 & 24545.71 & 3.12e5 & 147048 $\times$ 20 \\
\cline{1-5}
WSS & 1.69 & 10.74 & 25118.62 & 1.73e6 & \\
\hline
\end{tabular}
\label{time_stenosi}
\end{table}

\section{Conclusions and perspectives}
\label{sec:4}
In this work a machine learning-based reduced order modelling strategy is employed in order to investigate the hemodynamics in a CABG patient-specific configuration when an
isolated stenosis of the LMCA occurs. 

The use of artificial intelligence as a fast and accurate method for the development of ROM for CFD simulations in cardiovascular applications is a fast-growing research area, so we retain that this work can represent a further step forward in this direction. 
Furthermore, the choice of the FV method for the space discretization reveals a wide applicability and flexibility. 


After a computationally intensive offline stage, POD-ANN method has allowed to obtain accurate hemodynamic simulations of the problem at hand at a significantly reduced computational cost. This demonstrates that such technique would be able in perspective to allow real time simulations to be accessed in hospitals and
   operating rooms in a very efficient way.

However, several improvements are still feasible. It could be introduced a coupling with 0D models \cite{Infantino,Infantino1} at the aim to enforce more realistic boundary conditions, which represents a crucial step to obtain meaningful outcomes. 
In addition, it could be interesting to investigate the performance of ROM frameworks based on other deep learning approaches with the aim to furthermore improve both efficiency and accuracy of the method, such as physics-informed neural network~\cite{Chen,Hesthaven}.
\newenvironment{acknowledgements}%
    {\null\vfill\begin{center}%
    \bfseries Acknowledgements\end{center}}%
    {\vfill\null}
        \begin{acknowledgements}
We acknowledge the support provided by the European Research Council Executive Agency by the Consolidator Grant project AROMA-CFD "Advanced Reduced Order Methods with Applications in Computational Fluid Dynamics" - GA 681447, H2020-ERC CoG 2015 AROMA-CFD, the European Union's Horizon 2020 research and innovation program under the Marie Skłodowska-Curie Actions, grant agreement 872442 (ARIA), as well as developers and contributors of \openfoam.
\end{acknowledgements}

%
%



\cleardoublepage\thispagestyle{empty}

\end{document}